\documentclass[longauth]{aa}

\usepackage{graphicx}
\usepackage{txfonts}
\usepackage{hyperref}

\usepackage{tabularx}
\DeclareMathOperator{\e}{e}

\newcommand{\be}{\begin{equation}}
\newcommand{\ee}{\end{equation}}

\newcommand*\samethanks[1][\value{footnote}]{\footnotemark[#1]}

 {\everymath{\displaystyle\everymath{}}\array}%
 {\endarray}

\begin{document}

\title{Six transiting planets and a chain of Laplace resonances in TOI-178}
\subtitle{ }
\titlerunning{}

\author{
A. Leleu\thanks{CHEOPS Fellow}$^{1,2}$, 
Y. Alibert$^{2}$, 
N. C. Hara\samethanks$^{1}$, 
M. J. Hooton$^{2}$, 
T. G. Wilson$^{3}$, 
P. Robutel$^{4}$, 
J.-B. Delisle$^{1}$, 
J. Laskar$^{4}$, 
S. Hoyer$^{5}$, 
C. Lovis$^{1}$, 
E. M. Bryant$^{6,7}$
% ...
}
\authorrunning{A. Leleu et al}

\institute{
$^{1}$ Observatoire Astronomique de l'Universit\'e de Gen\`eve, Chemin Pegasi 51, Versoix, Switzerland\\
$^{2}$ Physikalisches Institut, University of Bern, Gesellsschaftstrasse 6, 3012 Bern, Switzerland\\
$^{3}$ Centre for Exoplanet Science, SUPA School of Physics and Astronomy, University of St Andrews, North Haugh, St Andrews KY16 9SS, UK\\
$^{4}$ IMCCE, UMR8028 CNRS, Observatoire de Paris, PSL Univ., Sorbonne Univ., 77 av. Denfert-Rochereau, 75014 Paris, France\\
$^{5}$ Aix Marseille Univ, CNRS, CNES, LAM, Marseille, France\\
$^{6}$ Department of Physics, University of Warwick, Gibbet Hill Road, Coventry CV4 7AL, UK\\
$^{7}$ Centre for Exoplanets and Habitability, University of Warwick, Gibbet Hill Road, Coventry CV4 7AL, UK\\
$^{8}$ Astrobiology Research Unit, Universit\'e de Li\`ege, All\'ee du 6 Ao\^ut 19C, B-4000 Li\`ege, Belgium\\
$^{9}$ Institute of Planetary Research, German Aerospace Center (DLR), Rutherfordstrasse 2, 12489 Berlin, Germany\\
$^{10}$ Space sciences, Technologies and Astrophysics Research (STAR) Institute, Universit{\'e} de Li{\`e}ge, All{\'e}e du 6 Ao{\^u}t 19C, 4000 Li{\`e}ge, Belgium\\
$^{11}$ School of Physics and Astronomy, University of Leicester, Leicester LE1 7RH, UK\\
$^{12}$ Instituto de Astrof\'isica e Ci\^encias do Espa\c{c}o, Universidade do Porto, CAUP, Rua das Estrelas, 4150-762 Porto, Portugal\\
$^{13}$ Centro de Astrof\'{\i}sica da Universidade do Porto, Rua das Estrelas, 4150-762 Porto, Portugal\\
$^{14}$ Departamento de F\'isica e Astronomia, Faculdade de Ci\^encias, Universidade do Porto, Rua do Campo Alegre, 4169-007 Porto, Portugal\\
$^{15}$ Instituto de Astrof\'\i sica de Canarias, 38200 La Laguna, Tenerife, Spain\\
$^{16}$ IF/00852/2015, and project PTDC/FIS-OUT/29048/2017.\\
$^{17}$ Departamento de Astrof\'\i sica, Universidad de La Laguna, 38206 La Laguna, Tenerife, Spain\\
$^{18}$ Camino El Observatorio 1515, Las Condes, Santiago, Chile\\
$^{19}$ ETH Z\"urich, Institute for Particle Physics and Astrophysics\\
$^{20}$ Institut de Ci\`encies de l'Espai (ICE, CSIC), Campus UAB, Can Magrans s/n, 08193 Bellaterra, Spain\\
$^{21}$ Institut d'Estudis Espacials de Catalunya (IEEC), 08034 Barcelona, Spain\\
$^{22}$ ESTEC, European Space Agency, 2201AZ, Noordwijk, NL\\
$^{23}$ Depto. de Astrofísica, Centro de Astrobiologia (CSIC-INTA), ESAC campus, 28692 Villanueva de la Cãda (Madrid), Spain\\
$^{24}$ Space Research Institute, Austrian Academy of Sciences, Schmiedlstrasse 6, A-8042 Graz, Austria\\
$^{25}$ Center for Space and Habitability, Gesellsschaftstrasse 6, 3012 Bern, Switzerland\\
$^{26}$ Université Grenoble Alpes, CNRS, IPAG, 38000 Grenoble, France\\
$^{27}$ Department of Astronomy, Stockholm University, AlbaNova University Center, 10691 Stockholm, Sweden\\
$^{28}$ Institute of Optical Sensor Systems, German Aerospace Center (DLR), Rutherfordstrasse 2, 12489 Berlin, Germany\\
$^{29}$ Department of Earth, Atmospheric and Planetary Science, Massachusetts Institute of Technology, 77 Massachusetts Avenue, Cambridge, MA 02139, USA\\
$^{30}$ Admatis, Miskok, Hungary\\
$^{31}$ Université de Paris, Institut de physique du globe de Paris, CNRS, F-75005 Paris, France\\
$^{32}$ CFisUC, Department of Physics, University of Coimbra, 3004-516 Coimbra, Portugal\\
$^{33}$ INAF - Osservatorio Astronomico di Trieste, via G. B. Tiepolo 11, I-34143 Trieste, Italy\\
$^{34}$ INAF - Osservatorio Astrofisico di Torino, via Osservatorio 20, 10025 Pino Torinese, Italy\\
$^{35}$ Lund Observatory, Dept. of Astronomy and Theoretical Physics, Lund University, Box 43, 22100 Lund, Sweden\\
$^{36}$ INAF, Istituto di Astrofisica e Planetologia Spaziali, via del Fosso del Cavaliere 100, 00133 Roma, Italy\\
$^{37}$ European Southern Observatory, Alonso de Co\'ordova 3107, Vitacura, Regio\'on Metropolitana, Chile\\
$^{38}$ Leiden Observatory, University of Leiden, PO Box 9513, 2300 RA Leiden, The Netherlands\\
$^{39}$ Department of Space, Earth and Environment, Chalmers University of Technology, Onsala Space Observatory, 43992 Onsala, Sweden\\
$^{40}$ Dipartimento di Fisica, Universit\`a degli Studi di Torino, via Pietro Giuria 1, I-10125, Torino, Italy\\
$^{41}$ Center for Astronomy and Astrophysics, Technical University Berlin, Hardenberstrasse 36, 10623 Berlin, Germany\\
$^{42}$ Astronomy Unit, Queen Mary University of London, Mile End Road, London E1 4NS, UK\\
$^{43}$ Cavendish Laboratory, JJ Thomson Avenue, Cambridge CB3 0HE, UK\\
$^{44}$ University of Vienna, Department of Astrophysics, Türkenschanzstrasse 17, 1180 Vienna, Austria\\
$^{45}$ Department of Physics and Kavli Institute for Astrophysics and Space Research, Massachusetts Institute of Technology, Cambridge, MA 02139, USA\\
$^{46}$ Departamento de Astronom\'ia, Universidad de Chile, Camino El Observatorio 1515, Las Condes, Santiago, Chile\\
$^{47}$ Centro de Astrof\'isica y Tecnolog\'ias Afines (CATA), Casilla 36-D, Santiago, Chile\\
$^{48}$ Facultad de Ingenier\'ia y Ciencias, Universidad Adolfo Ib\'{a}\~{n}ez, Av.\ Diagonal las Torres 2640, Pe\~{n}alol\'{e}n, Santiago, Chile\\
$^{49}$ Millennium Institute for Astrophysics, Chile\\
$^{50}$ Konkoly Observatory, Research Centre for Astronomy and Earth Sciences, 1121 Budapest, Konkoly Thege Miklós út 15-17, Hungary\\
$^{51}$ Brorfelde Observatory, Observator Gyldenkernes Vej 7, DK-4340 T\o{}ll\o{}se, Denmark\\
$^{52}$ DTU Space, National Space Institute, Technical University of Denmark, Elektrovej 327, DK-2800 Lyngby, Denmark\\
$^{53}$ Institut d'astrophysique de Paris, UMR7095 CNRS, Université Pierre \& Marie Curie, 98bis blvd. Arago, 75014 Paris, France\\
$^{54}$ INAF, Osservatorio Astronomico di Padova, Vicolo dell'Osservatorio 5, 35122 Padova, Italy\\
$^{55}$ Astrophysics Group, Keele University, Staffordshire, ST5 5BG, United Kingdom\\
$^{56}$ Department of Physics, University of Warwick, Coventry, UK\\
$^{57}$ INAF - Osservatorio Astronomico di Palermo, Piazza del Parlamento 1, 90134, Palermo, Italy\\
$^{58}$ IFPU, Via Beirut 2, 34151 Grignano Trieste\\
$^{59}$ Instituto de Astronom\'ia, Universidad Cat\'olica del Norte, Angamos 0610, Antofagasta, Chile\\
$^{60}$ Instituto de Astrof\'isica e Ci\^encias do Espa\c{c}o, Faculdade de Ci\^encias da Universidade de Lisboa, Campo Grande, PT1749-016 Lisboa, Portugal\\
$^{61}$ Department of Astrophysics, University of Vienna, Tuerkenschanzstrasse 17, 1180 Vienna, Austria\\
$^{62}$ INAF, Osservatorio Astrofisico di Catania, Via S. Sofia 78, 95123 Catania, Italy\\
$^{63}$ Dipartimento di Fisica e Astronomia "Galileo Galilei", Università degli Studi di Padova, Vicolo dell'Osservatorio 3, 35122 Padova, Italy\\
$^{64}$ Space Science Data Center, ASI, via del Politecnico snc, 00133 Roma, Italy\\
$^{65}$ Department of Physics, University of Warwick, Gibbet Hill Road, Coventry CV4 7AL, United Kingdom\\
$^{66}$ Fundaci\'on G. Galilei -- INAF (Telescopio Nazionale Galileo), Rambla J. A. Fern\'andez P\'erez 7, E-38712 Bre\~na Baja, La Palma, Spain\\
$^{67}$ INAF - Osservatorio Astronomico di Brera, Via E. Bianchi 46, I-23807 Merate, Italy\\
$^{68}$ Institut für Geologische Wissenschaften, Freie Universität Berlin, 12249 Berlin, Germany\\
$^{69}$ Paris Observatory, LUTh UMR 8102, 92190 Meudon, France\\
$^{70}$ School of Physics \& Astronomy, University of Birmingham, Edgbaston, Birmingham, B15 2TT, UK\\
$^{71}$ ELTE Eötvös Loránd University, Gothard Astrophysical Observatory, 9700 Szombathely, Szent Imre h. u. 112, Hungary\\
$^{72}$ MTA-ELTE Exoplanet Research Group, 9700 Szombathely, Szent Imre h. u. 112, Hungary\\
$^{73}$ INAF, Osservatorio Astronomico di Roma, via Frascati 33, 00078 Monte Porzio Catone, Roma, Italy\\
$^{74}$ Institute of Astronomy, University of Cambridge, Madingley Road, Cambridge, CB3 0HA, United Kingdom\\
$^{75}$ Centro de Astrobiología (CSIC-INTA), Crta. Ajalvir km 4, E-28850 Torrejón de Ardoz, Madrid, Spain\\
}

\abstract
{
Determining the architecture of multi-planetary systems is one of the cornerstones of understanding planet formation and evolution. Resonant systems are especially important as the fragility of their orbital configuration ensures that no significant scattering or collisional event has taken place since the earliest formation phase when the parent protoplanetary disc was still present. In this context, TOI-178 has been the subject of particular attention since the first {\it TESS} observations hinted at the possible presence of a near 2:3:3 resonant chain. Here we report the results of observations from {\it CHEOPS}, ESPRESSO, NGTS, and SPECULOOS with the aim of deciphering the peculiar orbital architecture of the system. We show that TOI-178 harbours at least six planets in the super-Earth to mini-Neptune regimes, with radii ranging from $1.152_{-0.070}^{+0.073}$ to $2.87_{-0.13}^{+0.14}$ Earth radii and periods of 1.91, 3.24, 6.56, 9.96, 15.23, and 20.71\,days. All planets but the innermost one form a 2:4:6:9:12 chain of Laplace resonances, and the planetary densities show important variations from planet to planet, jumping from $1.02^{+0.28}_{-0.23}$ to $0.177^{+0.055}_{-0.061}$ times the Earth's density between planets $c$ and $d$. Using Bayesian interior structure retrieval models, we show that the amount of gas in the planets does not vary in a monotonous way, contrary to what one would expect from simple formation and evolution models and unlike other known systems in a chain of Laplace resonances. The brightness of TOI-178 (H=8.76 mag, J=9.37 mag, V=11.95 mag) allows for a precise characterisation of its orbital architecture as well as of the physical nature of the six presently known transiting planets it harbours. The peculiar orbital configuration and the diversity in average density among the planets in the system will enable the study of interior planetary structures and atmospheric evolution, providing important clues on the formation of super-Earths and mini-Neptunes.
}


\maketitle

\section{Introduction}

Since the discovery of the first exoplanet orbiting a Sun-like star by \cite{MaQue1995}, the diversity of observed planetary systems has continued to challenge our understanding of their formation and evolution. As an ongoing effort to understand these physical processes, observational facilities strive to get a full picture of exoplanetary systems by looking for additional candidates to known systems and by better constraining the orbital architecture, radii, and masses of the known planets.

In particular, chains of planets in mean-motion resonances  (MMRs) are `Rosetta Stones' of the formation and evolution of planetary systems. Indeed, our current understanding of planetary system formation theory implies that such configurations are a common outcome of protoplanetary discs: Slow convergent migration of a pair of planets in quasi-circular orbits leads to a high probability of capture in first-order MMRs -- the period ratio of the two planets is equal to $(k+1)/k$, with $k$ an integer \citep{LePe2002,CoDeLa2018}. As the disc strongly damps the eccentricities of the protoplanets, this mechanism can repeat itself, trapping the planets in a chain of MMRs and leading to very closely packed configurations \citep{Kepler11}. However, resonant configurations are not the most common orbital arrangements \citep{Fabrycky2014}. As the protoplanetary disc dissipates, the eccentricity damping lessens, which can lead to instabilities in packed systems \citep[see, for example,][]{TePa2007,PuWu2015,Izidoro2017}. 

For planets that remained in resonance and are close enough to their host stars, tides become the dominant force that affects the architecture of the systems, which can then lead to a departure of the period ratio from the exact resonance \citep{HenLe1983,PaTe2010,DeLaCoBo2012}.
In some near-resonant systems, such as HD\,158259,  the tides seem to have pulled the configuration entirely out of resonance \citep{Hara2020}. However, through gentle tidal evolution it is possible to retain a resonant state even with null eccentricities through three-body resonances \citep{Morbidelli02,Papaloizou2015}. Such systems are too fine-tuned to result from scattering events and hence can be used to constrain the outcome of protoplanetary discs \citep{Mills16}.

A Laplace resonance, in reference to the configuration of Io, Europa, and Ganymede, is a three-body resonance where each consecutive pair of bodies is in, or close to, two planet MMRs. To date, only a few systems have been observed in a chain of Laplace resonances: GJ 876 \citep{Rivera2010}, Kepler-60 \citep{Gozdziewski2016}, Kepler-80 \citep{MacDonald2016}, Kepler-223 \citep{Mills16}, 
Trappist-1 \citep{Gillon2017,LugerTrappist1-h}, and K2-138 \citep{Christiansen2018,Lopez2019}. Out of these six systems, only K2-138 has so far been observed by both transit and radial velocity (RV), mainly due to the relative faintness of the other host stars in the visible (V magnitudes greater than $\sim$14). However, TTVs could also be used to estimate the mass of their planets (see, for example, \cite{Agol2020} for the case of Trappist-1).


In this study, we present photometric and RV observations of  TOI-178, a  V = 11.95\,mag,  K-type  star that was first observed by {\it TESS}\footnote{Transiting Exoplanet Survey Satellite} in its Sector 2. We jointly analyse the photometric data of {\it TESS}, two nights of NGTS\footnote{Next Generation Transit Survey} and SPECULOOS\footnote{Search for habitable Planets EClipsing ULtra-cOOl Stars} data, and 285 hours of \textit{CHEOPS}\footnote{CHaracterising ExOPlanet Satellite} observations, along with 46 ESPRESSO\footnote{Echelle Spectrograph for Rocky Exoplanet and Stable Spectroscopic Observations} RV points. This follow-up effort allows us to decipher the architecture of the system and demonstrate the presence of a chain of Laplace resonances between the five outer planets. 

We begin in Sect. \ref{sec:rationale} by presenting the rationale that led to the {\it CHEOPS} observation sequence (two visits totalling  11\,d followed by two shorter visits). In Sect. \ref{sec:star}, we describe the parameters of the star. In Sect. \ref{sec:Data}, we present the photometric  and RV data we use in the paper. In Sect. \ref{sec:detections}, we show how these data led us to the detection of six planets -- the outer five  of which  are in a chain of Laplace resonances -- and to constrain their parameters. In Sect. \ref{sec:dynamics}, we explain the resonant state of the system, discuss its stability, and describe the transit timing variations (TTVs) that this system could potentially exhibit in the coming years. Finally, we discuss the internal structure of the planets in Sect. \ref{sec:internal_structure}, and conclusions are presented in Sect. \ref{sec:discussion}.

\section{CHEOPS observation strategy}
\label{sec:rationale}

\begin{figure*}
  \includegraphics[width =\textwidth]{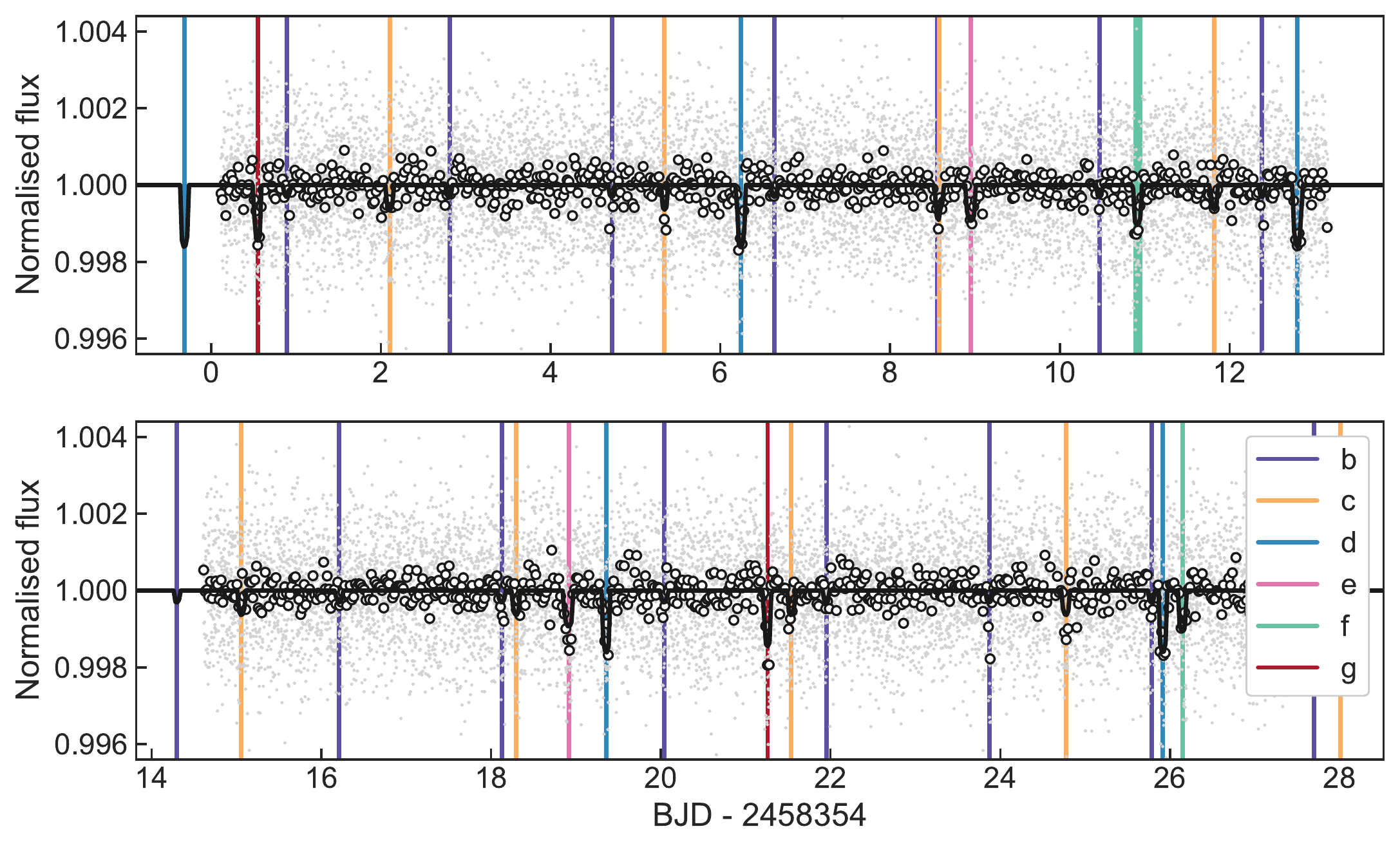}
  \caption{Light curves from TESS Sector 2 described in Sects. \ref{sec:rationale} and \ref{sec:TESS}. Unbinned data are shown as grey points, and data in 30-minute bins are shown as black circles. The best fitting transit model for the system is shown in black; the associated parameter values are shown in Tables \ref{table:TOI178bcd} and \ref{table:TOI178efg}. The positions of the transits are marked with lines coloured according to the legend. The photometry before and after the mid-sector gap are shown in the top and bottom panels, respectively. As the first transit of planet $f$ (thick teal line) occurred precisely between the two transits of the similarly sized planet $g$ (period of 20.71\,d - red lines), the three transits were originally thought to have arisen due to a single planet, which was originally designated TOI-178.02 with a period of 10.35\,d (see Sect. \ref{sec:rationale}).}
  \label{fig:tess_full}
\end{figure*}

The CHEOPS observation consisted of one long double visit (11 days) followed by two short visits (a few hours each) at precise dates. We explain in this section how we came up with this particular observation strategy. Details on all the data used and acquired, as well as their analysis, are presented in the following sections.

The first release of candidates from the {\it TESS} alerts of Sector 2 included three planet candidates in TOI-178 with periods of $6.55$\,d, $10.35$\,d, and $9.96$\,d. Based on these data, TOI-178 was identified as a potential co-orbital system \citep{Leleu2019} with two planets oscillating around the same period. This prompted ESPRESSO RV measurements and two sequences of simultaneous ground-based photometric observations with NGTS and SPECULOOS. From the latter, no transit was observed for TOI-178.02 ($P=10.35$\,d) in September 2019; however, a transit ascribed to this candidate was detected one month later by NGTS and SPECULOOS. The abovementioned absence of a TOI-178.02 transit combined with the three transits observed by {\it TESS} at high S/N (above 10) was interpreted as an additional sign of the strong TTVs expected in a co-orbital configuration. This solution was supported by RV data that were consistent with the horseshoe orbits of objects with similar masses \citep{LeRoCo2015}.

A continuous 11\,d {\it CHEOPS} observation (split into two visits for scheduling reasons) was therefore performed in August 2020 in order to confirm the orbital configuration of the system; as the instantaneous period of both members of the co-orbital pair will always be smaller than 11\,d, at least one transit of both targets should thus be detectable. Analysis of this light curve led to the confirmation of the presence of two of the planets already discovered by {\it TESS} (in this study denoted as planets $d$ and $e$, with periods of 6.55\,d and 9.96\,d, respectively) and the detection of two new inner transiting planets (denoted planets $b$ and $c$, with periods of 1.9\,d and 3.2\,d, respectively). However, one of the planets belonging to the proposed co-orbital pair (with a period of 10.35\,d) was not apparent in the light curve. A potential hypothesis was that the first and third transit of the TOI-178.02 candidate ($P=10.35$\,d) during {\it TESS} Sector 2 belonged to a planet of twice the period (20.7\,d), while the second transit belonged to another planet of unknown period. This scenario was supported by the two ground-based observations mentioned above since a $P=20.7$\,d planet would not transit during the September 2019 observation window. Using the ephemerides from fitting the TESS, NGTS, and SPECULOOS data, we predicted the mid-transit of the 20.7-day candidate to occur between UTC 14:06 and 14:23 on September 7, 2020 with 98\% certainty. A third visit of {\it CHEOPS} observed the system around this epoch for 13.36\,h and confirmed the presence of this new planet ($g$) at the predicted time with consistent transit parameters. Further analysis of all available photometric data found two possibilities for the unknown period of the additional planet: $\sim~12.9\,$d or $\sim~15.24\,$d.

Careful analysis of the whole system additionally revealed that planets $c$, $d$, $e$, and $g$ were in a Laplace resonance. In order to fit in the resonant chain, the unknown period of the additional planet could have only two values: $P = 13.4527\,$d or $P = 15.2318\,$d (see Appendix \ref{ap:Px}, Fig. \ref{fig:Pgineq}), the latter value being more consistent with the RV data. A fourth {\it CHEOPS} visit was therefore scheduled for 3 October, and it detected the transit for the additional planet at a period of $15.231915_{-0.000095}^{+0.000115}$ day.

As we will detail in the following sections, we can confirm the detection of a 6.56-day and a 9.96-day period planet by {\it TESS} using the new observations presented in this paper. Furthermore, we can announce the detection of planets with periods of 1.91, 3.24, 15.23, and 20.71\,d (see Tables \ref{table:TOI178bcd} and \ref{table:TOI178efg} for the complete parameters of the system).

\begin{figure}
  \includegraphics[width = 9cm]{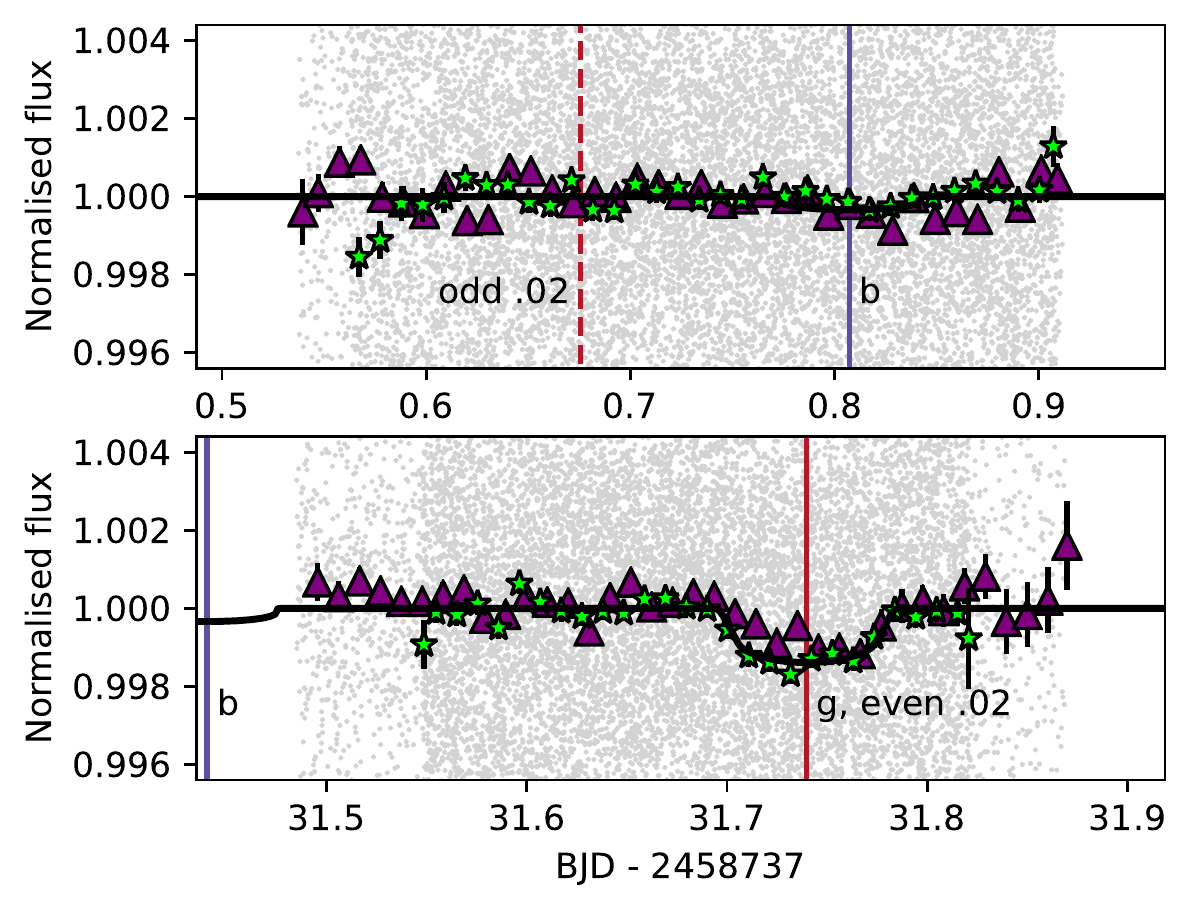}
  \caption{Light curves from simultaneous observations of TOI-178 by NGTS (green stars) and SPECULOOS-South (purple triangles), described in Sects. \ref{sec:NGTS} and \ref{sec:SPECULOOS}, respectively. Unbinned data are shown as grey points, and data in 15-minute bins are shown as green stars (NGTS) and purple triangles (SPECULOOS). The observations occurred on September 11, 2019 (top panel) and October 12, 2019 (bottom panel). The transit model is shown in black. The position of the odd transit of candidate TOI-178.02 is shown with a dashed red line, the transit of planet $g$ (which corresponds to an even transit of TOI-178.02) is shown with the solid red line, and the transits of planet $b$ are shown with purple lines.}
  \label{fig:NGTS+SPEC}
\end{figure}

\section{TOI-178 stellar characterisation}
\label{sec:star}

\begin{table}
\caption{Stellar properties of TOI-178, including the methods used to derive them.}
\label{tab:stellarParam}      
\centering                          
\begin{tabular}{lll}        
\hline\hline                 
\multicolumn{3}{c}{TOI-178} \\    
\hline                        
2MASS & \multicolumn{2}{l}{J00291228-3027133} \\
Gaia DR2 & \multicolumn{2}{l}{2318295979126499200} \\
TIC & \multicolumn{2}{l}{251848941} \\
TYC & \multicolumn{2}{l}{6991-00475-1} \\ 
\hline
Parameter & Value & Note \\
\hline
   $\alpha$ [J2000] & 00$^{h}$29$^{m}$12.30$^{s}$ & 1 \\
   $\delta$ [J2000] & -30$^{\circ}$27$^{'}$13.46$^{\arcsec}$ & 1 \\
   $\mu_{\alpha}$ [mas/yr] & 149.95$\pm$0.07 & 1 \\
   $\mu_{\delta}$ [mas/yr] & -87.25$\pm$0.04 & 1 \\
   $\varpi$ [mas] & 15.92$\pm$0.05 & 1 \\
   RV [km~s$^{-1}$] & 57.4$\pm$0.5 & 1 \\
\hline
   $V$ [mag] & 11.95 & 2 \\
   $G$ [mag] & 11.15 & 1 \\
   $J$ [mag] & 9.37 & 3 \\
   $H$ [mag] & 8.76 & 3 \\
   $K$ [mag] & 8.66 & 3 \\
   $W1$ [mag] & 8.57 & 4 \\
   $W2$ [mag] & 8.64 & 4 \\
\hline
   $T_{\mathrm{eff}}$ [K] & 4316$\pm$70 & spectroscopy \\
   $\log{g}$ [cgs]      & 4.45$\pm$0.15 & spectroscopy \\\relax
   [Fe/H] [dex] & -0.23$\pm$0.05 & spectroscopy \\\relax
   $v \sin i_\star$ [km~s$^{-1}$] & 1.5$\pm$0.3 & spectroscopy \\\relax
   $R_{\star}$ [$R_{\odot}$] & 0.651$\pm$0.011 & IRFM \\
   $M_{\star}$ [$M_{\odot}$] & 0.650$_{-0.029}^{+0.027}$ & isochrones \\
   $t_{\star}$ [Gyr]        & 7.1$_{-5.3}^{+6.1}$ & isochrones \\
   $L_{\star}$ [$L_{\odot}$] & 0.132$\pm$0.010 & from $R_{\star}$ and $T_{\mathrm{eff}}$\\
   $\rho_{\star}$ [$\rho_\odot$] & 2.35$\pm$0.17 & from $R_{\star}$ and $M_{\star}$ \\
\hline                                   
\end{tabular}
\tablefoot{
[1] \cite{GaiaCollaboration2018}, [2] \cite{Hog2000}, [3] \cite{Skrutskie2006}, [4] \cite{Wright2010}
}
\end{table}

Forty-six ESPRESSO observations (see Sect. \ref{sec:RVdata}) of TOI-178, a V = 11.95\,mag K-dwarf, have been used to determine the stellar spectral parameters. These observations were first shifted and stacked to produce a combined spectrum. We then used the publicly available spectral analysis package \href{http://www.stsci.edu/~valenti/sme.html}{{ \tt{SME}}} \citep[Spectroscopy Made Easy;][]{ValentiPiskunov96, pv2017} version 5.22 to model the co-added ESPRESSO spectrum. We selected the ATLAS12 model atmosphere grids \citep{Kurucz2013} and atomic line data from \href{http://vald.astro.uu.se}{VALD} to compute synthetic spectra, which were fitted to the observations using a $\chi^2$-minimising procedure. We modelled different spectral lines to obtain different photospheric parameters starting with the line wings of H$\alpha,$ which are particularly sensitive to the stellar effective temperature $T_\mathrm{eff}$. We then proceeded with the metal abundances and the projected rotational velocity $v \sin i_\star$, which were modelled with narrow lines between 5900 and 6500~\AA. We found similar values for [Fe/H], [Ca/H], and [Na/H]. The macro-turbulent velocity was modelled and found to be 1.2$\pm$0.9~km~s$^{-1}$, and micro-turbulent velocity was fixed to 0.91~km~s$^{-1}$ following the formulation in \citealt{Bruntt2010}. The surface gravity $\log g$ was constrained from the line wings of the \ion{Ca}{I} triplet (6102, 6122, and 6162\,\AA) and the \ion{Ca}{I}  6439~\AA~line with a fixed $T_\mathrm{eff}$ and Ca abundance.

We checked our model with the \ion{Na}{I} doublet that is sensitive to both $T_\mathrm{eff}$ and $\log g$, and, finally, we also tested the MARCS 2012 \citep{Gustafsson08} model atmosphere grids. The measured parameters are listed in Table~\ref{tab:stellarParam}. The {\tt SME} results are in agreement with the empirical {\tt SpecMatch-Emp} \citep{2017ApJ...836...77Y} code, which fits the stellar optical spectra to a spectral library of 404 M5 to F1 stars, resulting in $T_\mathrm{eff} = 4316 \pm 70$~K, $\log g = 4.45 \pm 0.15$, and $\mathrm{[Fe/H]} = -0.29 \pm 0.05$ dex. We also used ARES+MOOG \citep{Sousa2014, Sousa2015, Sneden1973} to conduct the spectroscopic analysis on the same combined ESPRESSO spectra, and, although we derived consistent parameters ($T_\mathrm{eff} = 4500 \pm 230$~K, $\log{g} = 4.38 \pm 0.62$, and $\mathrm{[Fe/H]} = -0.34 \pm 0.10$), {the large errors are indicative of the difficulties in using equivalent width methods with colder stars: Spectral lines are more crowded in the spectrum.}

Using these precise spectral parameters as priors on stellar atmospheric model selection, we determined the radius of TOI-178 using the infrared flux method (IRFM; \citealt{Blackwell1977}) in a Markov chain Monte Carlo (MCMC) approach. {The IRFM computes the stellar angular diameter and effective temperature by comparing observed broadband optical and infrared fluxes and synthetic photometry obtained from convolution of the considered filter throughputs, using the known zero-point magnitudes, with the stellar atmospheric model, with the stellar radius then calculated using the parallax of the star. For this study, we retrieved the {\it Gaia} G, G$_{\rm BP}$, and G$_{\rm RP}$, 2MASS J, H, and K, and {\it WISE} W1 and W2 fluxes and relative uncertainties from the most recent data releases \citep[][respectively]{Skrutskie2006,Wright2010,GaiaCollaboration2018}, and utilised the stellar atmospheric models from the \textsc{atlas} Catalogues \citep{Castelli2003}, to obtain $R_{\star}=0.651\pm0.011\, R_{\odot}$, and $T_\mathrm{eff} = 4352\pm52 $~K, in agreement with the spectroscopic $T_\mathrm{eff}$ used as a prior.} 


{We inferred the mass and age of TOI-178 using stellar evolutionary models, using as inputs $T_{eff}$, $R_{\star}$ and Fe/H, with evolutionary tracks and isochrones generated by two grids of models separately, the PARSEC\footnote{Padova and Trieste Stellar Evolutionary Code\\ {\tt http://stev.oapd.inaf.it/cgi-bin/cmd}.} v1.2S code \citep{marigo17} and the CLES code \citep[Code Li\'egeois d'Évolution Stellaire;][]{scuflaire08}, with the reported values representing a careful combination of results from both sets of models.} This was done because the sets of models differ slightly in their approaches (reaction rates, opacity and overshoot treatment, and helium-to-metal enrichment ratio), and thus by comparing masses and ages derived from both grids it is possible to include systematic uncertainties within the modelling of the position of TOI-178 on evolutionary tracks and isochrones. A detailed discussion of combining the PARSEC and CLES models to determine masses and ages can be found in \cite{Bonfanti2021}. For this study, we derived $M_{\star}$ = 0.647$_{-0.032}^{+0.035}\,M_{\odot}$ and $t_{\star}$ = 7.1$_{-5.4}^{+6.2}$\,Gyr. All stellar parameters are shown in Table~\ref{tab:stellarParam}.

\section{Data}
\label{sec:Data}

\subsection{Photometric data}
\label{sec:LC}
\label{sec:LC_data}

In order to determine the orbital configuration of the TOI-178 planetary system, we obtained photometric time series observations from multiple telescopes, as detailed below.

\subsubsection{TESS}
\label{sec:TESS}

Listed as TIC 251848941 in the {\it TESS} Input Catalog (TIC; \citealt{Stassun2018,Stassun2019}), TOI-178 was observed by {\it TESS} in Sector 2, by camera 2, from August 22, 2018 to September 20, 2018. The individual frames were processed into 2-minute cadence observations and reduced by the Science Processing Operations Center (SPOC; \citealt{Jenkins2016}) into light curves made publicly available at the Mikulski Archive for Space Telescopes (MAST). For our analysis, we retrieved the Presearch Data Conditioning Single Aperture Photometry (PDCSAP) light curve data, using the default quality bitmask, which have undergone known systematic correction \citep{Smith2012,Stumpe2014}. Lastly, we rejected data points flagged as of bad quality by SPOC ({\tt QUALITY} > 0) and those with `Not-a-Number' flux or flux error values. After these quality cuts, the {\it TESS} light curve of TOI-178 contained 18,316 data points spanning 25.95\,d. The full dataset with the transits of the six identified planets is shown in Fig. \ref{fig:tess_full}.

\subsubsection{CHEOPS}
\label{sec:CHEOPS}

{\it CHEOPS}, the first ESA small-class mission, is dedicated to the observation of bright stars ($V\lesssim12$ mag) that are known to host planets and performs ultra high precision photometry, with the precision limited by stellar photon noise of 150\,ppm/min for a $V$\,=\,9 magnitude star. 
The {\it CHEOPS} instrument is composed of an f/8 Ritchey-Chretien on-axis telescope ($\sim$30\,cm diameter) equipped with a single frame-transfer back-side illuminated charge-coupled device (CCD) detector. The satellite was successfully launched from Kourou (French Guiana) into a $\sim$700\,km altitude Sun-synchronous orbit on December 18, 2019. {\it CHEOPS} took its first image on February 7, 2020, and, after it passed the in-orbit commissioning (IOC) phase, routine  operations started on March 25, 2020. More details on the mission can be found in \citet{Benz2020}, and the first results have recently been presented in \citet{Lendl2020}.

The versatility of the \textit{CHEOPS} mission allows for space-based follow-up photometry of planetary systems identified by the \textit{TESS} mission. This is particularly useful for completing the inventory of multi-planetary systems whose outer transiting planets have periods beyond $\sim$10 days. 

\begin{figure}
\includegraphics[width=9cm]{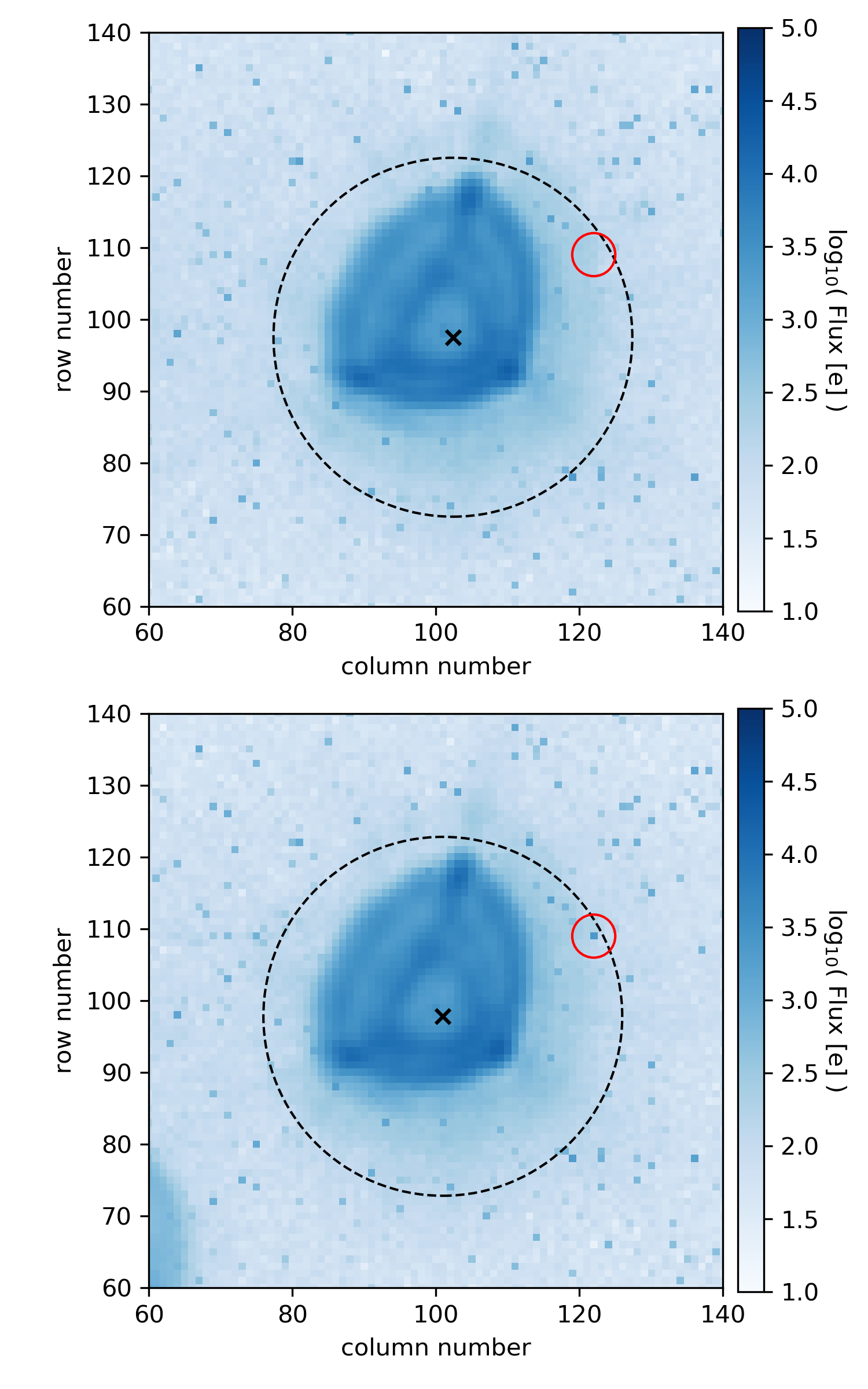}
        \caption{Extraction of 80$\times$80 arcsec of the {\it CHEOPS} field-of-view for two different data frames at the beginning (top) and the end (bottom) of the second visit (see Table \ref{tab:obs_log}). The TOI-178 PSF is shown at the centre, with the DEFAULT DRP photometric aperture represented by the dashed black circles. The telegraphic pixel location that appeared close to the end of the observation is marked by the red circle.  }
        \label{fig:telgpix1}
\end{figure}

We obtained four observation runs (or visits) of TOI-178 with {\it CHEOPS} between 4 August 2020 and 3 October 2020 as part of the Guaranteed Time, totalling 11.88 days on target with the observing log shown in Table~\ref{tab:obs_log}. The majority of this time was spent during the first two visits (with lengths of 99.78\,h and 164.06\,h), which were conducted sequentially beginning on August 4,  2020 and ending on August 15, 2020; as such, we achieved a near-continuous 11-day photometric time series. 
The runs were split due to scheduling constraints, with a 0.84 h gap between visits. 
The third and fourth visits were conducted to confirm the additional planets predicted in the scenario presented in Sect.~\ref{sec:rationale}. They took place on September 7, 2020 and October 3, 2020 and lasted for 13.36\,h and 8.00\,h, respectively.

Due to the low-Earth orbit of {\it CHEOPS}, the spacecraft-target line of sight was interrupted by Earth occultations and passages through the South Atlantic Anomaly (SAA), where no data were downlinked. This resulted in gaps in the photometry on {\it CHEOPS} orbit timescales (around 100 min). For our observations of TOI-178, this resulted in light curve efficiencies of 51\%, 54\%, 65\%, and 86\%. For all four visits, we used an exposure time of 60\,s.

These data were automatically processed with the latest version of the {\it CHEOPS} data reduction pipeline (DRP v12; \citealt{Hoyer2020}). This includes image calibration (bias, gain, non-linearity, dark current, and flat field) and instrumental and environmental corrections (cosmic rays and smearing trails of field stars and background). The DRP also performs aperture photometry for a set of three size-fixed apertures -- R=22.5\arcsec (RINF), 25\arcsec (DEFAULT), and 30\arcsec (RSUP) -- plus one extra aperture that minimises the contamination from nearby field stars (ROPT), which, in the case of TOI-178, was set at R=28.5\arcsec. The DRP estimates the level of contamination by simulating the field-of-view of TOI-178 using the GAIA star catalogue \citep{GaiaCollaboration2018} to determinate the location and flux of the stars throughout the duration of the visit. In the case of TOI-178, the mean contamination level was below 0.1\% and was mostly modulated by the rotation around the target of a nearby star of Gaia G=13.3\,mag at a projected sky distance of 60.8\arcsec  from the target. The contamination as a function of time (or, equivalently, as a function of the roll angle of the satellite) is provided as a product of the DRP for further de-trending (see Sect.\ref{sec:phot_analysys}). For the second visit, careful removal of one `telegraphic' pixel (a pixel with a non-stable abnormal behaviour during the visit) within the photometric aperture was needed. The location of this telegraphic pixel is shown in red in Fig.\ref{fig:telgpix1}, and details regarding the detection and correction are described in Appendix \ref{sec:data_inspection}. Following the reductions, we found that the light curves obtained using the DEFAULT aperture (R=25\arcsec) yielded the lowest RMS for all visits and, as such, are used for this study (see Appendix A, Fig.~\ref{fig:rawLC_1}). 

Lastly, it became apparent that -- due to the nature of the {\it CHEOPS} orbit and the rotation of the {\it CHEOPS} field around the target -- photometric, non-astrophysical short-term trends -- either from a varying background, a nearby contaminating source, or other sources -- can be found in the data on orbital timescales. These effects can be efficiently modelled by using a Gaussian process (GP) with roll angle as input \citep[e.g.][]{Lendl2020,Bonfanti2021}, as we will discuss in our data analysis (Sect. \ref{sec:photom_analysis}). Following this, the average noise over a 3\,h sliding window for the four visits of the G = 11.15\,mag target was found to be 63.9, 64.2, 66.3, and 75.8\,ppm. In all cases, this marginally improved upon the precision of the light curves that we had previously simulated for these observation windows using the \texttt{CHEOPSim} tool \citep{CHEOPSim}.


\begin{figure*}
  \includegraphics[width =\textwidth]{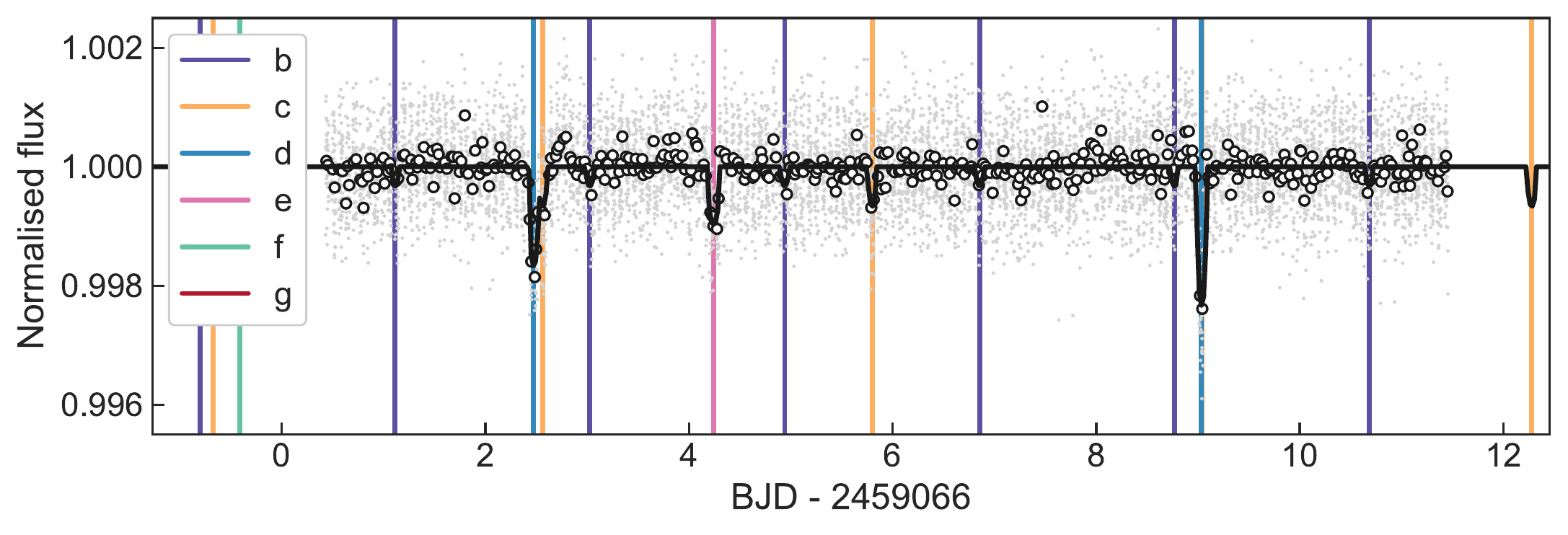}

  \includegraphics[width =\textwidth]{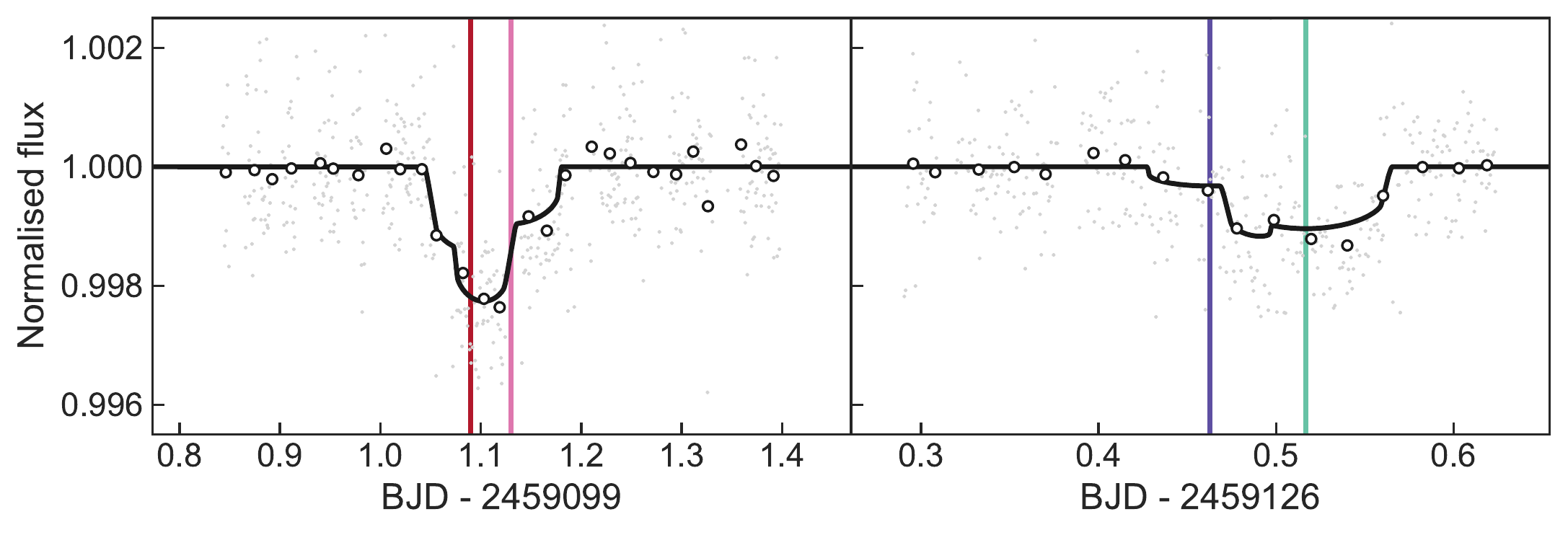}
  \caption{Similar to Fig. \ref{fig:tess_full}, but instead displaying data from the four CHEOPS visits described in Sect. \ref{sec:CHEOPS}. Top panel: 11 d observation. Transits of planets $c$ and $d$ at $\sim$ 2459075 BJD occur too close to each other for their corresponding lines to be individually visible in the figure. Bottom-left panel: Subsequent observation scheduled to confirm the presence of a planet with a$~20.7$ d period (planet g), which overlaps with a transit of planet $e$. Bottom-right panel: Final observation scheduled to confirm the presence of a planet fitting in the Laplace resonance (planet $f$, with a period of $\sim$15.23 d), which overlaps with a transit of planet $b$.}
  \label{fig:CHEOPS_11}
\end{figure*}

\begin{table*}
\caption{Log of {\it CHEOPS} observations of TOI-178.}             
\label{tab:obs_log}      
\centering   
\small
\begin{tabular}{c c c c c c c c}
\hline\hline
visit  & Identified & Start date & Duration & Data points & File key & Efficiency & Exp. Time  \\ 
 \# &planets & [UTC] & [h] & [\#] & & [\%] & [s] \\
\hline                    
 1 &   b,c,d,e & 2020-08-04T22:11:39 & 99.78 & 3030 & CH\_PR100031\_TG030201\_V0100 & 51 & 60 \\%
 2 &    b,c,d & 2020-08-09T02:48:39 & 164.04 & 5280 & CH\_PR100031\_TG030301\_V0100 & 54 & 60 \\%
3 &    e,g & 2020-09-07T08:06:44 & 13.36 & 521 & CH\_PR100031\_TG030701\_V0100 & 65 & 60 \\%
 4 &   b,f & 2020-10-03T18:51:46 & 8.00 & 413 & CH\_PR100031\_TG033301\_V0100 & 86 & 60 \\%
\hline                  
\end{tabular}
\end{table*}

\subsubsection{NGTS}
\label{sec:NGTS}
The NGTS \citep[][]{wheatley2018ngts} facility consists of twelve 20-cm diameter robotic telescopes and is situated at the ESO Paranal Observatory in Chile. The individual NGTS telescopes have a wide field-of-view of 8 square-degrees and a plate scale of 5\,\arcsec\,pixel$^{-1}$. The DONUTS auto-guiding algorithm \citep{mccormac13donuts} affords the NGTS telescopes sub-pixel level guiding. Simultaneous observations using multiple NGTS telescopes have been shown to yield ultra high precision light curves of exoplanet transits \citep{bryant2020multicam, smith2020multicam}.

TOI-178 was observed using NGTS multi-telescope observations on two nights. On UT September 10, 2019, TOI-178 was observed using six NGTS telescopes during the predicted transit event of TESS candidate TOI-178.02. However, the NGTS data for this night rule out a transit occurring during the observations. A second predicted transit event of TOI-178.02 was observed on the night of UT October 11, 2019 using seven NGTS telescopes, and on this night the transit event was robustly detected by NGTS. A total of 13,991 images were obtained on the first night, and 12,854 were obtained on the second. For both nights, the images were taken using the custom NGTS filter (520 - 890\,nm) with an exposure time of 10\,s. All NGTS observations of TOI-178 were performed with airmass $<\,2$ and under photometric sky conditions.

The NGTS images were reduced using a custom pipeline that uses the \texttt{SEP} library to perform source extraction and aperture photometry \citep{bertin96sextractor, Barbary2016}. A selection of comparison stars with brightnesses, colours, and CCD positions similar to those of TOI-178 were identified using the second GAIA data release (DR2) \citep{GaiaCollaboration2018}. More details on the photometry pipeline are provided in \citet{bryant2020multicam}.

The NGTS light curves are presented in Fig.~\ref{fig:NGTS+SPEC}. They show transit events for planet $b$ and planet $g$ on the nights of September 11, 2019 and October 1, 2019, respectively.

\subsubsection{SPECULOOS}
\label{sec:SPECULOOS}

The SPECULOOS Southern Observatory \citep[SSO;][]{Gillon2018, Burdanov2018, Delrez2018} is located at ESO’s Paranal Observatory in Chile and is part of the \href{https://www.speculoos.uliege.be/cms/c_4259452/en/speculoos}{SPECULOOS project}. The facility consists of a network of four robotic 1-m telescopes (Callisto, Europa, Ganymede, and Io). Each robotic SSO telescope has a primary aperture of 1\,m and a focal length of 8\,m, and is equipped with a 2k$\times$2k deep-depletion CCD camera whose 13.5 $\mu$m pixel size corresponds to 0.35" on the sky (field-of-view = 12$^\prime$x12$^\prime$). Observations were performed on the nights of October 10, 2019 (for $\simeq$ 8 hours) and  November 11, 2019 (for $\simeq$ 8 hours) with three SPECULOOS telescopes on sky simultaneously (SSO/Io, SSO/Europa, and SSO/Ganymede). These observations were carried out in a Sloan i' filter with  exposure times of 10\,s. A small de-focus was applied to avoid saturation as the target was too bright for SSO. Light curves were extracted using the SSO pipeline \citep{murray_photometry_2020} and are shown in purple in Fig. \ref{fig:NGTS+SPEC}. For each observing night, the SSO pipeline used the \texttt{casutools} software \citep{Irwin2004} to perform automated differential photometry and to correct for systematics caused by time-varying telluric water vapour.

\subsection{ESPRESSO data}
\label{sec:RVdata}

The RV data we analyse consist of 46 ESPRESSO points\footnote{The first 32 come from programme 0104.C-0873(A) and the last 14 from 1104.C-0350 (Guaranteed Time observations). }. Each measurement was taken in high resolution (HR) mode with an integration time of 20 min using a single telescope (UT) and slow read-out (HR 21). The source on fibre B is the Fabry-Perot interferometer. Observations were made with a maximum airmass of 1.8 and a minimum 30$^\circ$ separation from the Moon. 

\begin{figure}
        \includegraphics[width=9cm]{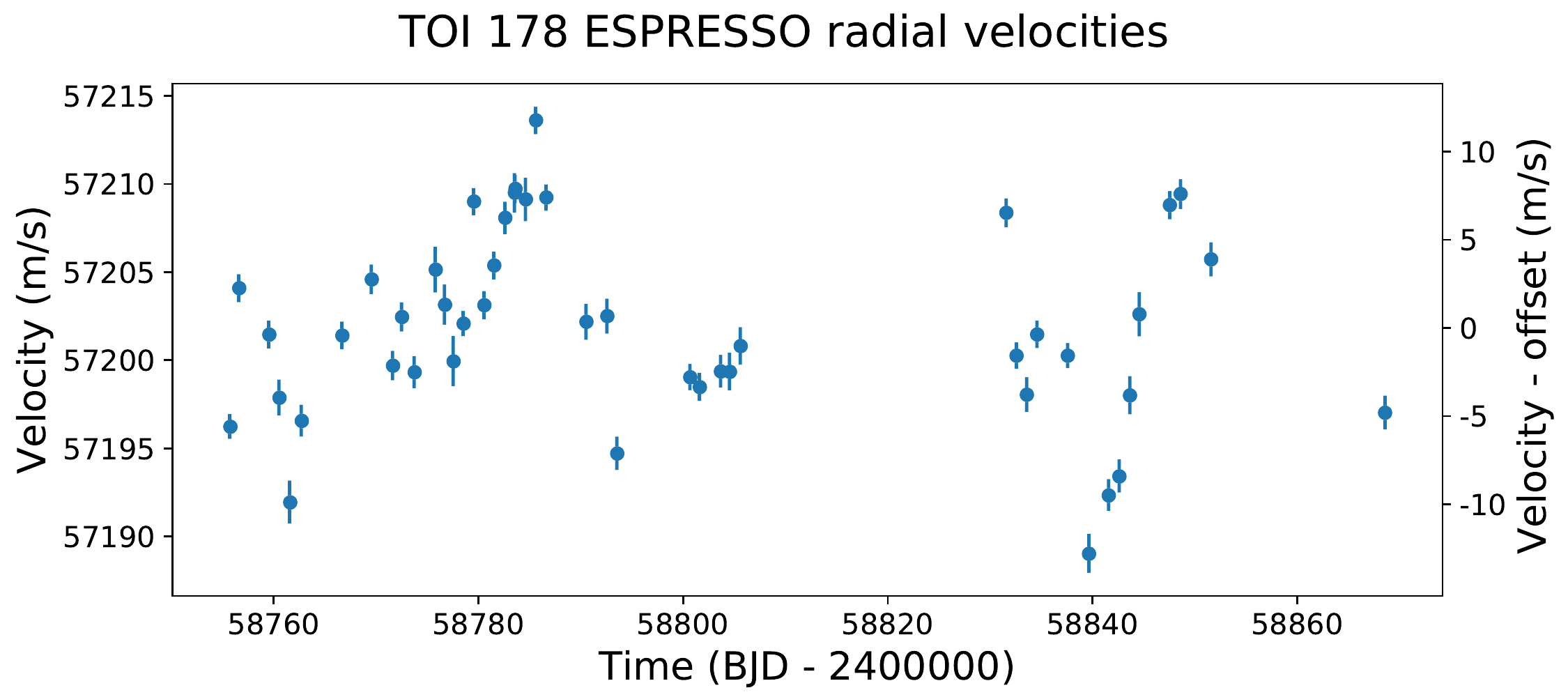}
        \caption{ESPRESSO RV data of  TOI-178. }
        \label{fig:RV}
\end{figure}

The measurements span from September 29, 2019 to January 20, 2020 and have an average nominal error bar of 93 cm/s. We also included the  time series of $H \alpha$ measurements, the full width half maximum (FWHM), and the $S$-index in our analysis. 
The velocity and ancillary indicators are extracted from the spectra with the standard ESPRESSO pipeline v 2.0.0~\citep{Pepe2020}. The RV time series with nominal error bars is shown in Fig.~\ref{fig:RV}.

\section{Detections and parameter estimations }
\label{sec:detections}

In this section, we analyse the photometric and spectral data in order to derive the orbital and planetary parameters of the six planets in the system.


\subsection{Analysis of the photometry}
\label{sec:phot_analysys}

\subsubsection{Identification of the solution}

The first release of candidates from the {\it TESS} alerts of Sector 2 included three planet candidates in TOI-178 with periods of $6.55$\,d, $10.35$\,d, and $9.96$\,d, which transited four, three, and two times, respectively. In addition, our analysis of this dataset with the DST \citep[D\'etection Sp\'ecialis\'ee de Transits][]{Cabrera2012} yielded two additional candidates: a clear $3.23$\,d signal and a fainter $1.91$\,d signal. Upon receiving visits 1 and 2 from {\it CHEOPS} (Table \ref{tab:obs_log}), a study of the {\it CHEOPS} data alone with successive applications of the boxed-least-square (BLS) algorithm \citep{kovacs2002} retrieved the $6.55$\,d, $3.23$\,d, and $1.91$\,d signals, in phase agreement with the {\it TESS} data. An additional dip, consistent in epoch with a transit of the $9.9$\,d candidate, was also identified; however, it could also marginally correspond to a transit of the $10.35$\,d candidate. The transit of one of these candidates was hence missing, pointing  to a mis-attribution of transits in the TESS Sector 2 data. 

\begin{figure}
  \includegraphics[width = 8.5cm]{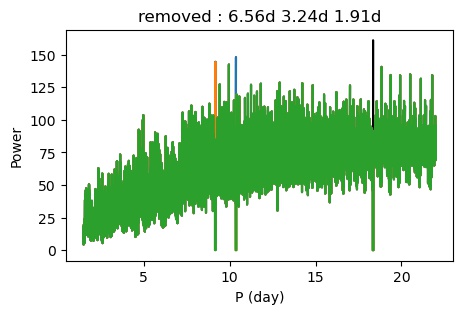}

  \includegraphics[width = 8.5cm]{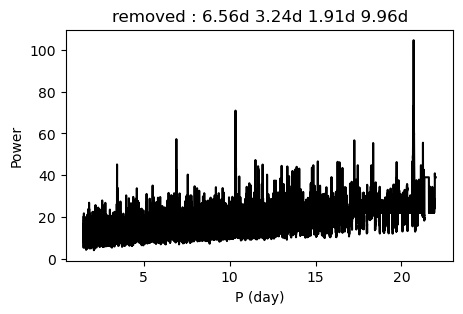}

    \includegraphics[width = 8.5cm]{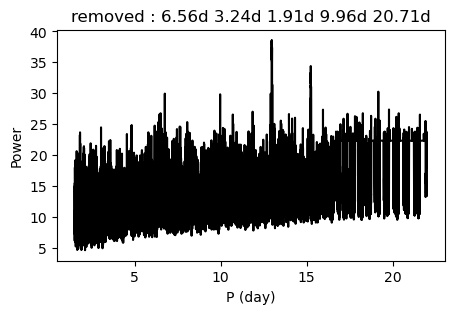}
  \caption{Successive use of the BLS algorithm to identify the new candidates. In the top panel, the green curve has its highest power at 9.96\,d once the three highest peaks ($i=3$) at 18.35\,d (black), 10.37\,d (blue), and 9.17\,d (orange) are ignored. The middle panel shows the BLS after the removal of the 9.96\,d signal in the light curve, with no peaks ignored ($j=0$), and the bottom panel  shows the BLS after the removal of the 20.71\,d signal as well.}
  \label{fig:BLS}
\end{figure}

In order to identify new possible solutions for the available data ({\it TESS} Sector 2, NGTS visits in September and October 2019, and {\it CHEOPS} visits 1 and 2), we individually pre-detrended each light curve and subtracted the signal of the $6.55$\,d, $3.23$\,d, and $1.91$\,d candidates. We then applied the BLS algorithm to this dataset for the first time. The resulting periodogram (Fig. \ref{fig:BLS}, top panel) had: several peaks of similar power due to the existence of multiple transits; dips of similar depths and durations; and, overall, a low number of transits per candidate spread along a long baseline. We hence explored different solutions by successively ignoring some of the highest peaks of the BLS periodogram, proceeding as follows: On the first periodogram, we saved the most likely candidate of period $P_{1}$ and removed the corresponding signal in the light curve; we then applied the BLS a second time to obtain a second candidate of period  $P_{2}$. That created a first pair of candidates, $c_{i,j}=c_{0,0}$, where $i$ and $j$ are the number of peaks that have been ignored in the first or second iteration of the BLS, respectively (in this case, no peaks have been ignored). We then repeated this process, but ignoring the result of the BLS for periods less than $0.2$\,d away from the highest peak of the periodogram (in black in the top panel of Fig. \ref{fig:BLS}). Since we ignored the black peak, the second highest is the blue one, which we assigned to period $P_1$, and we removed the associated signal in the light curve.  For this candidate, $i=1$  since we ignored one peak in the first BLS. For the second iteration of the BLS, we did not ignore any peaks and hence take the largest one ($j=0$), leading to the pair of candidates $c_{1,0}$. We repeated this process $25$ times, yielding 25 potential pairs of candidates $c_{0 \leq i \leq 4 ,0 \leq j \leq 4}$. For all of these potential solutions, we modelled the transits of the five candidates -- $1.91$ d, $3.23$ d, $6.55$ d, and $c_{i,j}$ -- using the {\ttfamily batman} package \citep{batman} and ran an MCMC on the pre-detrended light curve to estimate the relative likelihood of the different $c_{i,j}$.

The $c_{3,0}$ pair was favoured (its first BLS is shown with the green curve in the top panel of Fig. \ref{fig:BLS}), and the second iteration of the BLS (shown in the middle panel) yielded a$1.91$\,d, $3.23$\,d, $6.55$\,d, $9.96\,$d, and $20.71\,$d solution, which explained all the significant dips observed in the NGTS/SPECULOOS data (Fig. \ref{fig:NGTS+SPEC}) and the first visit of {\it CHEOPS} (Fig. \ref{fig:CHEOPS_11} - {\it top}). This solution was later confirmed by the predicted double transit observed by the third visit of {\it CHEOPS} (Fig. \ref{fig:CHEOPS_11} - {\it bottom left}).    

Applying the BLS algorithm to the residuals of the available photometric data (bottom panel of Fig. \ref{fig:BLS}), two mutually exclusive peaks appeared, $12.9\,$d and $15.24\,$d, which shared the odd transit of the previous TOI178.02 candidate. The $12.9\,$d signal was slightly favoured by the BLS analysis; however, the global fit of the light curve favoured the $\sim15.24\,$d signal. In addition, this solution was very close to the period that would fit the resonant structure of the system (see Sect. \ref{sec:dynamics}). 
The $15.23\,$d candidate was confirmed by a fourth {\it CHEOPS} visit (Fig. \ref{fig:CHEOPS_11} - {\it bottom right}). In the next section, we develop the characterisation of this six-planet solution: $1.91$\,d, $3.23$\,d, $6.55$\,d, $9.96\,$d, $15.23\,$d, and $20.71\,$d. 


\subsubsection{Determination of radii and orbital parameters}
\label{sec:photom_analysis}

To characterise the system, we performed a joint fit of the {\it TESS}, NGTS, and {\it CHEOPS} photometry. As the NGTS and SPECULOOS data presented in Fig.~\ref{fig:NGTS+SPEC} cover the same epoch of observations, we only included the NGTS data in our fit as they had a smaller RMS photometric scatter.

The fit was performed with the \texttt{juliet} package \citep{juliet}, which uses \texttt{batman} \citep{batman} for the modelling of transits and the nested sampling \texttt{dynesty} algorithm \citep{dynesty,dynesty2,dynesty3,dynesty4,dynesty5,dynesty6,dynesty7} for estimating Bayesian posteriors and evidence. In our analysis, the fitted parameters for each planet were: the planet-to-star radius ratio $R_p/R_\star$, the impact parameter $b$, the orbital period $P$, and the mid-transit time $T_0$. We also fitted for the stellar density $\rho_\star$, which, together with the orbital period $P$ of each planet, defines through Kepler's third law a value for the scaled semi-major axis $a/{R_\star}$ of each planet. We assumed a normal prior for the stellar density based on the value and uncertainty reported in Table \ref{tab:stellarParam} (Sect. \ref{sec:star}) and wide uniform priors for the other transit parameters. The orbits were assumed to be circular, as justified in Sect. \ref{sec:stability}. For each bandpass (\textit{TESS}, \textit{CHEOPS}, and NGTS), we fitted two quadratic limb-darkening coefficients, which were parametrised with the ($q_1$, $q_2$) sampling scheme of \cite{limbdark}. Normal priors were placed on these limb-darkening coefficients based on \citet{Claret2011}.

We modelled the correlated noise present in the light curves simultaneously with the planetary signals to ensure a full propagation of the uncertainties. We first performed individual analyses of each light curve in order to select for each of them the best correlated noise model based on Bayesian evidence. We explored a large range of models for the \textit{CHEOPS} light curves, consisting of first- to fourth-order polynomials in the recorded external parameters (most importantly: time, background level, position of the point-spread-function (PSF) centroid, spacecraft roll angle, and contamination), as well as GPs (\texttt{celerite} \citealt{celerite} and \texttt{george} \citealt{george}) against time, roll angle, or a combination of the two. We found that a Matérn 3/2 GP against roll angle was strongly favoured for all visits to account for the roll-angle-dependent photometric variations (cf. Sect. \ref{sec:CHEOPS}). First- to second-order polynomials in time and $x-y$ centroid position were also needed for the first three visits. For the \textit{TESS} light curve, we used a Matérn 3/2 GP against time, and we used a linear function of airmass for the NGTS light curves. For each light curve, we also fitted a jitter term, which was added quadratically to the error bars of the data points, to account for any underestimation of
the uncertainties or any excess noise not captured by our modelling.

{The results of our joint fit are displayed in Tables \ref{table:TOI178bcd} (planets $b$, $c$, and $d$) and \ref{table:TOI178efg} (planets $e$, $f$, and $g$). The transits of all detected planets are shown in Figs. \ref{fig:tess_full} (\textit{TESS}), \ref{fig:NGTS+SPEC} (NGTS), and \ref{fig:CHEOPS_11} (\textit{CHEOPS}), with the phase-folded transits in the {\it CHEOPS} data presented in Fig.~\ref{fig:cheops_fold}.

\begin{figure}
  \includegraphics[width = \columnwidth]{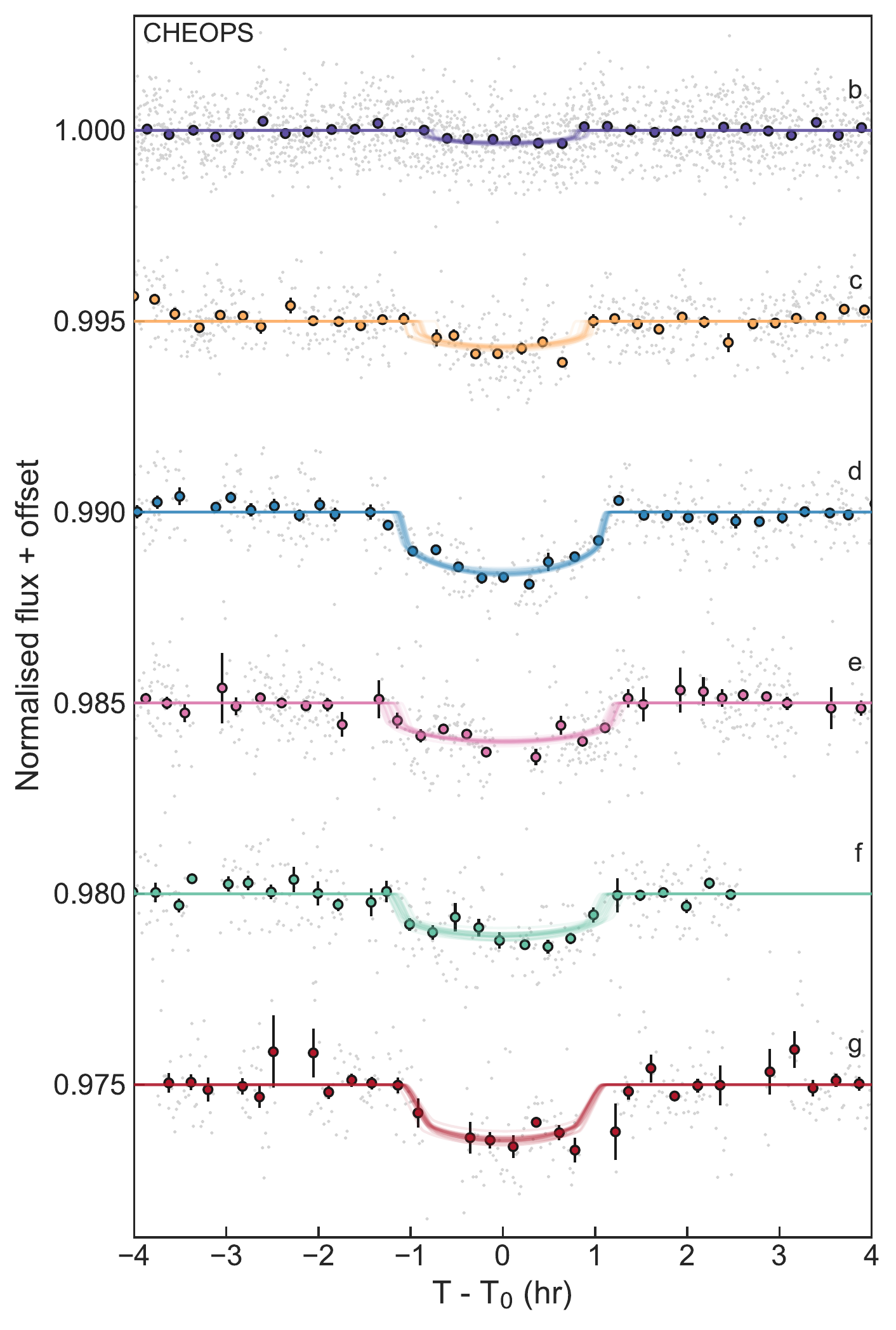}
  \caption{Detrended CHEOPS light curves phase-folded to the periods of each of the planets, with signals of the other planets removed. Unbinned data are shown with grey points, data in 15-minute bins are shown with coloured circles, and samples drawn from the posterior distribution of the global fit are shown with coloured lines.}
  \label{fig:cheops_fold}
\end{figure}

\begin{table*}
    \centering
    \caption{Fitted and derived results for planets $b, c$, and $d$ associated with the fits to the photometry and spectroscopy described in Sects. \ref{sec:phot_analysys} and \ref{sec:RVanalysis}, respectively. $ ^{a}$ $T_{\mathrm{eq}}= T_{\mathrm{eff}} \sqrt{R_\star / a} \,\, (f(1-A_{\rm B}))^{1/4}$, assuming an efficient heat redistribution ($f=1/4$) and a null Bond albedo ($A_{\rm B}=0$).}
    \label{table:TOI178bcd}
    \begin{tabular}{cccc}
 Parameter (unit) & b & c & d \\ 
\hline 
\multicolumn{4}{c}{\textit{Fitted parameters (photometry)}} \\ 
\hline 
$R_\mathrm{p} / R_\star$ & $0.01623\pm0.00097$ & $0.0235_{-0.0013}^{+0.0015}$ & $0.03623_{-0.00091}^{+0.00087}$ \\ 
$b$ ($R_\star$) & $0.17_{-0.13}^{+0.19}$  & $0.34_{-0.23}^{+0.30}$ & $0.485_{-0.060}^{+0.051}$ \\ 
$T_{0}$ (BJD-TBD) & $2458741.6365_{-0.0030}^{+0.0043}$ & $2458741.4783_{-0.0029}^{+0.0034}$ & $2458747.14623_{-0.00095}^{+0.00087}$ \\ 
$P$ (d) & $1.914558\pm0.000018$ & $3.238450_{-0.000019}^{+0.000020}$ & $6.557700\pm0.000016$ \\
$\rho_\star$ ($\rho_{\odot}$) & \multicolumn{3}{c}{$2.35\pm0.17$ (same for all planets)}\\
\hline 
\multicolumn{4}{c}{\textit{Fitted parameters (spectroscopy)}} \\ 
\hline 
$K$ ($\mathrm{ms}^{-1}$) & $1.05^{+0.25}_{-0.30}$ & $2.77^{+0.22}_{-0.33}$ & $1.34^{+0.31}_{-0.39}$\\ 
\hline 
\multicolumn{4}{c}{\textit{Derived parameters}} \\ 
\hline 
$\delta_\mathrm{tr}$ (ppm) & $263_{-30}^{+32}$ & $551_{-59}^{+68}$ & $1313_{-65}^{+64}$\\
detection SNR & 8.2 & 8.1 & 20.2 \\
$R_\star/a$ & $0.1161_{-0.0027}^{+0.0030}$ & $0.0818_{-0.0019}^{+0.0021}$ & $0.0511_{-0.0012}^{+0.0013}$  \\ 
$a/R_\star$ & $8.61_{-0.22}^{+0.21}$ & $12.23_{-0.31}^{+0.29}$ & $19.57_{-0.49}^{+0.47}$ \\ 
$R_\mathrm{p}$ ($\mathrm{R_{\oplus}}$) & $1.152_{-0.070}^{+0.073}$ & $1.669_{-0.099}^{+0.114}$ & $2.572_{-0.078}^{+0.075}$ \\ 
$a$ (AU) & $0.02607\pm0.00078$ & $0.0370\pm0.0011$ & $0.0592\pm0.0018$ \\ 
$i$ (deg) & $88.8_{-1.3}^{+0.8}$ & $88.4_{-1.6}^{+1.1}$ & $88.58_{-0.18}^{+0.20}$ \\ 
$t_\mathrm{14}$ (h) & $1.692_{-0.086}^{+0.056}$ & $1.95_{-0.25}^{+0.15}$ & $2.346_{-0.046}^{+0.047}$ \\ 
$T_\mathrm{eq}$ (K)$^{a}$ & $1040_{-21}^{+22}$ & $873\pm18$ & $690\pm14$ \\ 
$M_\mathrm{p}$ ($\mathrm{M_{\oplus}}$) & $1.50^{+0.39}_{-0.44}$ & $4.77^{+0.55}_{-0.68}$ & $3.01^{+0.80}_{-1.03}$\\ 
$\rho_\mathrm{p}$ ($\mathrm{\rho_{\oplus}}$) & $0.98^{+0.35}_{-0.31}$ & $1.02^{+0.28}_{-0.23}$ & $0.177^{+0.055}_{-0.061}$ \\ 
\hline



    \end{tabular}

\end{table*}

\begin{table*}
    \centering
    \caption{Fitted and derived results for planets $e, f$, and $g$ associated with the fits to the photometry and spectroscopy described in Sects. \ref{sec:phot_analysys} and \ref{sec:RVanalysis}, respectively. $ ^{a}$ $T_{\mathrm{eq}}= T_{\mathrm{eff}} \sqrt{R_\star / a} \,\, (f(1-A_{\rm B}))^{1/4}$, assuming an efficient heat redistribution ($f=1/4$) and a null Bond albedo ($A_{\rm B}=0$).}
    \label{table:TOI178efg}
    \begin{tabular}{cccc}

Parameter (unit) & e & f & g \\ 
\hline 
\multicolumn{4}{c}{\textit{Fitted parameters (photometry)}} \\ 
\hline 
$R_p / R_\star$ & $0.0311_{-0.0012}^{+0.0011}$ & $0.0322\pm0.0014$ & $0.0404_{-0.0018}^{+0.0019}$\\ 
$b$ ($R_\star$) & $0.583_{-0.066}^{+0.046}$ & $0.765_{-0.031}^{+0.027}$ & $0.866_{-0.019}^{+0.017}$ \\ 
$T_{0}$ (BJD-TBD) & $2458751.4658_{-0.0019}^{+0.0016}$ & $2458745.7178_{-0.0027}^{+0.0023}$ & $2458748.0302_{-0.0017}^{+0.0023}$ \\ 
$P$ (d) & $9.961881\pm0.000042$ & $15.231915_{-0.000095}^{+0.000115}$ & $20.70950_{-0.00011}^{+0.00014}$\\ $\rho_\star$ ($\rho_{\odot}$) & \multicolumn{3}{c}{$2.35\pm0.17$ (same for all planets)}\\
\hline 
\multicolumn{4}{c}{\textit{Fitted parameters (spectroscopy)}} \\ 
\hline 
$K$ ($\mathrm{ms}^{-1}$)  & $1.62^{+0.41}_{-0.34}$  & $2.76^{+0.46}_{-0.42}$ & $1.30^{+0.38}_{-0.59}$\\ 
\hline 
\multicolumn{4}{c}{\textit{Derived parameters}} \\ 
\hline 
$\delta_\mathrm{tr}$ (ppm) & $968_{-71}^{+69}$ & $1037_{-90}^{+94}$ & $1633_{-139}^{+157}$ \\ 
detection SNR & 13.6 & 11.0 & 10.4 \\
$R_\star/a$ & $0.03866_{-0.00090}^{+0.00100}$ & $0.02913_{-0.00068}^{+0.00075}$ & $0.02373_{-0.00056}^{+0.00061}$ \\ 
$a/R_\star$ & $25.87_{-0.65}^{+0.62}$ & $34.33_{-0.87}^{+0.82}$ & $42.13_{-1.06}^{+1.01}$ \\ 
$R_\mathrm{p}$ ($\mathrm{R_{\oplus}}$) & $2.207_{-0.090}^{+0.088}$ & $2.287_{-0.110}^{+0.108}$ & $2.87_{-0.13}^{+0.14}$\\ 
$a$ (AU) & $0.0783_{-0.0024}^{+0.0023}$ & $0.1039\pm0.0031$ & $0.1275_{-0.0039}^{+0.0038}$ \\ 
$i$ (deg) & $88.71_{-0.13}^{+0.16}$ & $88.723_{-0.069}^{+0.071}$ & $88.823_{-0.047}^{+0.045}$ \\ 
$t_\mathrm{14}$ (h) & $2.501_{-0.077}^{+0.106}$ & $2.348_{-0.087}^{+0.097}$ & $2.167_{-0.082}^{+0.090}$ \\ 
$T_\mathrm{eq}$ (K)$^{a}$ & $600\pm12$ & $521\pm11$ & $470\pm10$ \\ 
$M_\mathrm{p}$ ($\mathrm{M_{\oplus}}$) & $3.86^{+1.25}_{-0.94}$ & $7.72^{+1.67}_{-1.52}$ & $3.94^{+1.31}_{-1.62}$ \\ 
$\rho_\mathrm{p}$ ($\mathrm{\rho_{\oplus}}$) & $0.360^{+0.143}_{-0.097}$ & $0.65^{+0.21}_{-0.15}$ & $0.166^{+0.065}_{-0.068}$ \\ 
\hline

    \end{tabular}

\end{table*}

\begin{figure}
  \includegraphics[width = 9cm]{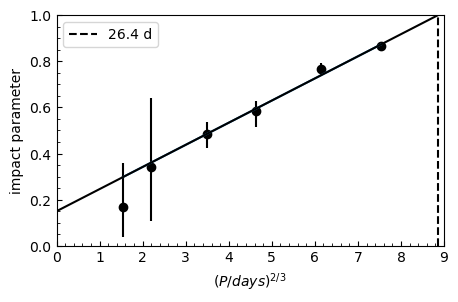}

  \includegraphics[width = 9cm]{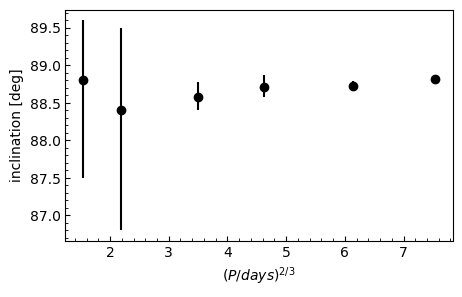}
  \caption{Impact parameter (\textit{top}) and inclination (\textit{bottom}) of the six planets of the TOI-178 system. In the top panel, the solid line shows the evolution of the impact parameter as a function of the period assuming that all planets are in the same plane (obtained by linear fit). The dashed line shows that an outer planet in the same plane will not transit if its period is above $26.4$\,d. }
  \label{fig:inclination}
\end{figure}

All planets appear to be in the super-Earth to mini-Neptune range, with the inner two planets falling to either side of the radius valley \citep{fulton2017}. The inclination of the planets are also worth mentioning: The projected inclination of the four outer planets differ only by about $0.1$\,deg. As discussed in \cite{Agol2020} in the case of TRAPPIST-1, it is unlikely that the underlying inclinations and ascending nodes are scattered given the clustering of the projected inclination; hence, the outer part of the system is probably quite flat. In addition, the uncertainty on the inclination of the inner planets allows for the near-coplanarity of the entire system, a feature that was also found in the TRAPPIST-1 system \citep{Agol2020}.

\subsection{Analysis of the radial velocities }
\label{sec:RVanalysis}
\subsubsection{Detections}

In this section, we consider the 46 ESPRESSO  data points only and look for potential planet detections.  Our analysis follows the same steps as in~\cite{Hara2020} and is described in detail in Appendix ~\ref{ap:rv}. To search for potential periodicities, we computed the $\ell_1$-periodogram\footnote{https://github.com/nathanchara/l1periodogram} of the RV as defined in~\cite{hara2017}. 
This method outputs a figure that has a similar aspect as a regular periodogram, but with fewer peaks due to aliasing. The peaks can be assigned a false alarm probability (FAP), whose interpretation is close to the FAP of a regular periodogram peak. 

A preliminary analysis of ancillary indicators $H \alpha$, FWHM, bisector span~\citep{Queloz2001}, and $\log R'_{HK}$~\citep{Noyes1984} revealed that they exhibit statistically significant periodicities at $\approx 36$ days and $\approx 16$ days, such that stellar activity effects are to be expected in the RVs, especially at these periods. In our analysis, stellar activity has been taken into account both with a linear model constructed with  activity indicators smoothed with a GP regression similarly to~\cite{haywood2014}, which we call the base model, and with a Gaussian noise model with a white, correlated (Gaussian kernel), and quasi-periodic component. 

Radial velocity signals found to be statistically significant might vary from one activity model to another. To confirm the robustness of our detections, we tested whether signal detections can be claimed for a variety of noise models, following~\cite{Hara2020}.
This approach consists of three steps. First, we computed the $\ell_1$-periodogram of the data on a grid of models. The
linear base models considered include an an offset and smoothed ancillary indicators ($H \alpha$, FWHM, neither, or both), where the smoothing is done with a GP regression with a Gaussian kernel. The noise models we considered are Gaussian with three components: white, red with a Gaussian kernel, and quasi-periodic. According to the analysis of the ancillary indicators, the quasi-periodic term of the noise is fixed at 36.5 days. We considered a grid of values for each noise component (amplitude and decay timescale) and computed the $\ell_1$-periodograms. Secondly, we ranked the noise models with a procedure based on cross-validation (CV).  Finally, we examined the distribution of FAPs of each signal in the 20\% highest ranked models. 

Here we will succinctly describe the conclusions of our analysis and refer the reader to Appendix~\ref{ap:rv} for a more detailed presentation. The $\ell_1$-periodogram corresponding to the highest ranked model is represented in Fig.~\ref{fig:l1perio_best}. The periods at which the peaks occur are shown in red and account for most of the signals that might appear with varying assumptions on the noise (or activity) model. More precisely, we find the following results. We note that these results stem from an analysis of over 1300 noise models and that they are not all portrayed in Fig.~\ref{fig:l1perio_best}, which only shows the results from the highest ranked noise model. 

First, our analysis yields a consistent, significant detection of signals close to 3.2, $\approx$ 36, and $\approx$ 16 days. Radial velocity then allows an independent  detection of planet $c$. Signals at $\approx$ 36 and $\approx$ 16 d appear in ancillary indicators, and as such we attributed these apparent periodicities in RV to stellar activity. The 16-day signal is, however, very likely partly due to planet $f$. Indeed,  when the activity signals close to 40 and 16 days are modelled and the signal of transiting planets is removed, one finds a residual signal at 15.1 or 15.2 days, even though the 16 and 15.2-day signals are very close. 

Second, we find signals, though not statistically significant ones, at 6.5 and 9.8 days (consistent with planets $d$ and $e$) and at 2.08 days, which is the one-day alias~\citep[see][]{dawsonfabricky2010} of 1.91 days (planet $b$). 
Third, we do not consistently find a candidate near 20.7 days. However, a 20.6 d signal appears in the highest ranked model, and a stellar activity signal might hide the signal corresponding to TOI-178 g.

Fourth, the signal at 43.3 days appearing in Fig.~\ref{fig:l1perio_best} might be a residual effect of an imperfect correction of the activity. However, we have not  strictly ruled out the possibility that it stems from a planetary companion at 45 days. As will be discussed in the conclusion, this period corresponds to one of the possible ways to continue the Laplace resonance beyond planet $g$. Finally, depending on the assumptions, hints at 1.2 or 5.6 days (aliases of each other) can appear.

For comparison, we performed the RV analysis with an iterative periodogram approach. This analysis is able to show  signals corresponding to 15.2, 3.2, and 6.5 days; however, it is unable to clearly establish the significance of the 3.2 d signal and fails to unveil candidates at 1.91 and 9.9 days. 

 The photometric data allow us to independently detect six planets at 1.91, 3.24, 6.56, 9.96, 15.23, and 20.71 days. As detailed in Appendix~\ref{ap:rv}, we find that the phases of RV and photometric signals are consistent within $2\sigma$. We phase-folded the RV data at the periods given in  Tables~\ref{table:TOI178bcd} and~\ref{table:TOI178efg}, which are shown in Fig.~\ref{fig:phasefoldedrv} with periods increasing from top to bottom. The variations at 3.24 and 15.2 days, corresponding to the planets with the most significant RV signals, are the clearest. As a final remark, the signals corresponding to the transiting planets have been fitted, and the strongest periodic signals occur at 38 and 16.3 days, which is compatible with the activity periods seen in the ancillary indicators.  



\subsubsection{Mass and density estimations}

To estimate the planetary masses, we fitted circular orbits to the radial velocities. As shown in Sect.~\ref{sec:dynamics}, for the system to be stable, eccentricities cannot be greater than a few percent. We set the posterior distributions obtained from the fits to photometric data from Tables~\ref{table:TOI178bcd} and \ref{table:TOI178efg} as priors on $T_c$ and period. This approach, as opposed to a joint fit, is justified by the fact that, here, RVs brings very little information to the parameters constrained by photometry, and vice versa.  
Activity signals are clearly present in the RV data, and, depending on the activity model used, mass estimates may vary. To assess the model dependency of mass estimates, we used two different activity models. Both include as a linear predictor the smoothed $H \alpha$ time series, as described above. In the first model, we represented activity as a sum of two sine functions at 36 and 16 days and a correlated noise with an exponential kernel. This was motivated by the fact that both periodicities appear in the ancillary indicators, but with different phases. The correlated noise models low-frequency variations, which can also be anticipated from the analysis of the indicators. The second model of activity consists of a correlated Gaussian noise with a quasi-periodic kernel. We computed the posterior distributions of the orbital parameters with an adaptive MCMC, as in~\cite{Delisle2018}.
This analysis is presented in detail in Appendix~\ref{ap:mass1}.

The mass estimates for each model are given in Tables~\ref{table:massRV},~\ref{table:TOI178bcd}, and~\ref{table:TOI178efg}, respectively. The 1$\sigma$ intervals obtained with the two methods all have a large overlap. In Tables \ref{table:TOI178bcd} and \ref{table:TOI178efg}, we give the mass and density intervals in a conservative manner: The lower and upper bounds are respectively taken as the minimum lower bound and maximum upper bound we obtained with the two estimation methods.  The mass estimates are given as the mean of the estimates obtained with the two methods, which are posterior medians. We took this approach and did not select the error bars of one model or another since model comparisons depend heavily on the prior chosen, which, in our case, would be rather arbitrary. 
We find planet masses and densities in the ranges of $1.50^{+0.39}_{-0.44}$ $7.72^{+1.67}_{-1.52}$ $M_\oplus$ and $0.177^{+0.055}_{-0.061}$ $1.02^{+0.28}_{-0.23}$  $\rho_\oplus$.

It appears that the mass of planet $f$ (at 15.24 days) is the highest. This result might seem untrustworthy since there are activity signals close to 16 days in the ancillary indicators and since $1/(1/15.23-1/16) = 316$\,days, which is greater than the observation time span. However, the two activity models considered here -- one of which includes a signal at 16 days -- yield a similar mass estimates, and the convergence of MCMC was ensured by computing the number of effective samples. As a consequence, we deemed the mass estimate appropriate.  Nonetheless, more stellar activity models can be considered, and the mass intervals might still evolve as more data come along and the results become less model-dependent. A longer baseline would be suitable, so that the difference between 15.2 and 16 would be greater than the frequency resolution.

\begin{figure}
\centering
\includegraphics[width=\linewidth]{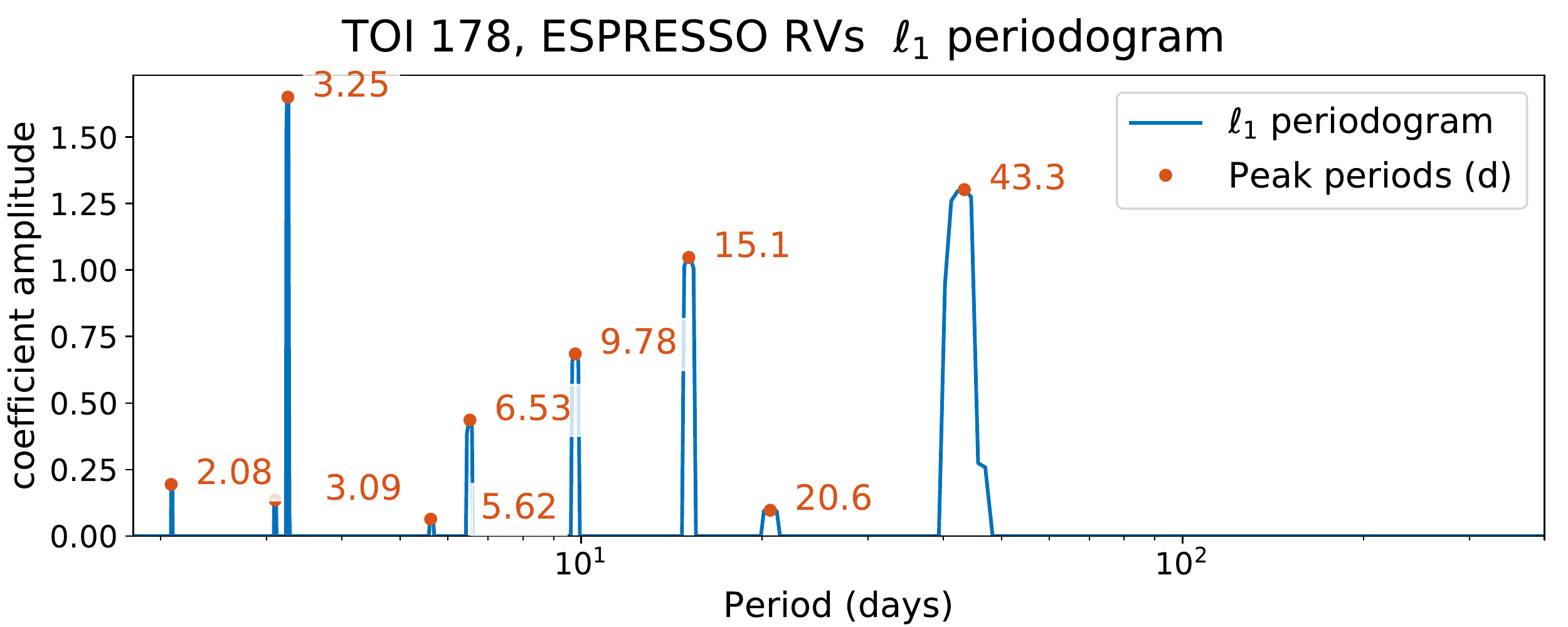}
\caption{$\ell_1$-periodogram corresponding to the best noise models in terms of the CV score, computed on a grid of frequencies from 0 to 0.55 cycles per day.}
\label{fig:l1perio_best}
\end{figure}

\begin{figure}
        \includegraphics[width=1\linewidth]{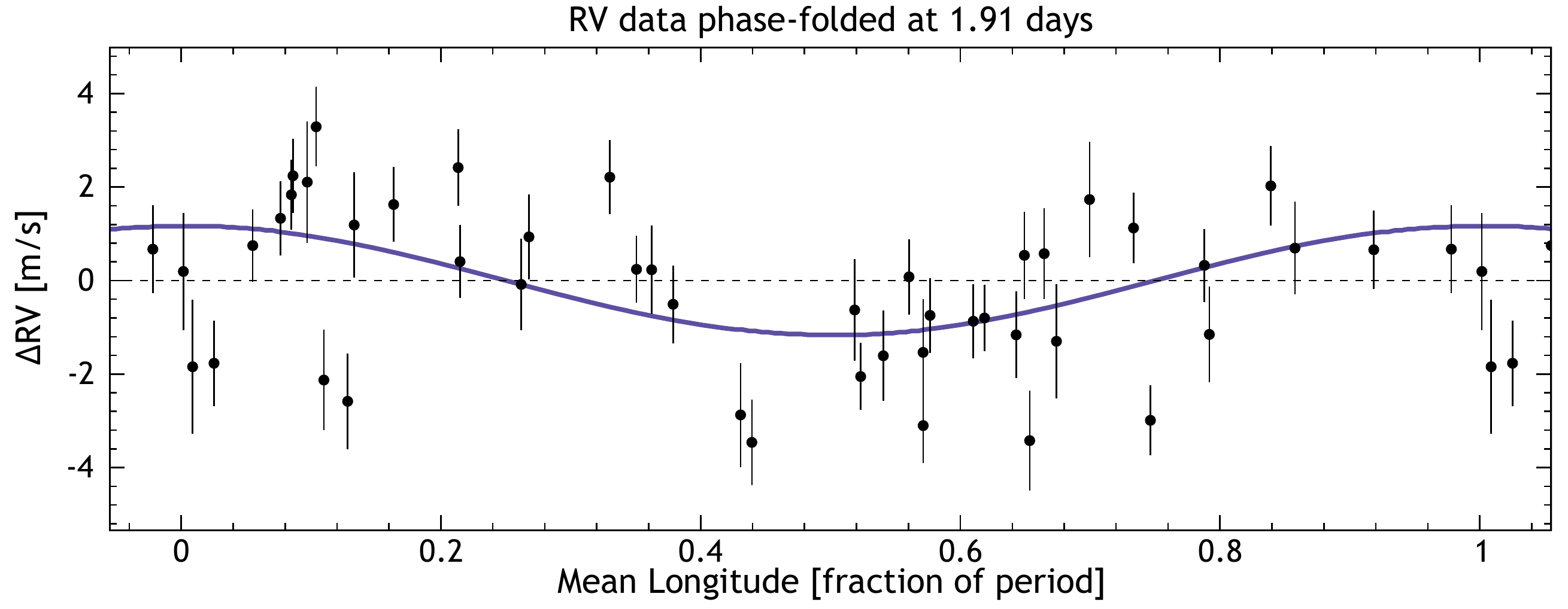}
    \includegraphics[width=1\linewidth]{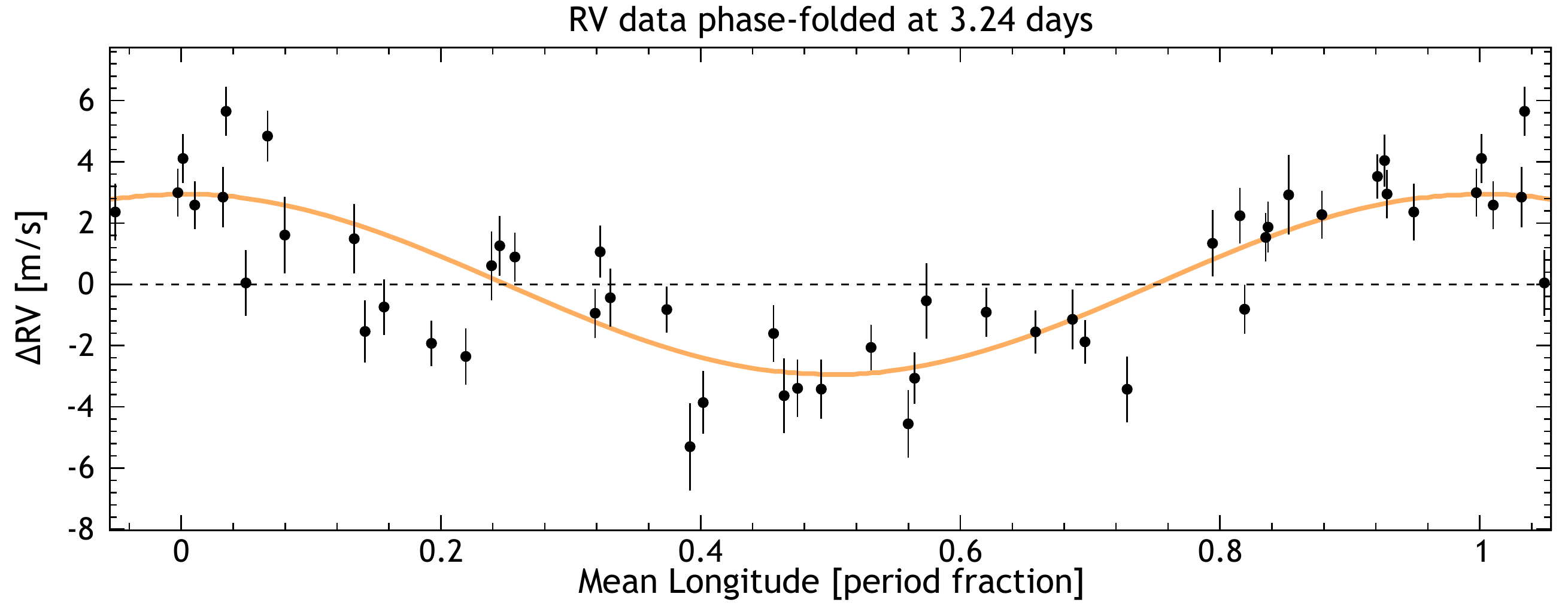}
    \includegraphics[width=1\linewidth]{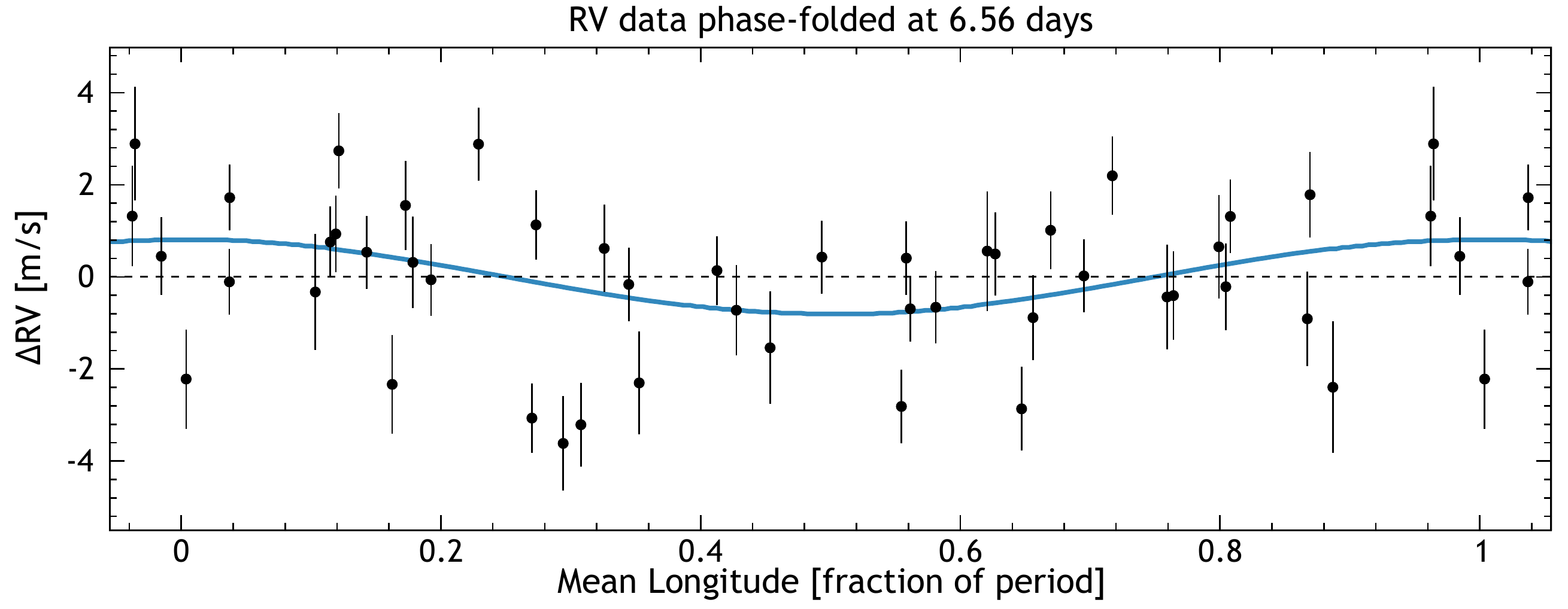}
    \includegraphics[width=1\linewidth]{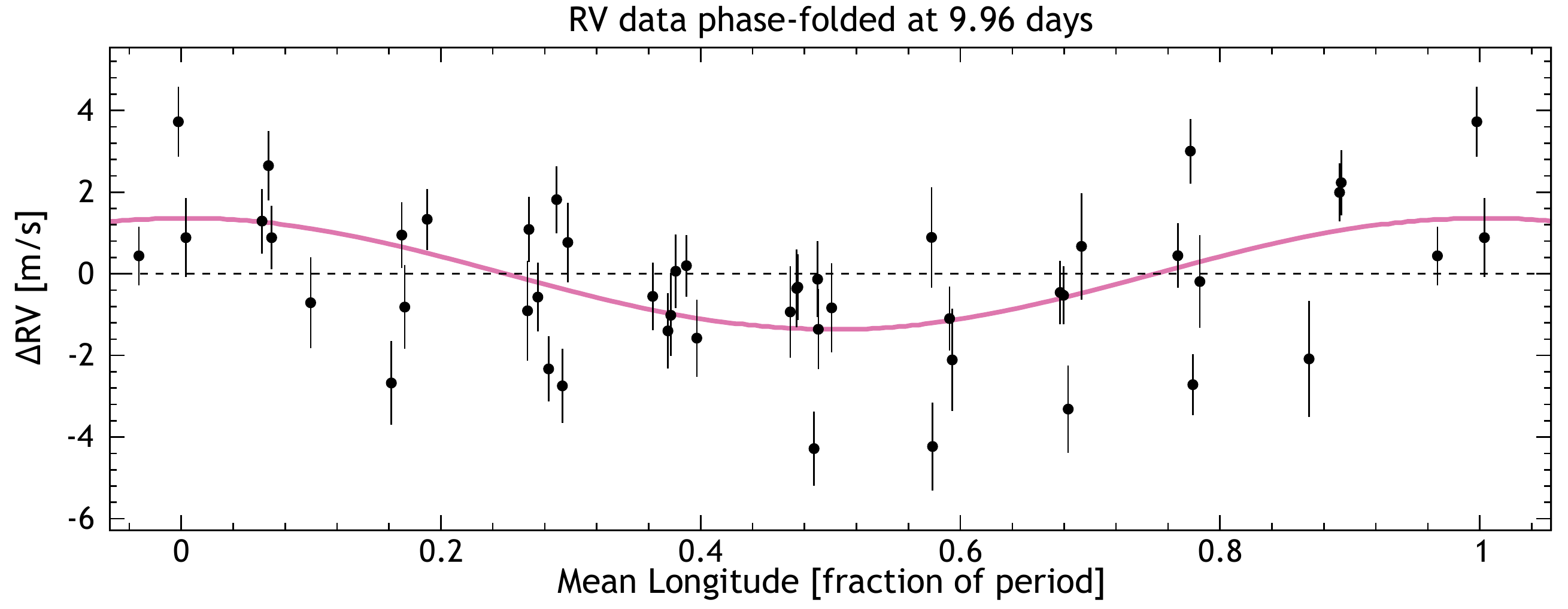}
    \includegraphics[width=1\linewidth]{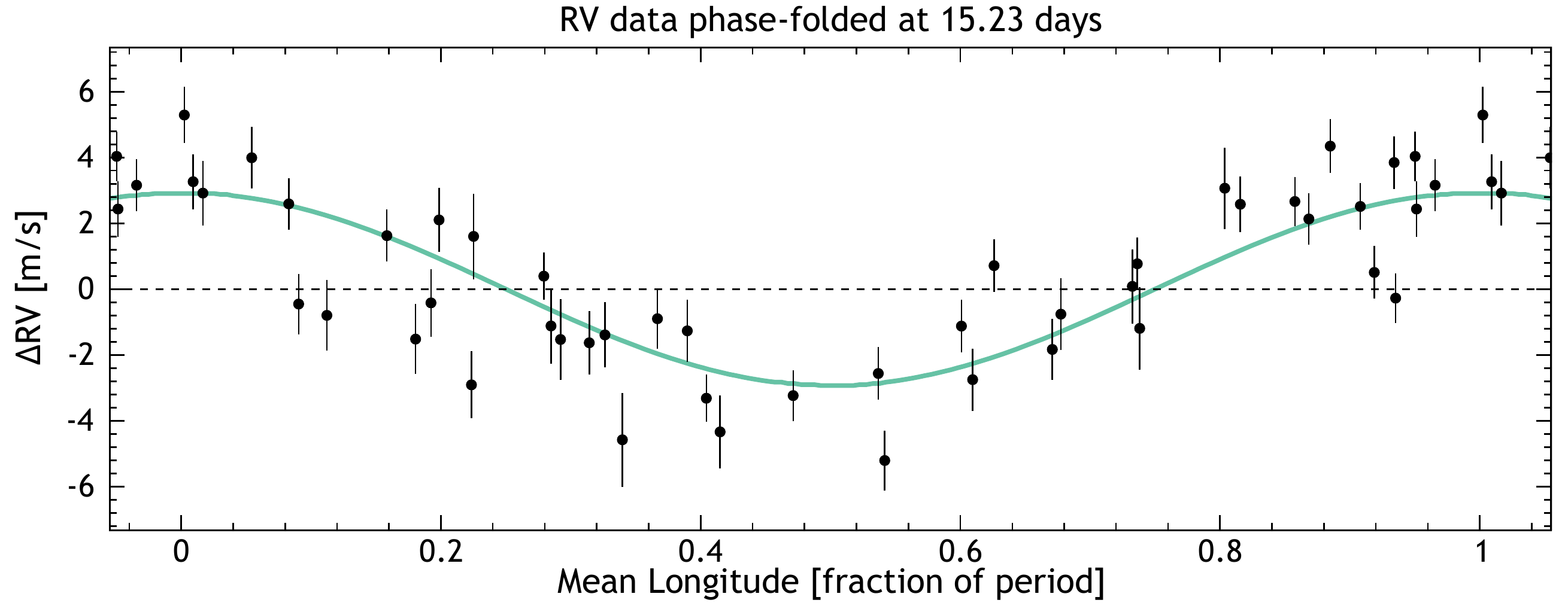}
    \includegraphics[width=1\linewidth]{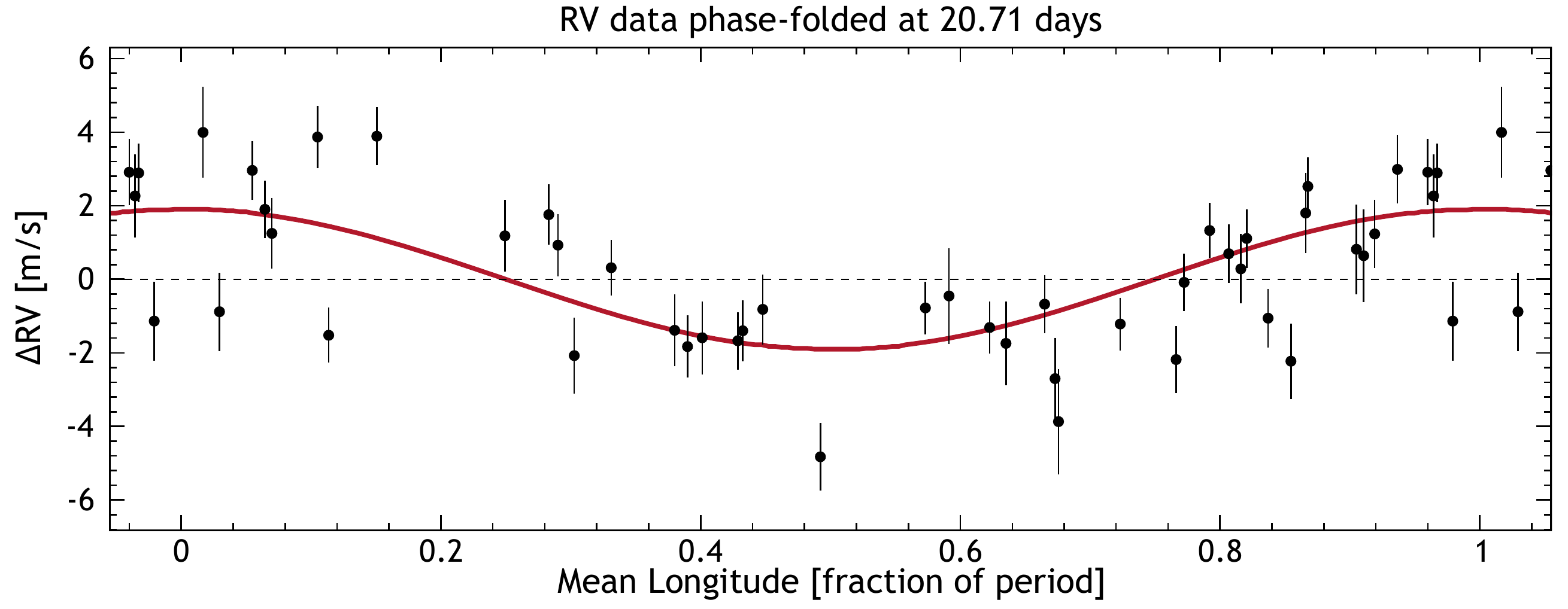} 
        \caption{Phase-folded RV. Error bars correspond to nominal errors.}
        \label{fig:phasefoldedrv}
\end{figure}



\section{Dynamics}
\label{sec:dynamics}
\subsection{A Laplace resonant chain}

\begin{table}
\caption{Instantaneous distances from resonances for the six planets of TOI-178, expressed in terms of near-resonant angles $\phi$. The $\lambda$ symbol indicates the mean longitude of the planet specified in the subscript. }
\label{table:ineq}
\centering
\setlength{\extrarowheight}{2pt}
\begin{tabular}{l c  c  c c}
$ \phi_i$ & $d \phi_i / dt$ [deg/day] & super-period [day] \\
\hline\hline
$ \phi_0=3\lambda_b-5\lambda_c$ & $8.2811_{-0.0062}^{+0.0065}$ & $43.472_{-0.034}^{+0.033}$ \\
$ \phi_1=1\lambda_c-2\lambda_d$ & $1.36886_{-7.6e-04}^{+7.8e-04}$ & $262.99_{-0.15}^{+0.15}$ \\
$ \phi_2=2\lambda_d-3\lambda_e$ & $1.38188_{-9.7e-04}^{+9.4e-04}$ & $260.52_{-0.18}^{+0.18}$ \\
$ \phi_3=2\lambda_e-3\lambda_f$ & $1.37131_{-7.8e-04}^{+8.0e-04}$ & $262.52_{-0.15}^{+0.15}$ \\
$ \phi_4=3\lambda_f-4\lambda_g$ & $1.37050_{-6.0e-04}^{+6.0e-04}$ & $262.68_{-0.11}^{+0.12}$ \\

\end{tabular}
\end{table}

\begin{table*}
\caption{Estimated values of the Laplace angles of planets $c$ to $g$ of TOI-178 towards the beginning of {\it TESS} Sector 2 at 2458350.0$\,$BJD (August 2018). The derivative of the angles are averaged values from between August 2018 and August 2020, based on the solution presented in Tables \ref{table:TOI178bcd} and \ref{table:TOI178efg}. The equilibrium values of the Laplace angles are discussed in Sect. \ref{sec:LapEq}.}
\label{table:Laplace}
\centering
\setlength{\extrarowheight}{2pt}
\begin{tabular}{l c c c c}
$ \psi_j$ & value [deg] & $d \psi_j / dt$ [deg/year] & equilibrium [deg] & distance from eq. [deg]\\
\hline\hline
$ \psi_1=\phi_1-\phi_2=1\lambda_c-4\lambda_d+3\lambda_e$ & $166.70_{-0.82}^{+0.82}$ & $-4.74_{-0.48}^{+0.47}$ & $180$ & $-13.30_{-0.82}^{+0.82}$\\
$ \psi_2=\phi_2-\phi_3=2\lambda_d-5\lambda_e+3\lambda_f$ & $157.57_{-1.06}^{+1.16}$ & $3.86_{-0.60}^{+0.58}$  & $168.94 \pm 7.79$ & $-11.37_{-7.86}^{+7.88}$\\
$ \psi_3= \frac{1}{2}(\phi_3-\phi_4)=1\lambda_e-3\lambda_f+2\lambda_g$ & $71.98_{-0.48}^{+0.38}$ & $0.15_{-0.22}^{+0.23}$  & $78.90 \pm 1.23$ & $-6.92_{-1.32}^{+1.29}$\\
\end{tabular}
\end{table*}


Mean-motion resonances are orbital configurations where the period ratio of a pair of planets is equal to, or oscillates around, a rational number of the form $(k+q)/k$, where $k$ and $q$ are integers. In TOI-178, candidates $b$ and $c$ are close to a second-order MMR ($q=2$): $P_c/P_b=1.6915 \approx 5/3$, while $c$, $d$, $e$, $f$, and $g$ are close, pairwise, to first-order MMRs ($q=1$): $P_d/P_c = 2.0249 \approx 2/1$,  $P_e/P_d = 1.5191 \approx 3/2$, $P_f/P_e =1.5290 \approx 3/2$, and $P_g/P_f = 1.3595 \approx 4/3$. Pairs of planets lying just outside MMRs are common occurrences in systems observed by transit \citep{Fabrycky2014}. To study such pairs of planets, a relevant quantity is the distance to the exact resonance. Taking the pair of $c$ and $d$ as an example, the distance to the $2/1$ MMR in the frequency space is given by $1 n_c-2 n_d$, where $n_c=2 \pi / P_c$ is the mean motion of planet $c$. Following \cite{Lithwick2012}, we name the associated timescale the `super-period':
\be
\label{eq:superperiod}
P_{c,d} \equiv \frac{1}{|(k+q)/P_d - k /P_c|}\, .
\ee

The values of these quantities are given in Table \ref{table:ineq} for planets $b$ to $g$, along with the expression of the associated angles $\phi_i$. Transit timing variations  are expected over the super-period, with amplitudes depending on the distance to the resonance, the mass of the perturbing planet, and the eccentricities of the pair \citep{Lithwick2012}. The fact that the super-periods of all three of these pairs are close to the same value from planet $c$ outwards has additional implications : The difference between the angles $\phi_i$ is evolving very slowly. In other words, there is a Laplace relation between consecutive triplets:
\begin{equation}
\begin{aligned}
d\psi_1 /dt & = d (\phi_1 - \phi_2)/dt =   1n_c-4n_d+3n_e \approx 0\, , \\
d\psi_2 /dt & = d (\phi_2 - \phi_3 )/dt = 2 n_d-5 n_e+3n_f\approx 0\, , \\
d\psi_3 /dt & = \frac{1}{2} d (\phi_3 - \phi_4 )/dt = 1n_e-3 n_f+2n_g\approx 0\, , 
\end{aligned}
\label{eq:LaplaceRelations}
\end{equation}

\noindent implying that the system is in a 2:4:6:9:12 Laplace resonant chain. The values of the Laplace angles $\psi_j$ and derivatives are given in Table \ref{table:Laplace}. The values of the $\psi_j$ are instantaneous and computed at the date 2458350.0 BJD, which is towards the beginning of the observation of TESS Sector 2. As no significant TTVs were determined over the last two years, the derivatives of the $\psi_j$ are average values over that period. The Laplace relations described in Eq. (\ref{eq:LaplaceRelations}) do not extend towards the innermost triplet of the system: According to Eqs. (\ref{eq:LapRel}) and (\ref{eq:DAlembert}), $P_{b,c}$ should be equal to half of $P_{c,d}$ for the {$b$-$c$-$d$} triplet to form a Laplace relation, which is not the case (see Table \ref{table:ineq}). To continue the chain, planet $b$ would have needed a period of $\sim 1.95\,$d. Its current period of $1.91\,$d could indicate that it was previously in the chain but was pulled away, possibly by tidal forces.

Figure \ref{fig:Laplace6p} shows the evolution of the Laplace angles when integrating the nominal solution given in Tables \ref{table:TOI178bcd} and \ref{table:TOI178efg}, starting at the beginning of the observation of TESS Sector 2. The three angles librate over the integrated time for the selected initial conditions, with combinations of periods ranging from a few years to several decades. The exact periods and amplitudes of these variations depend on the masses and eccentricities of the involved planets. The theoretical equilibria of the resonant angles are discussed in Sect.  \ref{sec:LapEq}, while the long-term stability of this system is discussed in Sect. \ref{sec:stability}.

\begin{figure}
  \includegraphics[width = 9cm]{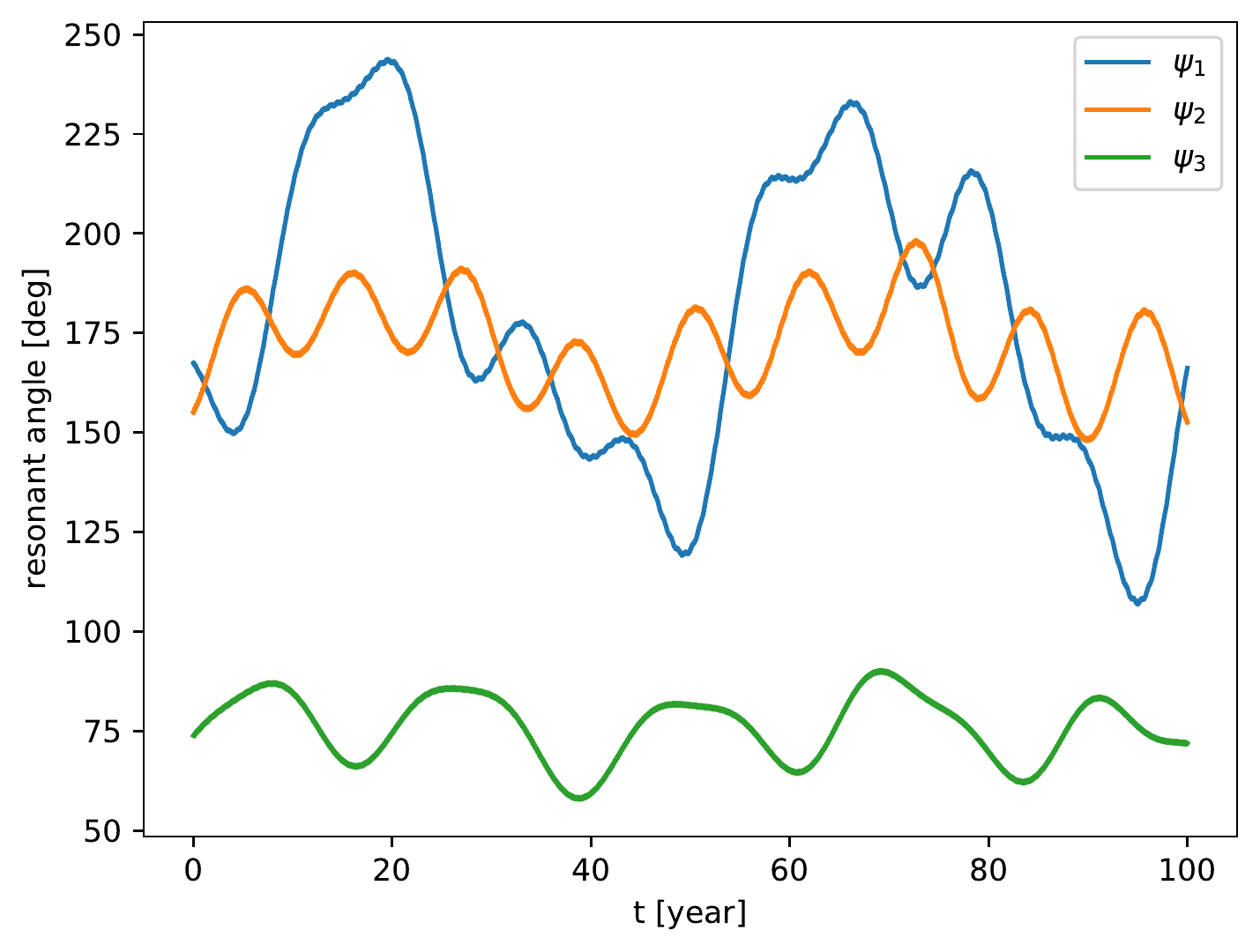}
  \caption{Example of the evolution of the Laplace angles over 100 years, starting from TESS observation of Sector 2, using the masses and orbital parameters from Tables \ref{table:TOI178bcd} and \ref{table:TOI178efg}.}
  \label{fig:Laplace6p}
\end{figure}

\subsection{Equilibria of the resonant chain }
\label{sec:LapEq}

For a given resonant chain, there might exist several equilibrium values around which the Laplace angles could librate \citep{Delisle2017}.
For instance, the four planets known to orbit Kepler-223 are observed to librate around one of the six possible equilibria predicted by theory \citep{Mills16,Delisle2017}.

We used the method described in \cite{Delisle2017} to determine the position of the possible equilibria for the Laplace angles of TOI-178.
The five external planets ($c$ to $g$) orbiting TOI-178 are involved in a 2:4:6:9:12 resonant chain.
All consecutive pairs of planets in the chain are close to first-order MMRs (1:2, 2:3, 2:3, 3:4).
Moreover, as in the Kepler-223 system, there are also strong interactions between non-consecutive planets.
Indeed, planet $e$ and planet $g$ (which are non-consecutive) are also close to a 1:2 MMR.
As explained in \cite{Delisle2017}, this breaks the symmetry of the equilibria (i.e. the equilibrium is not necessarily at 180~deg), and the position of the equilibria for the Laplace angles depends on the planets' masses.

We solved for the position of these equilibria using the masses given in Tables \ref{table:TOI178bcd} and \ref{table:TOI178efg}.
We also propagated the errors to estimate the uncertainty on the Laplace angle equilibria.
We find two possible equilibria for the system, which are symmetric with respect to 0~deg.
We provide the values of the Laplace angles corresponding to the first equilibrium in Table~\ref{table:Laplace} (the second is simply obtained by taking $\psi_j \rightarrow -\psi_j$ for each angle).

It should be noted that these values correspond to the position of the fixed point around which the system is expected to librate. Depending on the libration amplitude, the instantaneous values of the Laplace angles can significantly differ from the equilibrium.
For instance, in the case of Kepler-223, the amplitude of libration could be determined and is about $15$\,deg for all Laplace angles \citep{Mills16}.
In Table~\ref{table:Laplace}, we observe that all instantaneous values of Laplace angles
(as of August 2018) are also found within $15$\,deg of the expected equilibrium. We note that $\psi_1$ was moving away from the equilibrium throughout the two years of observations (see the averaged derivatives in Table~\ref{table:Laplace}).
This would imply a final (as of September 2020) distance from equilibrium of about 23~deg.
On the other hand, $\psi_2$ was moving closer to the equilibrium and $\psi_3$ was only slowly evolving during this two-year span.
A more precise determination of the evolution of Laplace angles (with more data and the detection of TTVs) would be needed to estimate the amplitude of libration of each angle.
However, it is unlikely to find values so close to the equilibrium for each of the three angles by chance alone.
Therefore, these results provide strong evidence that the system is indeed librating around the Laplace equilibrium.

\subsection{Stability}
\label{sec:stability}
Planets $c$ to $g$ are embedded in a resonant chain, which seems to greatly
stabilise this planetary system.
The distance between the planets being quite small, their eccentricities may become a major
source of instability. This point is illustrated by Fig. \ref{fig:stab_ae_g}, which
shows a section of
the system's phase space. The initial conditions and masses of the planets are displayed in Tables \ref{table:TOI178bcd} and \ref{table:TOI178efg}, with the exception of those of planet $f$, for which the initial period and eccentricity vary; all other planets start on circular orbits.
The colour code corresponds to a stability index based on
the diffusion of the main frequencies of the system defined as \citep{Laskar1990,Laskar1993}:
\be
\log_{10} \left\vert n_f^{(1)} - n_f^{(2)}\right\vert  \, ,
\label{eq:diffusion}
\ee
where $n_f^{(1)}$ and $n_f^{(2)}$ are the proper mean motion of planet $f$ in degree per year, computed over the first half and the second half of the integration, respectively, implying that the stability
increases from red to black \citep[for more details, see Sect. 4.1 of][]{Petit2018}. This map shows that the eccentricity of
planet $f$ must not exceed a few hundredths if the system is to remain stable. The
same stability maps (not reproduced here) made for each of the planets of the system lead to the same result: The planetary eccentricities have to be small in order to
guarantee the system's stability. 
This constrain is verified if the system starts with sufficiently small eccentricities. In particular, we verified that starting a set of numerical integrations on circular orbits with masses and initial conditions close to those of the nominal solution and integrating the system over 100 000 years 
(i.e. about 19 million orbits of planet $b$ or 1.3 million orbits of planet $g$)
does not excite the eccentricities above 1/100 in most cases.

All the $P_i$ versus $e_i$ stability maps, such as the one in the top panel of Fig. \ref{fig:stab_ae_g}, show long quasi-vertical structures that contain very stable regions
in their central parts and less stable regions at their edges. These are mainly MMRs between two or three planets. In particular, the black area crossed by the dashed white line in the top panel of Fig. \ref{fig:stab_ae_g} corresponds to the resonant
chain where the nominal system is located.

The bottom panel of Fig. \ref{fig:stab_ae_g} shows a different section of the phase space where
$P_f$ and $P_g$ vary and all eccentricities are initialised at 0. This reveals part of the resonant structure of the
system. The black regions are stable, while the green to red areas mark the instability
caused by the resonance web. This figure still highlights the stability island in which
the planetary system is located. This narrow region is surrounded by resonances: the $n_e
- 3n_f + 2n_g=0$ Laplace resonance (central diagonal  strip), high-order two-body MMRs
between planets $f$ and $g$ (horizontal strips), and high-order two-body MMRs between planets
$e$ and $f$ (vertical lines).

  The extent of this resonant chain versus the initial period and mass of planet $f$ is shown in Fig.~\ref{fig:stab_Pm_f}.  On both panels, the x-axis corresponds to the orbital period of planet $f$, while the y-axis corresponds to the planetary mass.  The figure presents two
different, but complementary, aspects of the dynamics of TOI-178.
The bottom panel indicates whether the Laplace angles $\psi_1$, $\psi_2$, and $\psi_3$
librate or not. More precisely, the colour code corresponds to the number of angles that
librate during the first 200 years of integration. The top panel present the same stability indicator as Fig. \ref{fig:stab_ae_g}. Three regions stand out clearly from this figure, each with a different dynamical regime.

  The central region (yellow in the bottom panel), where the three Laplace angles librate
simultaneously (see also Fig. \ref{fig:Laplace6p}), shows the heart of the 2:4:6:9:12
resonant chain, where the nominal system is located. On the stability map (top), its dark
black colour reveals a very low diffusion rate and therefore a very long-term stability.
This central region seems not to depend strongly on the mass of  planet $f$.

  On the other hand, the modification of the orbital period within a fraction of a day changes the dynamics of the system.
Indeed, outside of the central region where the three Laplace angles librate, which is
about 0.015 days wide, chaotic layers are present (red in the top panel panel, dark blue in the
bottom) where none of the three Laplace angles librate. Here, the red colour of the stability
index corresponds to a significant, but moderate,  diffusion rate. Thus, although in this
region the trajectories are not quasi-periodic, the chaos has limited consequences (it is bounded) and
probably does not lead to the destruction of the system.

Outside of these layers lies a very stable (quasi-periodic) region. The blue on the Laplace
angle map shows that only one Laplace angle librates, while the others circulate.
Although the 2:4:6:9:12 chain is broken, planets $c, d, e$, and $g$ remain inside the
2:4:6:12 resonant chain, independently of the orbital period of planet $f$. The stability map indicates a very strong regularity of
the whole region. Nevertheless, one can notice the presence of some narrow zones where the
diffusion is more important; they are induced by high-order orbital resonances but do not have any
significant consequence on the stability of the system.

The study described in this section gives only a very partial picture of the
structure of the phase space (parameter space) of the problem.\ Nevertheless, it can be seen that as long
as the system is in the complete resonant chain, which is the case for the nominal derived parameters and for the bulk of the posterior given in Tables \ref{table:TOI178bcd} and \ref{table:TOI178efg}, it remains stable.


\begin{figure}
\hfill \includegraphics[width = 8cm]{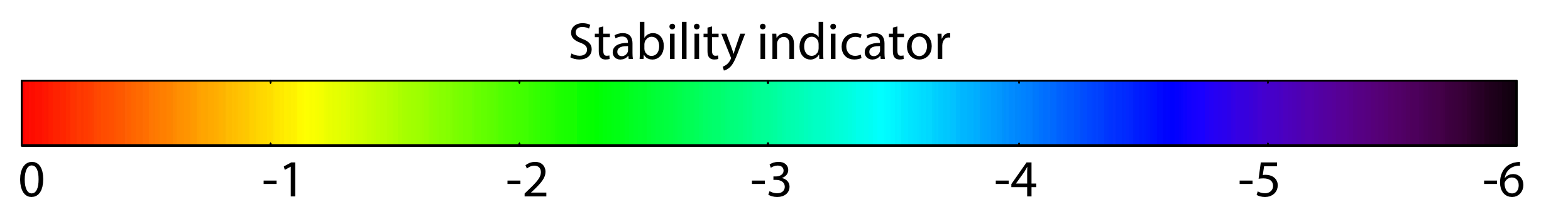}\\
  \includegraphics[width = 9cm]{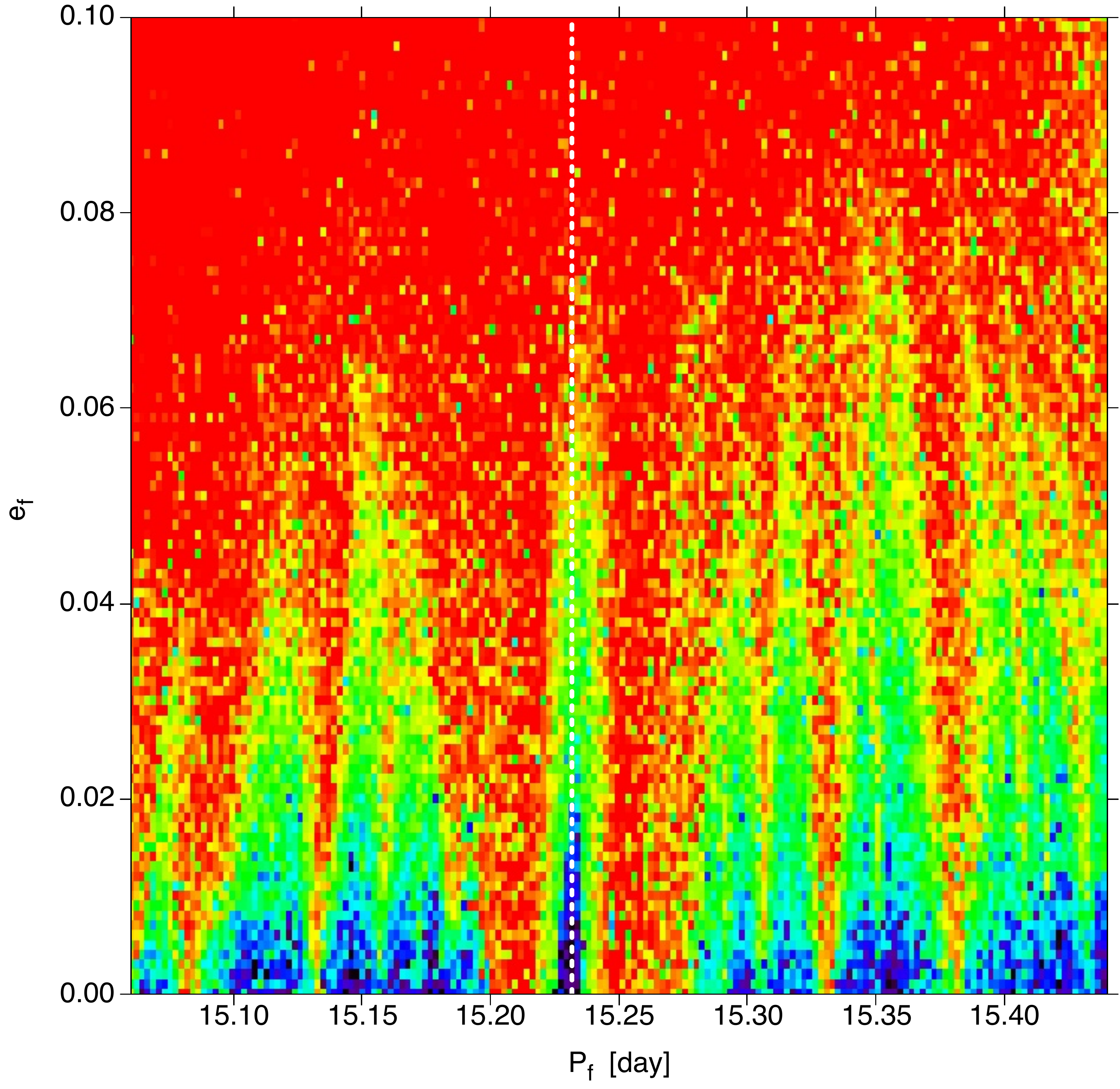}\\
  \includegraphics[width = 9cm]{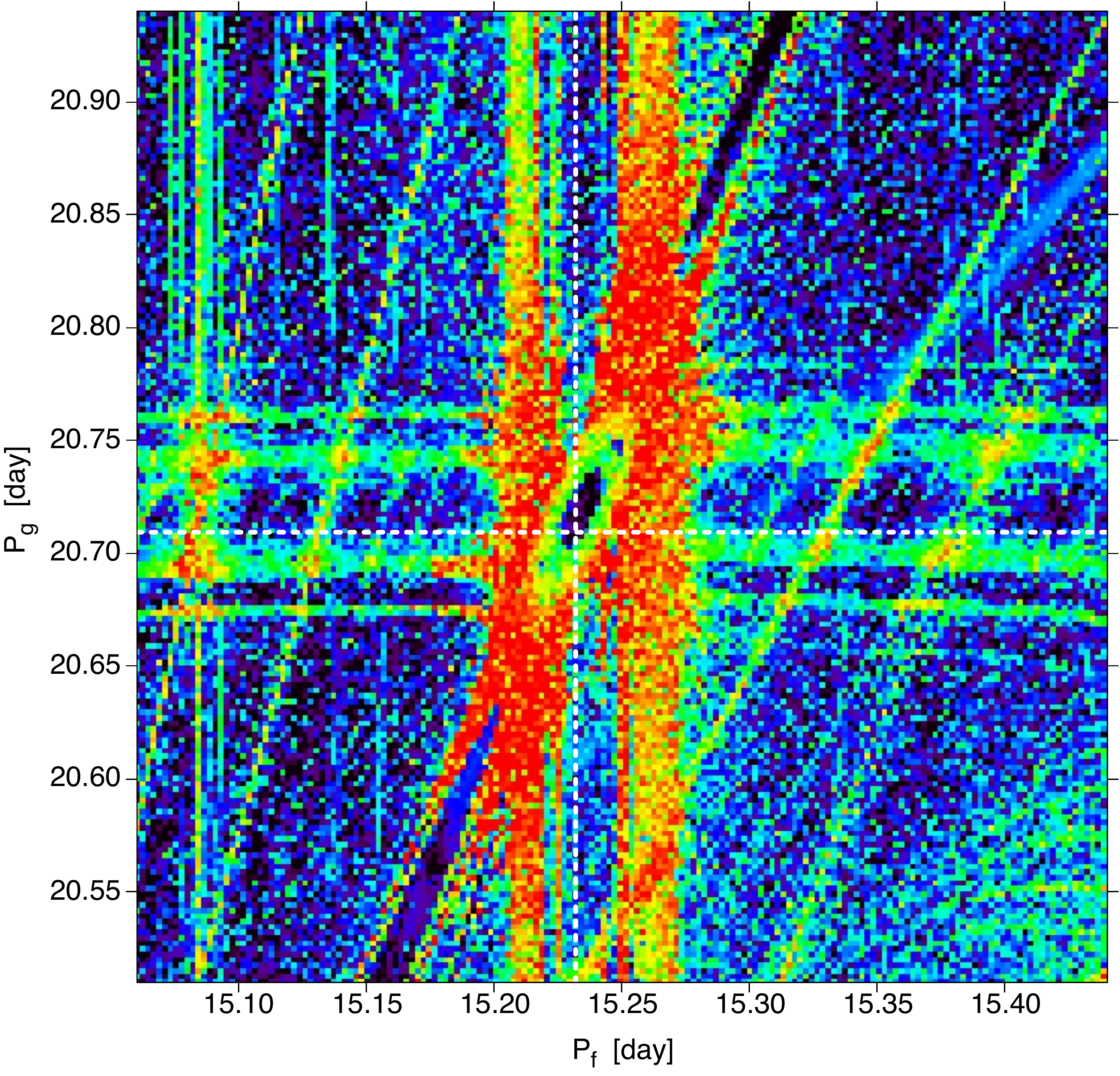}
  \caption{Stability indicator (defined in Eq. \ref{eq:diffusion}) for TOI-178 as a function of the periods and eccentricity of planet $f$ (\textit{top}). The red areas show the initial conditions of unstable trajectories, while black shows stable (quasi-periodic) trajectories. The \textit{bottom} panel shows the same stability criterion with respect to the initial periods of planets $f$ and $g$. The dashed white lines show the observed periods reported in Table \ref{table:TOI178efg}.}
  \label{fig:stab_ae_g}
\end{figure}

\begin{figure}
  \includegraphics[width=9cm]{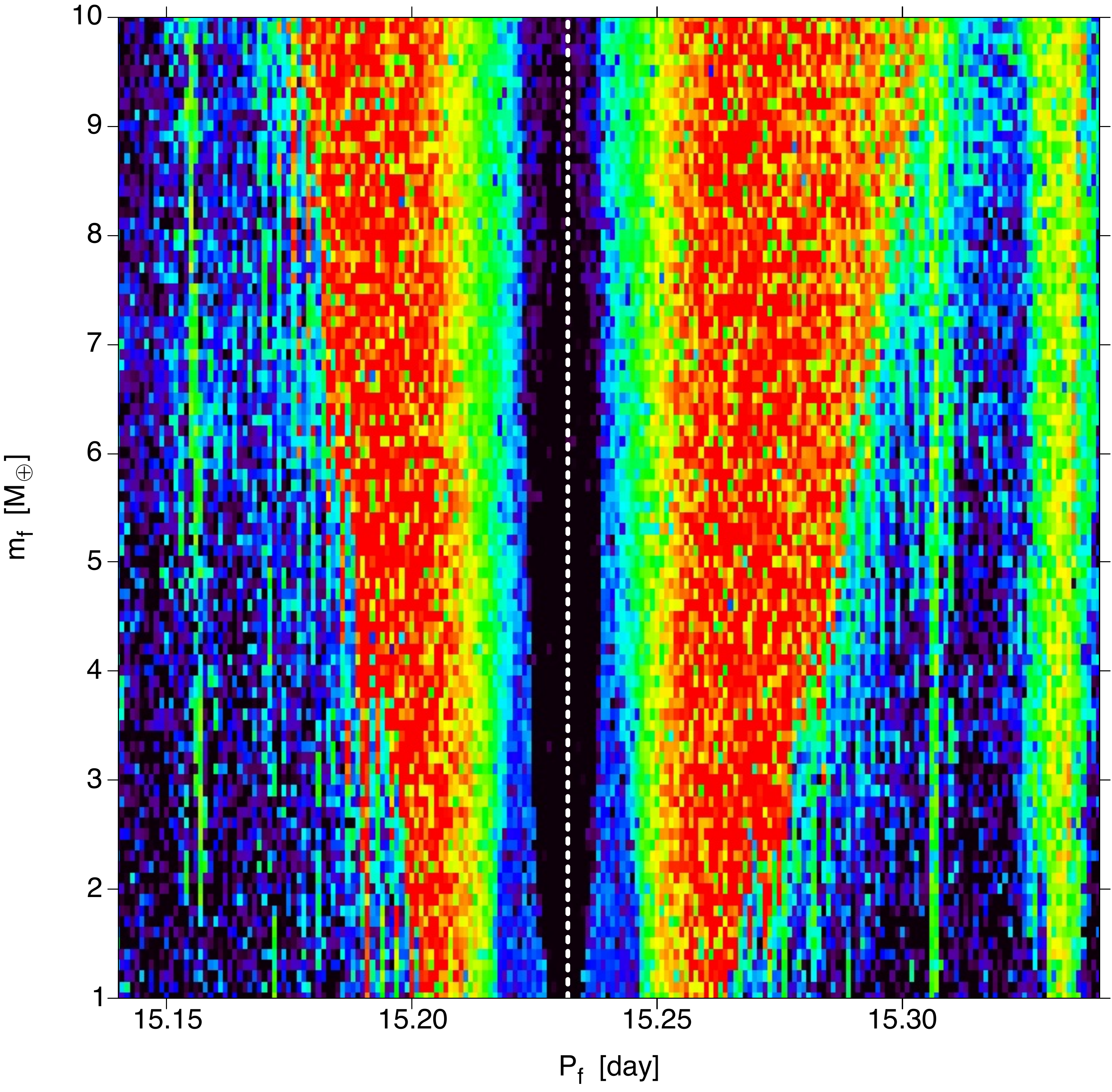}

  \includegraphics[width=9cm]{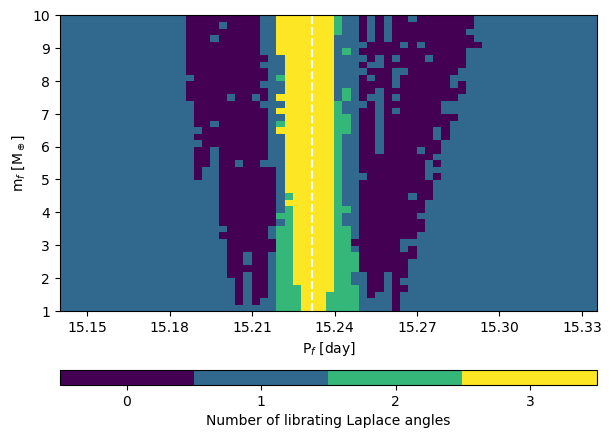}
  \caption{Same stability indicator as in Fig. \ref{fig:stab_ae_g} (\textit{top}) and the number of librating Laplace angles (\textit{bottom}) for TOI-178 as a function of the period and mass of planet $f$. The dashed white line shows the observed period of planet $f$  reported in Table \ref{table:TOI178efg}. }
  \label{fig:stab_Pm_f}
\end{figure}

%
%

\subsection{Expected TTVs}
\label{sec:TTVs}

\begin{figure*}
  \includegraphics[width=\linewidth]{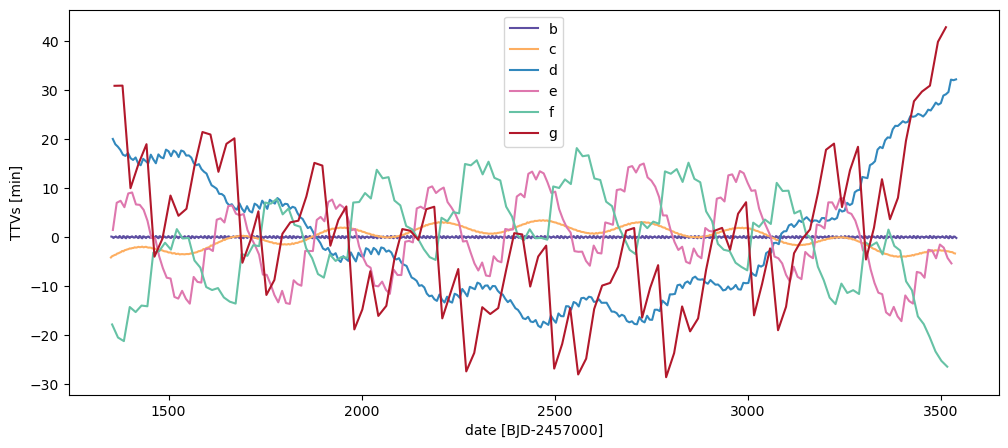}
  \caption{Example of TTVs of the six planets of TOI-178, starting from the TESS observation of Sector 2 and spanning 6 years, using the masses and orbital parameters from Tables \ref{table:TOI178bcd} and \ref{table:TOI178efg}.}
  \label{fig:TTVs6p}
\end{figure*}

Since planets $c$, $d$, $e$, $f$, and $g$ are, pairwise, close to first-order MMRs, we expect TTVs to occur over the super-period  \citep[Eq. \ref{eq:superperiod}, see also][]{Lithwick2012}, which is roughly 260 days for all these pairs (Table \ref{table:ineq}). The amplitude of these TTVs depends on the masses and eccentricities of each planet of a pair. Since the stability analysis concludes that the system is more stable with eccentricities close to 0, we integrated the initial conditions described in Tables \ref{table:TOI178bcd} and \ref{table:TOI178efg} over 6 years, starting during the observation of TESS Sector 2 using the {\ttfamily rebound} package \citep{rebound,rebound2}. The resulting TTVs are shown in Fig. \ref{fig:TTVs6p}. In addition to the terms coming from the super-periods, the six-year evolution of the TTVs shows hints of the long-term evolution of the Laplace angles. 

As the exact shape of the TTVs depends mainly on the masses of the involved planets, the sampling of the 260-day signal over a few years, in addition to a longer monitoring of the evolution of the Laplace angles, can provide precise constrains on the orbital configuration and masses of this system, as well as the presence of other planets in the chain. The detailed study of the TTVs of this system will be the subject of a forthcoming paper.

\section{Internal structure}
\label{sec:internal_structure}

\subsection{Minimum mass of the protoplanetary disc}

The total mass of the planets detected in TOI-178 
is approximately $24.8^{+5.96}_{-6.23} M_\oplus$. Assuming a mass fraction of H and He of maximum $\sim 20 \%$ (similar to those of Uranus and Neptune), the amount of heavy elements in the planets is at least $14.85 M_\oplus$ (16th quantile). This number can be compared with the mass of heavy elements one would expect in a protoplanetary disc similar to the Minimum Mass Solar Nebula (MMSN) but around TOI-178. Assuming a disc mass equal to 1\% of the stellar mass, and using the metallicity of TOI-178 ($[Fe/H] = -0.23$, corresponding to a metal content of $Z\sim 0.0061$)}, the mass of heavy elements in such a disc would equal $\sim 13 M_\oplus$, which is remarkably similar to the minimum mass of heavy elements in planets, as mentioned above. The concept of the MMSN, scaled appropriately to reflect the reduced mass and metallicity of TOI-178, seems therefore to be as applicable for this system as for the Solar System.
We note, however, that in \citep{Hayashi1981} the MMSN model is assumed to be based on the \textit{in situ} formation of planets, whereas here we compare the mass of planets with the whole mass of solids in the inner parts.
Another implication of this comparison is that a problem in term of available mass would appear should the planetary masses be revised to higher values or should another massive planet be detected in the same system. In such a case, the TOI-178 system would point towards a formation channel similar to the one envisioned for the Trappist-1 system \citep{schoonenberg19}.

\subsection{Mass-radius relation}

The six planets we detected in the TOI-178 system are in the super-Earth to mini-Neptune range, with radii ranging from $1.152_{-0.070}^{+0.073}$ to $2.87_{-0.13}^{+0.14}$ $R_\oplus$. Although the mass determination is limited by the extent of the available spectroscopic dataset, planets $b$ and $c$ appear to have roughly terrestrial densities of $0.98^{+0.35}_{-0.31} \rho_\oplus$ and $1.02^{+0.28}_{-0.23}$, respectively, where $ \rho_\oplus$ is the density of the Earth.  The outer planets seem to have significantly lower densities; in particular, we estimate the density of planet $d$ to be $0.177^{+0.055}_{-0.061} \rho_\oplus$.

\begin{figure*}
        \includegraphics[width=\linewidth]{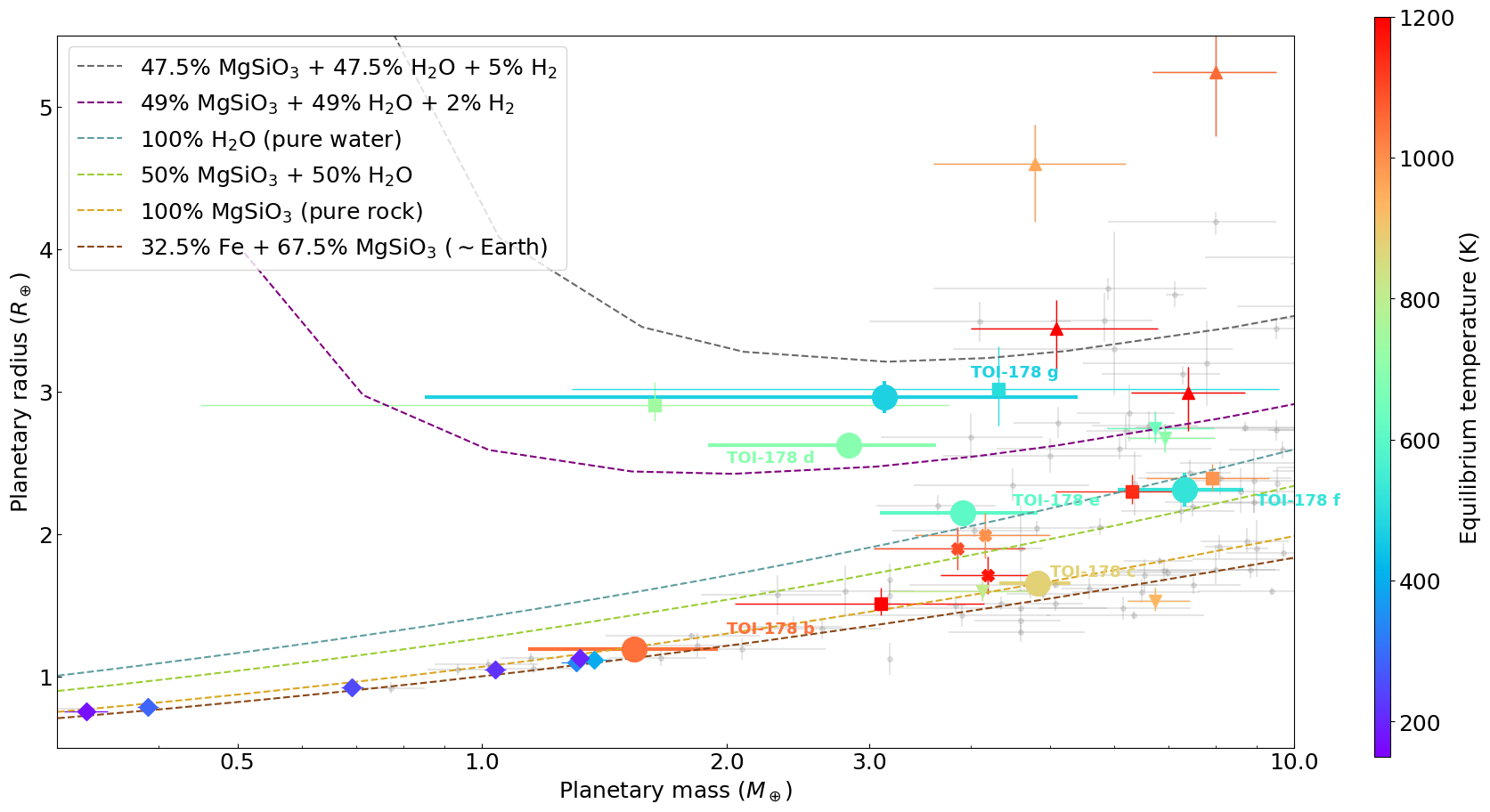}
        \caption{ TOI-178 planets compared to other known transiting exoplanets with radius and mass uncertainties less than 40$\%$ (grey) and other systems known to harbour a Laplace resonance. Data on known exoplanets were taken from the NASA Exoplanet Archive
        on 18 September 2020. The dashed lines show theoretical mass-radius curves for some idealised compositions \citep{Zeng19}. The six planets orbiting around TOI-178 are indicated; the colour of the points and error bars give the equilibrium temperature. The seven planets orbiting Trappist-1 are shown with diamonds, and the parameters are taken from \citet{Agol2020}. The three planets orbiting Kepler-60 are shown with 'X' marks, and the parameters are taken from \citet{K60}. The six planets orbiting K2-138 are shown with squares, and the parameters are taken from \citet{Lopez2019}. The four planets orbiting Kepler-80 are shown with inverse triangles, and the parameters are taken from \citet{MacDonald2016}. The four planets orbiting Kepler-223 are shown with regular triangles, and the parameters are taken from \citet{Mills16}.}
        \label{fig:MR}
\end{figure*}

Figure \ref{fig:MR} shows the position of the six planets in a mass-radius diagram, in comparison with planets with mass and radius uncertainties less than 40$\%$ (light grey). Planets belonging to five systems in Laplace resonance are indicated in the same diagram: Trappist-1, K2-138, Kepler-60, Kepler-80, and Kepler 223. The diversity of planetary composition in TOI-178 is clearly visible in the diagram: The two inner planets have a radius compatible with a  gas-free structure, whereas the others contain water and/or gas. This is similar to the Kepler-80 system, where the two innermost planets are compatible with a gas-free structure and the two outermost ones likely contain gas. Planets in Kepler-223, Kepler-60, and Trappist-1 seem to have a more homogeneous structure: All planets in Kepler-223  have a gas envelope, and the planets in Kepler-60 and Trappist-1 have small gas envelopes (below $\sim.01\%$ of the total mass). 

Considering the TOI-178 system in greater detail, planets $d$, $e$, and $ g$ are located above the pure water line and definitely contain a non-zero gas mass fraction. Planet $d$ and, depending on its mass, planet $g$ are located in a part of the diagram where no other planets exist (at least no planets with mass uncertainties smaller than 40$\%$) and must contain a large gas fraction.

\begin{figure}
  \includegraphics[width = 9cm]{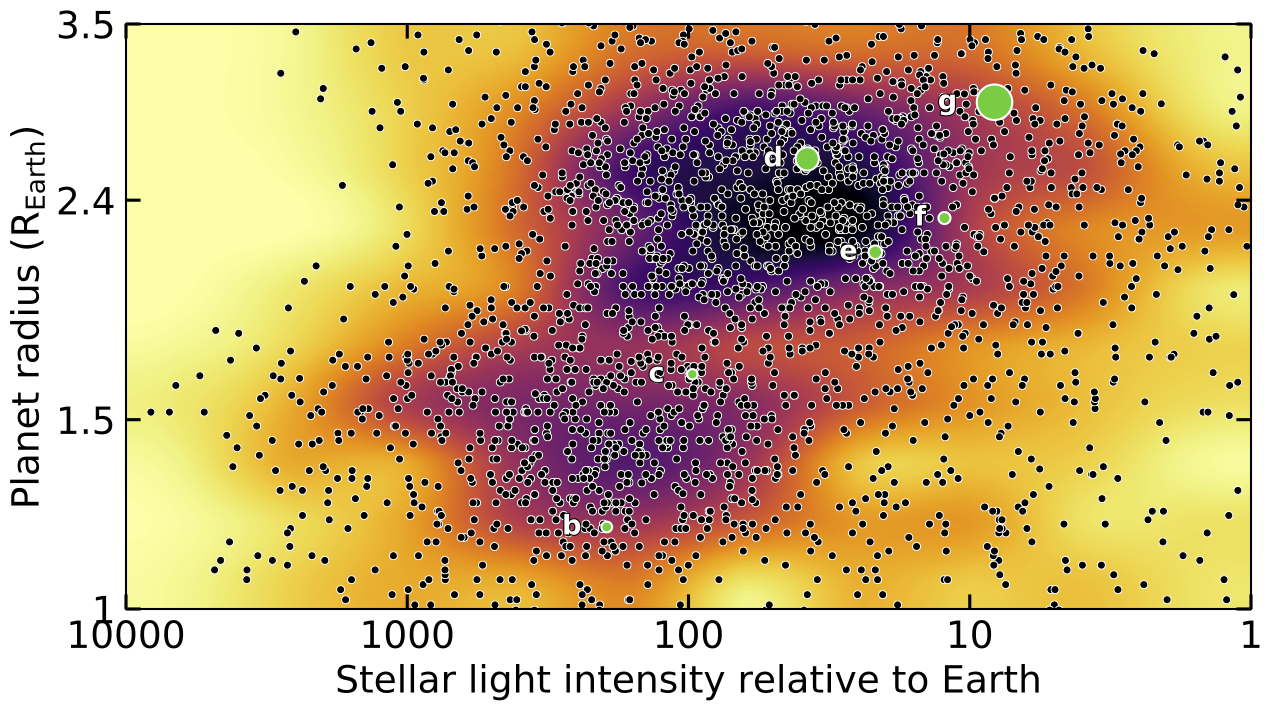}
   \caption{Diagram of the position of the TOI-178 planets in a stellar light intensity (relative to Earth) \textit{versus} planetary radius. The marker size is inversely proportional to the density, and the colour code gives the density of exoplanets, from yellow for empty regions of the diagram to violet for high-density and/or highly populated regions. }
  \label{fig:evaporation}
\end{figure}

 Figure \ref{fig:evaporation}  is a diagram showing the location of the TOI-178 planets in stellar insolation versus the planetary radius. The colour code illustrates the density of exoplanets. This diagrams clearly shows the so-called evaporation valley\footnote{We note that if the presence of a valley seems robust, it could be due to effects that are not related to evaporation, e.g. core cooling \citep{schlichting19}, or from combined formation and evolution effects \citep{julia20}.}. Planets $b$ and $c$ are located below the valley, and their high densities could result from the evaporation of a primordial envelope. The outer planets  are located above the valley and have probably preserved (part of) their primordial gas envelope. 

\subsection{Comparison with other systems in Laplace resonance}

\begin{figure*}
          \includegraphics[width=\linewidth]{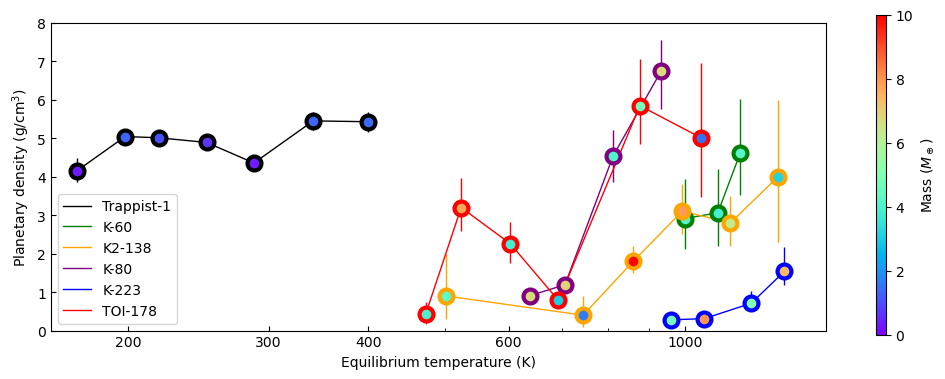}
        \caption{Densities of planets in the TOI-178, Trappist-1, K2-138, Kepler-60, Kepler-80, and Kepler-223 systems as a function of their equilibrium temperatures. 
         The error bars give the $16\%$ and $84 \%$ quantiles,
        and the marker is located at the median of the computed distribution. The colour code gives the planetary masses in Earth masses. The parameters of the planets are taken from the references mentioned in Fig. \ref{fig:MR}.}        
        \label{fig:density}
\end{figure*}

The difference between the TOI-178 system and other systems in Laplace resonance can clearly be seen in Fig. \ref{fig:density}, where we show the density of planets as a function of their equilibrium temperature for the same systems as in Fig. \ref{fig:MR}. In Kepler-60, Kepler-80, and Kepler-223, the density of the planets decreases when the equilibrium temperature decreases. This can be understood as the effect of evaporation that removes part of the primordial gaseous envelope, this effect being stronger for planets closer to their stars. The densities of K2-138 are also 1-$\sigma$ consistent with such behaviour.  In Trappist-1, the density of planets is always higher than 4 g/cm$^3$ and increases (with the exception of Trappist-1 f) with decreasing equilibrium temperature. This is likely due to the presence of more ices in planets far from their stars \citep{Agol2020}. {Contrary to the three Kepler systems, in the TOI-178 system the density of the planets is not a growing function of the equilibrium temperature. Indeed, TOI-178f has a density higher than that of planet $e$, and TOI-178d has a density smaller than that of planet $e$.}

Planet $f$ is substantially more massive than all the other planets in the system. From a formation perspective, one would expect this planet to have a density smaller than the other planets, in particular planet $e$, at the end of the formation phase, similar to what we observe, for example, for Jupiter, Saturn, Uranus, and Neptune. Since this planet is farther away from the star compared to planet $e$, evaporation should have been less effective for planet $f$ compared to planet $e$. The combined effect of formation and evolution should therefore lead to a smaller density for planet $f$ compared to planet $e$. Similarly, planet $d$ is smaller than planet $e$ and is located closer to the star. Using the same arguments, the combined effect of formation and evolution should have led to planet $d$ having a density larger than that of planet $e$. The TOI-178 system therefore seems at odds with the general understanding of planetary formation and evaporation, where one would expect the density to decrease when the distance to the star increases, or when the mass of the planet increases. 

\subsection{Internal structure modelling}

We used a Bayesian analysis to compute the posterior distribution of the internal planetary structure parameters. The method we used closely follows the one in \citet{Dorn15} and \citet{Dorn17} and has already been used in \citet{Mortier20} and Delrez (2020 - \textit{submitted}). Here we review the main physical assumptions of the model.

The model is split into two parts. The first is the forward model, which provides the planetary radius as a function of the internal structure parameters, and the second is the Bayesian analysis, which provides the posterior distribution of the internal structure parameters, given the observed radii, masses, and stellar parameters (in particular their composition).

For the forward model, we assume that each planet is composed of four layers: an iron-sulfur inner core, a mantle, a water layer, and a gas layer. We used the equation of state (EOS) for \citet{Hakim} for the core, the EOS from \citet{Sotin} for the silicate mantle, and the  EOS from \citet{Haldemann_20} for the water. These three layers constitute the `solid' part of the planets. The thickness of the gas layer (assumed to be made of pure H/He) is computed as a function of the stellar age, mass, and radius of the solid part as well as irradiation from the star, using the formulas in \citet{LopezFortney14}. The internal structure parameters of each planet are therefore the iron molar fraction in the core, the Si and Mg molar fraction in the mantle, the mass fraction of all layers (inner core, mantle, and water), the age of the planet (equal to the age of the star), and the irradiation from the star. More technical details regarding the calculation of the forward model are given in Appendix \ref{ap:internal_structure}.

In the Bayesian analysis part of the model, we proceeded in two steps. We first generated 150000 synthetic stars, taking their masses, radii, effective temperatures, and ages at random following the stellar parameters computed in Sect. \ref{sec:star}. The Fe/Si/Mg bulk molar ratios in the star are assumed to be solar, with an uncertainty of 0.05 (uncertainty on [Fe/H], see Sect. \ref{sec:star}). For each of these stars, we generated 1000 planetary systems, varying the internal structure parameters of all planets and assuming that the bulk Fe/Si/Mg molar ratios are equal to the stellar ones. We then computed the transit depth and RV semi-amplitude for each of the planets and retained the models that fitted the observed data within the error bars. 
By doing so, we included the fact that all synthetic planets orbit a star with exactly the same parameters. Indeed, planetary masses and radii are correlated by the fact that the fitted quantities are the transit depth and RV semi-amplitude, which depend on the stellar radius and mass. In order to take this correlation into account, it is therefore important to fit the planetary system at once, rather than fitting each planet independently.

For the Bayesian analysis, we assumed the following two priors: first, that the mass fraction of the gas envelope is uniform in log; and second, that the mass fraction of the inner core, mantle, and water layer are uniform on the simplex (the surface on which they add up to one) for the solid part. We also assumed: the mass fraction of water to be smaller than 50$\%$ \citep{Thiabaud,Marboeuf}; the molar fraction of iron in the inner core to be uniform between 0.5 and 1; and the molar fraction of Si, Mg, and Fe in the mantle to be uniform on the simplex (they also add up to one). In order to compare TOI-178 with other systems with a Laplace resonance, we also performed a Bayesian analysis for the K2-138, Kepler-60, Kepler-80, and Kepler-223 systems to compute the probability distribution of the gas mass in the different planets. The parameters for all systems are taken from the references mentioned above, and for all systems we have assumed that [Si/H]=[Mg/H]=[Fe/H]. We did not consider the Trappist-1 system in this comparison as it is likely that the variations in the densities of these planets result from variations in their ice contents \citep{Agol2020}. 

\begin{figure*}
        \includegraphics[width=\linewidth]{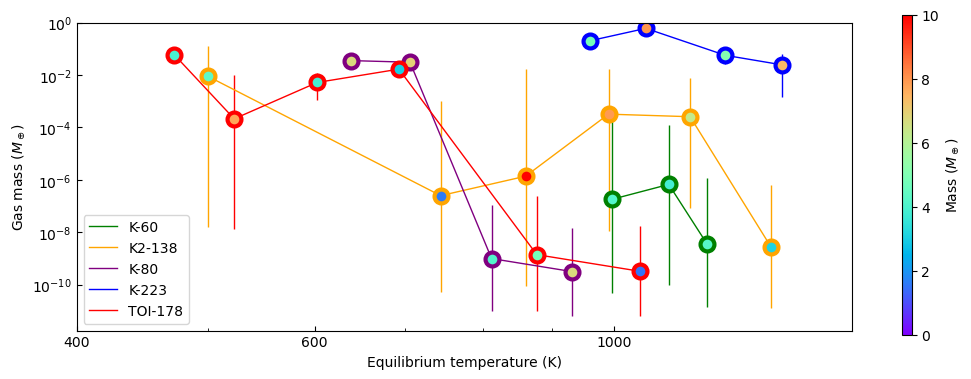}
        \caption{Gas mass in the planets of the TOI-178, K2-138, Kepler-60, Kepler-80, and Kepler-223 systems as a function of their equilibrium temperatures. 
        The error bars give the $16\%$ and $84 \%$ quantiles 
        and the marker is located at the median of the computed distribution. The colour code gives the planetary masses in Earth masses. The parameters of the planets are taken from the references mentioned in Fig. \ref{fig:MR}.}
        \label{fig:gasmass}
\end{figure*}

The posteriors distributions of the two most important parameters (mass fractions and composition of the mantle) of each planet in TOI-178 are shown in Appendix \ref{ap:internal_structure}, Figs. \ref{fig:cornerb} to \ref{fig:cornerg}. We focus here on the mass of gas in each planet, and we plot the mass of the gaseous envelope for each planet as a function of their equilibrium temperature in Fig. \ref{fig:gasmass}. 

In all three Kepler systems, the gas mass in planets generally decreases when the equilibrium temperature decreases, as can be seen in Fig. \ref{fig:gasmass}. One exception to this tendency is the mass of the envelope of Kepler-223d, which is larger than that of planet $e$ in the same system. Kepler-223d is, however, more massive than planets $c$ and $e$ in the same system, and one would expect from formation models that the mass of the primordial gas envelope is a growing function of the total planetary mass. In K2-138, the gas fraction is compatible (within 1-sigma error bars) with a monotonous function of the equilibrium temperature. In the case of TOI-178, the mass of gas also globally increases when the equilibrium temperature decreases, with the notable exception of planet $d$. Indeed, a linear interpolation would provide a gas mass for planet $d $ of the order of $10^{-6} - 10^{-5} M_\oplus$, whereas the interior structure modelling gives values of  $9.66 \times 10^{-3} M_\oplus$ and $2.56 \times 10^{-2} M_\oplus$ for the $16 \%$ and $84 \%$ quantiles. Even more intriguing, our results show that the amount of gas in planet $d$ is larger than in planet $e$, the latter being both more massive and at a larger distance from the star. Indeed, from the joint probability distribution of all planetary parameters as provided by the Bayesian model, the probability that planet $d$ has more gas than planet $e$ is  $92 \%$. 

From a formation point of view, one would generally expect that the mass of gas is a growing function of the core mass. From an evolution perspective, one would also expect that evaporation is more effective for planets closer to the star. Both would point towards a gas mass in planet $d$ that should be smaller than that of planet $e$. The large amount of gas in planet $d$ and, more generally, the apparent irregularity in the planetary envelope masses are surprising in view of the apparent regularity of the orbital configuration of the system, which was presented in Sect. \ref{sec:dynamics}.

\section{Discussion and conclusions}
\label{sec:discussion}

In this study, we presented new observations of TOI-178 by CHEOPS, ESPRESSO, NGTS, and SPECULOOS. Thanks to this follow-up effort, we were able to determine the architecture of the system: Out of the three previously announced candidates at 6.56\,d, 9.96\,d, and 10.3\,d, we confirm the first two (6.56\,d and 9.96\,d) and re-attribute the transits of the third to 15.23\,d and 20.7\,d planets, in addition to the detection of two new inner planets at 1.91\,d and 3.24\,d; all of these planets were confirmed by follow-up observations. In total, we therefore announce six planets in the super-Earth to mini-Neptune range, with orbital periods from 1.9\,d to  20.7\,d, all of which (with the exception of the innermost planet) are in a 2:4:6:9:12 Laplace resonant chain. All the orbits appear to be quasi-coplanar, with projected mutual inclinations between the outer planets estimated at about $0.1$\,deg, a feature that is also visible in other systems with three-body resonances, such as Trappist-1 \citep{Agol2020} and the Galilean moons. Current ephemeris and mass estimations indicate that the TOI-178 system is very stable, with Laplace angles librating over decades.
In the TESS Sector 29 data, made available during the referee process of this paper, we recovered the transits of planets $b$, $c$, $d$, $e$, and $f$ at the predicted dates. The transit of planet $g$ was not observed as it transited during the gap and was observed during the third CHEOPS visit (see Fig. \ref{fig:CHEOPS_11}).

As there is no theoretical reason for the resonant chain to stop at 20.7 days, and the current limit probably comes from the duration of the available photometric and RV datasets, we give in Table \ref{table:Px} the periods that would continue the resonant chain for first-order ($q=1$) and second-order ($q=2$) MMRs. These periods result from the equations detailed in Appendix \ref{ap:Px}. In the TOI-178 system, as well as in similar systems in Laplace resonance, planet pairs are virtually all nearly first-order MMRs. The most likely of the periods shown in Table \ref{table:Px} are therefore 
the first-order solutions with low $k$, hence 45.00, 32.35, or 28.36 days; however, additional planets with such periods would not be transiting if they were in the same orbital plane as the others (see Fig. \ref{fig:inclination}).
We  note that, for a star such as TOI-178, the inner boundary of the habitable zone lies around 0.2 AU, or at a period of the order of $~40$ days. Additional planets in the Laplace resonance could therefore orbit inside, or very close to, the habitable zone.\\
 
\begin{table}
\caption{Potential periods that would continue the resonant chain in a near $(k+q):k$ MMR with planet $g$, taking the super-period at 260 days (see  equations in Appendix \ref{ap:Px}). Changing the super-period by a few days typically changes the resulting period by less than 0.1 day.}
\label{table:Px}
\centering
\setlength{\extrarowheight}{2pt}
\begin{tabular}{l c  c}
$k,q$ &Period [day]\\
\hline\hline
$1,1$ & 45.0028 \\
$2,1$ & 32.3522 \\
$3,1$ & 28.3653 \\
$4,1$ & 26.4124 \\
$5,1$ & 25.2533 \\
$3,2$ & 36.4508 \\
$5,2$ & 29.9470 \\
$7,2$ & 27.2461 \\
\end{tabular}
\end{table}

The brightness of TOI-178 allows for further characterisation of the system by photometric measurements, radial velocities, and transit spectroscopy. These measurements will be essential for further constraining the system, not only on its orbital architecture but also for the physical characterisation of the different planets.

As discussed above, the current mass and radius determinations show significant differences between the different planets. It appears that the two innermost planets are likely to be rocky, which may be due to the fact that they have lost their primary, hydrogen-dominated atmospheres through escape \citep{kubyshkina2018,kubyshkina2019}, whereas all the other planets may have retained part of their primordial gas envelope. In this respect, the different planets of the TOI-178 system lie on both sides of the radius valley \citep{fulton2017}. Therefore, reconstructing the past orbital and atmospheric history of this planet may provide clues regarding the origin of the valley. The system is located at a declination in the sky that makes it observable by most ground-based observatories around the globe. Furthermore, the host star is bright enough and the radii of the outer planets large enough to make them potentially amenable to optical and infrared transmission spectroscopy observations from both ground and space, particularly by employing the upcoming E-ELT and James Webb Space Telescope (JWST) facilities. Indeed, it has been shown that the JWST has the capacity to perform transmission spectroscopy of planets with radii down to 1.5 $R_\oplus$ \citep{samuel14}.

Planets $d$ and $f$ are particularly interesting as their densities are very different from those of their neighbours, and they depart from the general tendency of planetary density (i.e. decreasing for decreasing equilibrium temperatures). The densities of planets $d$ and $f$, in the context of the general trend seen in TOI-178, are difficult to understand in terms of formation and evaporation process, and they could be difficult to reproduce with planetary system formation models \citep{Mordasini12,Alibert13,NGPPS1}. We stress that even though two different analyses yielded similar estimates of masses and densities, they were performed with only 46 data points. As such, these estimates need to be confirmed by further RV measurements, which would provide, in particular, a better frequency resolution and confirm the mass estimate of planet $f$. 

The orbital configuration of TOI-178 is too fragile to survive giant impacts, or even significant close encounters: Fig. \ref{fig:stab_ae_g} shows that a sudden change in period of one of the planets of less than a few $\sim .01\,$d can render the system chaotic, while Fig. \ref{fig:stab_Pm_f} shows that modifying a single period axis can break the resonant structure of the entire chain.
Understanding,  in a single framework, the apparent disorder in terms of planetary density on one side and the high level of order seen in the orbital architecture on the other side will be a challenge for planetary system formation models. Additional observations with \textit{CHEOPS} and RV facilities will allow further constraining of the internal structures of all planets in the system, in particular the (lack of) similarity between {the water fraction and gas mass} fraction between planets.

Follow-up transit observations should also unveil TTVs for all but the innermost planet of the TOI-178 system (see Fig. \ref{fig:TTVs6p}), with two timescales: a $\sim$~$260$\,d signal with an amplitude of several minutes to a few tens of minutes and a larger signal over several years or decades. Future observations of this system can hence be used to measure the planetary masses and estimate eccentricities directly from TTVs as well as compare the results with masses derived from RV measurements. 

Finally, the innermost planet, $b$, lies just outside the 3:5 MMR with planet $c$; {however, it is too far to be part of the Laplace chain}, which would require a period of $\sim 1.95$d. Since the formation of the Laplace resonant chain probably results from a slow drift from a chain of two-planet resonances due to tidal effects \citep{PaTe2010,DeLaCoBo2012,Papaloizou2015,MacDonald2016}, the current state of the system might constrain the dissipative processes that tore apart the innermost link of the chain while the rest of the configuration survived. 

The TOI-178 system, as revealed by the recent observations described in this paper, contains a number of very important features: Laplace resonances, variation in densities from planet to planet, and a stellar brightness that allows a number of follow-up observations (photometric, atmospheric, and spectroscopic). It is therefore likely to become one of the Rosetta Stones for understanding planet formation and evolution, even more so if additional planets continuing the chain of Laplace resonances is discovered orbiting inside the habitable zone. 

\vspace{2cm}



Software list:
\begin{itemize}

\item \texttt{ellc} \citep{ellc}.
\item \texttt{emcee} \citep{emcee}.
\item \texttt{lmfit} \citep{lmfit}.
\item \texttt{tqdm} \citep{tqdm}.
\item \texttt{Tensorflow}
\item \texttt{Keras}
\item \texttt{pycheops}\footnote{https://github.com/pmaxted/pycheops} Maxted et al., \textit{in prep.}.

\end{itemize}

\begin{acknowledgements}
The authors acknowledge support from the Swiss NCCR PlanetS and the Swiss National Science Foundation. 
 YA and MJH  acknowledge  the  support  of  the  Swiss  National  Fund  under  grant 200020\_172746. 
ACC and TW acknowledge support from STFC consolidated grant number ST/M001296/1. 
This work was granted access to the HPC resources of MesoPSL financed by the Region Ile de France and the project Equip@Meso (reference ANR-10-EQPX-29-01) of the programme Investissements d’Avenir supervised by the Agence Nationale pour la Recherche. 
SH acknowledges CNES funding through the grant 837319. 
Based on data collected under the NGTS project at the ESO La Silla Paranal Observatory. The NGTS facility is operated by the consortium institutes with support from the UK Science and Technology Facilities Council (STFC) project ST/M001962/1. 
The Belgian participation to CHEOPS has been supported by the Belgian Federal Science Policy Office (BELSPO) in the framework of the PRODEX Program, and by the University of Li{\`e}ge through an ARC grant for Concerted Research Actions financed by the Wallonia-Brussels Federation. 
V.A. acknowledges the support from FCT through Investigador FCT contract nr.  IF/00650/2015/CP1273/CT0001. 
We acknowledge support from the Spanish Ministry of Science and Innovation and the European Regional Development Fund through grants ESP2016-80435-C2-1-R, ESP2016-80435-C2-2-R, PGC2018-098153-B-C33, PGC2018-098153-B-C31, ESP2017-87676-C5-1-R, MDM-2017-0737 Unidad de Excelencia “María de Maeztu”- Centro de Astrobiología (INTA-CSIC), as well as the support of the Generalitat de Catalunya/CERCA programme. The MOC activities have been supported by the ESA contract No. 4000124370. 
S.C.C.B. acknowledges support from FCT through FCT contracts nr. IF/01312/2014/CP1215/CT0004. 
XB, SC, DG, MF and JL acknowledge their role as ESA-appointed CHEOPS science team members. 
ABr was supported by the SNSA. 
A.C. acknowledges support by CFisUC projects (UIDB/04564/2020 and UIDP/04564/2020), GRAVITY (PTDC/FIS-AST/7002/2020), ENGAGE SKA (POCI-01-0145-FEDER-022217), and PHOBOS (POCI-01-0145-FEDER-029932), funded by COMPETE 2020 and FCT, Portugal. 
This work was supported by FCT - Fundação para a Ciência e a Tecnologia through national funds and by FEDER through COMPETE2020 - Programa Operacional Competitividade e Internacionalização by these grants: UID/FIS/04434/2019; UIDB/04434/2020; UIDP/04434/2020; PTDC/FIS-AST/32113/2017 \& POCI-01-0145-FEDER- 032113; PTDC/FIS-AST/28953/2017 \& POCI-01-0145-FEDER-028953; PTDC/FIS-AST/28987/2017 \& POCI-01-0145-FEDER-028987. 
O.D.S.D. is supported in the form of work contract (DL 57/2016/CP1364/CT0004) funded by national funds through FCT. 
B.-O.D. acknowledges support from the Swiss National Science Foundation (PP00P2-190080). 
MF and CMP gratefully acknowledge the support of the Swedish National Space Agency (DNR 65/19, 174/18). 
DG gratefully acknowledges financial support from the CRT foundation under Grant No. 2018.2323 ``Gaseousor rocky? Unveiling the nature of small worlds''. 
EG gratefully acknowledges support from the David and Claudia Harding Foundation in the form of a Winton Exoplanet Fellowship. 
M.G. is an F.R.S.-FNRS Senior Research Associate. 
J.I.G.H. acknowledges financial support from Spanish Ministry of
Science and Innovation (MICINN) under the 2013 Ram\'on y Cajal 
programme RYC-2013-14875. 
J.I.G.H., A.S.M., R.R., and C.A.P. acknowledge financial support 
from the Spanish MICINN AYA2017-86389-P.
A.S.M. acknowledges financial support from the Spanish Ministry of
Science and Innovation (MICINN) under the 2019 Juan de la Cierva
Programme. 
MNG ackowledges support from the MIT Kavli Institute as a Juan Carlos Torres Fellow. 
JH acknowledges the support of the Swiss National Fund under grant 200020\_172746. 
KGI is the ESA CHEOPS Project Scientist and is responsible for the ESA CHEOPS Guest Observers Programme. She does not participate in, or contribute to, the definition of the Guaranteed Time Programme of the CHEOPS mission through which observations described in this paper have been taken, nor to any aspect of target selection for the programme. 
JSJ acknowledges support by FONDECYT grant 1201371, and partial support from CONICYT project Basal AFB-170002. 
AJ  acknowledges support from ANID – Millennium Science Initiative – ICN12\_009 and from FONDECYT project 1171208. 
PM acknowledges support from STFC research grant number ST/M001040/1. 
N.J.N is supported by the contract and exploratory project
IF/00852/2015, and projects UID/FIS/04434/2019, PTDC/FIS-OUT/29048/2017,
COMPETE2020: POCI-01-0145-FEDER-028987 \& FCT: PTDC/FIS-AST/28987/2017. 
N.J.N is supported by the contract and exploratory project IF/00852/2015, and project PTDC/FIS-OUT/29048/2017. 
This work was also partially supported by a grant from the Simons Foundation (PI Queloz, grant number 327127). 
Acknowledges support from the Spanish Ministry of Science and Innovation and the European Regional Development Fund through grant PGC2018-098153-B- C33, as well as the support of the Generalitat de Catalunya/CERCA programme. 
S.G.S. acknowledge support from FCT through FCT contract nr. CEECIND/00826/2018 and POPH/FSE (EC). 
This work has made use of data from the European Space Agency (ESA) mission {\it Gaia} (\url{https://www.cosmos.esa.int/gaia}), processed by the {\it Gaia} Data Processing and Analysis Consortium (DPAC, \url{https://www.cosmos.esa.int/web/gaia/dpac/consortium}). Funding for the DPAC has been provided by national institutions, in particular the institutions participating in the {\it Gaia} Multilateral Agreement. 
This  project  has  been  supported  by  the  Hungarian National Research, Development and Innovation Office (NKFIH) grants GINOP-2.3.2-15-2016-00003, K-119517,  K-125015, and the City of Szombathely under Agreement No.\ 67.177-21/2016. 
This research received funding from the MERAC foundation, from the European Research Council under the European Union’s Horizon 2020 research and innovation programme (grant agreement nº 803193/ BEBOP, and from the Science and Technology Facilites Council (STFC, grant nº ST/S00193X/1). 
V.V.G. is an F.R.S-FNRS Research Associate. 
JIV acknowledges support of CONICYT-PFCHA/Doctorado Nacional-21191829. 
M. R. Z. O. acknowledges financial support from projects AYA2016-79425-C3-2-P and PID2019-109522GB-C51 from the Spanish Ministry of Science, Innovation and Universities.
\end{acknowledgements}

\bibliographystyle{aa}
\bibliography{biblio_TOI178}

\appendix




\section{Inspection of the CHEOPS data}
\label{sec:data_inspection}

\begin{figure}
\includegraphics[width=9cm]{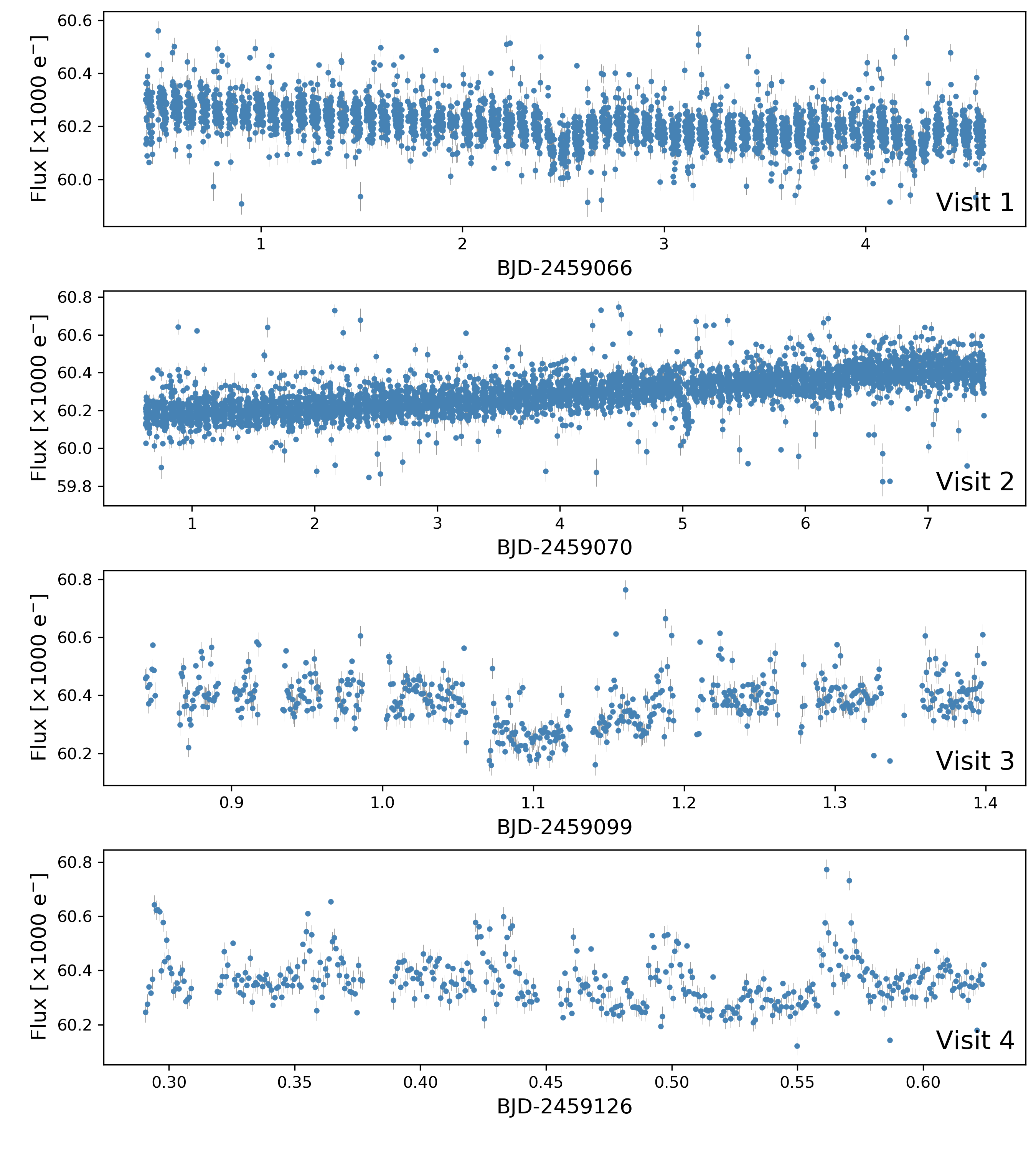}
        \caption{ DRP-produced light curves (DEFAULT aperture) of the four {\it CHEOPS} visits to TOI-178 presented in this work. Three-$\sigma$ outliers have been removed for better visualisation.} 
        \label{fig:rawLC_1}
\end{figure}

The four {\it CHEOPS} visits were automatically processed through the DRP with individual frames undergoing various calibrations and corrections; aperture photometry was subsequently conducted for four aperture radii, as highlighted in Sect.~\ref{sec:CHEOPS} and covered in detailed in \citealt{Hoyer2020}. The light curves produced for all runs in this study -- which are often referred to as `raw' in order to indicate no post-processing detrending has taken place and which were obtained with the DEFAULT aperture -- are shown in Fig.~\ref{fig:rawLC_1}. For the first, third, and fourth visits, standard data processing within the DRP was performed; however, for the second run, careful treatment of telegraphic pixels was needed.

\begin{figure}
\includegraphics[width=9cm]{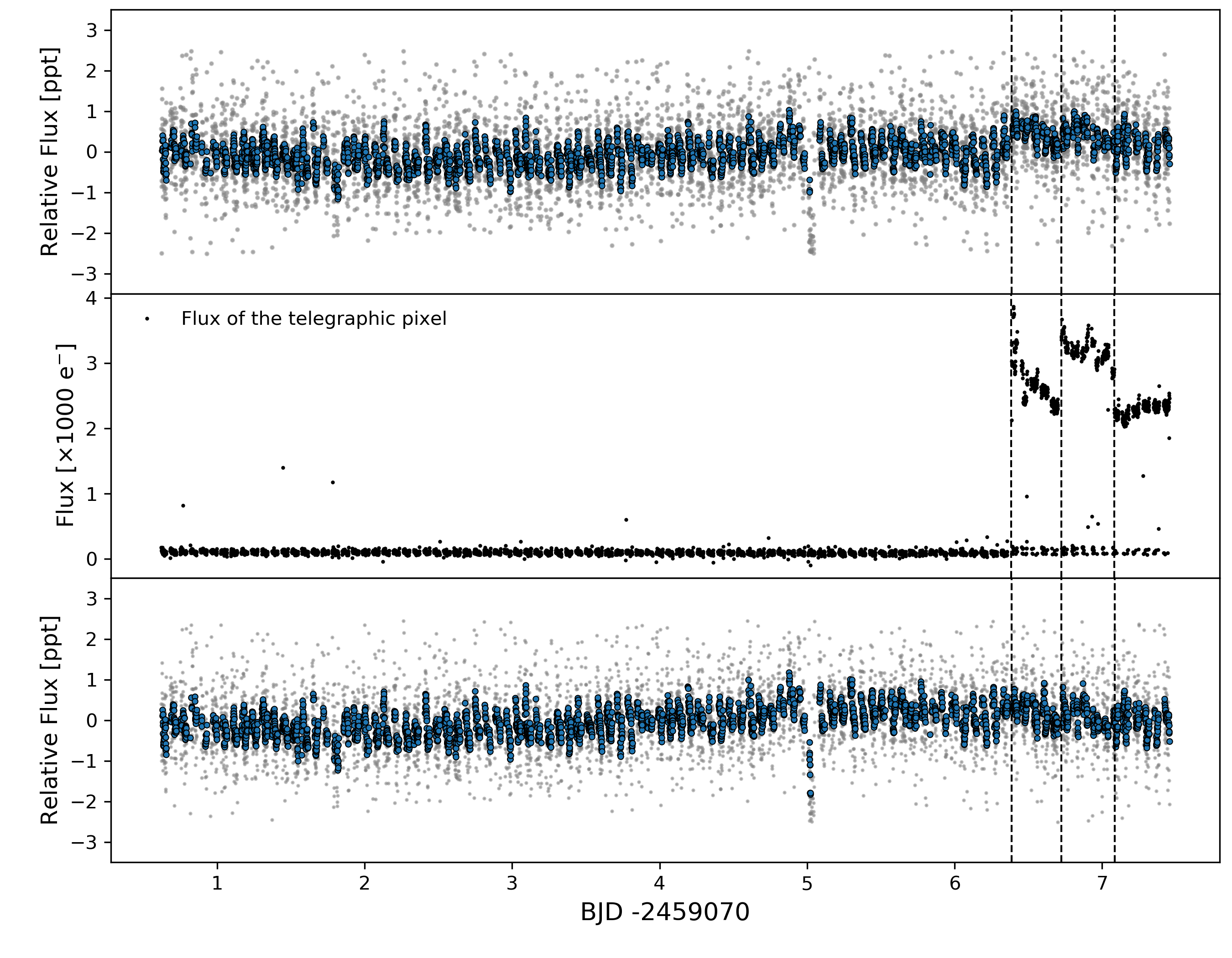}
        \caption{TOI-178 normalised light curve of the second visit (grey symbols)  shown in the top panel with its 10\,min smoothed version overplotted (blue symbols). The flux jumps produced by the appearance of a new hot pixel are marked with the dashed vertical lines. The light curve of the detected telegraphic pixel is presented in the middle panel, showing its anomalous behaviour at the end of the visit. The light curve extracted from the corrected data is shown in the bottom panel. For better visualisation, the light curves are presented after a 3-$\sigma$ clipping and corrected by a second-order polynomial in time. } 
        \label{fig:telgpix2}
\end{figure}

In the {\it CHEOPS} CCD, there is a large number of hot pixels (see, for example, Fig. \ref{fig:telgpix1}). Moreover, the behaviour of some normal pixels can change to an abnormal state within the duration of a visit. For example, a pixel can become `hot' after an SAA crossing of the satellite. These pixels are called telegraphic due to their unstable response during the observations, and they can disturb the photometry if located inside the photometric aperture. To rule out the possibility that the detected transit events in the light curves correspond to the effect of telegraphic pixels, the data frames were carefully inspected and compared to the detection map of hot pixels delivered by the {\it CHEOPS} DRP \citep[see details in][]{Hoyer2020}. By doing this, one telegraphic pixel was detected inside the DEFAULT aperture at the end of the second TOI-178 visit . The exact CCD location of this abnormal pixel is shown in Fig. \ref{fig:telgpix1}. The effect of this pixel in the photometry is shown in the form of a jump in flux in the light curve of the visit (top panel, Fig. \ref{fig:telgpix2}) at BJD$\sim$2\,459\,076.5, which corresponds to the flux increase of the pixel (middle panel, Fig. \ref{fig:telgpix2}). After correcting the data by simply cancelling the flux of this pixel through the full observation and repeating the photometry extraction with the same aperture (R=25$\arcsec$), we removed the flux jump in the light curve (bottom panel, Fig. \ref{fig:telgpix2}). No telegraphic pixels were detected in the other TOI-178 visits.

\section{Analysis of the radial velocity data}
\label{ap:rv}

\subsection{Method}

In this appendix, we describe the analysis of the RV data alone.
To search for planet detections, we computed the $\ell_1$-periodogram of the RV, as defined in~\cite{hara2017}. This tool is based on a sparse recovery technique called the basis pursuit algorithm~\citep{chen1998}. It aims to find a representation of the RV time series as a sum of a small number of sinusoids whose frequencies are in the input grid.  The $\ell_1$-periodogram  has the advantage, over a regular periodogram, of searching for several periodic components at the same time, therefore drastically reducing the impact of aliasing~\citep{hara2017}.


The $\ell_1$-periodogram has three inputs: a frequency grid, onto which the signal will be decomposed in the Fourier domain; a noise model in the form of a covariance matrix; and a base model.  
 The base model represents offsets, trends, or activity models and can be understood as follows. The principle behind the $\ell_1$-periodogram is to consider the signal in the Fourier domain and to minimise the sum of the absolute value of the Fourier coefficients on a discretised frequency grid (their $\ell_1$ norm) while ensuring that the inverse Fourier transform is close enough to the model in a certain, precise sense. Due to the $\ell_1$ norm penalty, frequencies `compete' against one another to have non-zero coefficients. However, one might assume that certain frequencies, and more largely certain signals, are in the data by default  and should not be penalised by the $\ell_1$ norm. We can define a linear model whose column vectors are not penalised, which we call the base model.

The signals found to be statistically significant might vary depending on the frequency grid, base, and noise models. To explore this aspect,  as in~\cite{Hara2020}, we computed the $\ell_1$-periodogram of the data with different assumptions regarding the noise covariance. We then ranked the covariance models via CV. 
That is, we fixed a frequency grid. Secondly, for every choice of base and noise models, we recorded which detections are announced. We assessed the score of the detections+noise models via CV. The data were separated randomly into a training set and a test set that contain, respectively, 70 and 30\% of the data. The model was fitted onto the training set, and we computed the likelihood of the data on the test set. This operation was repeated 250 times, and we attributed the median of the 250 scores to the triplet base model, covariance model, and signal detected. 
 To determine if a signal at a given period was significant, we studied the distribution of its FAP among the highest ranked models.

\subsection{Definition of the alternative models}

\begin{figure}
        \includegraphics[width=1\linewidth]{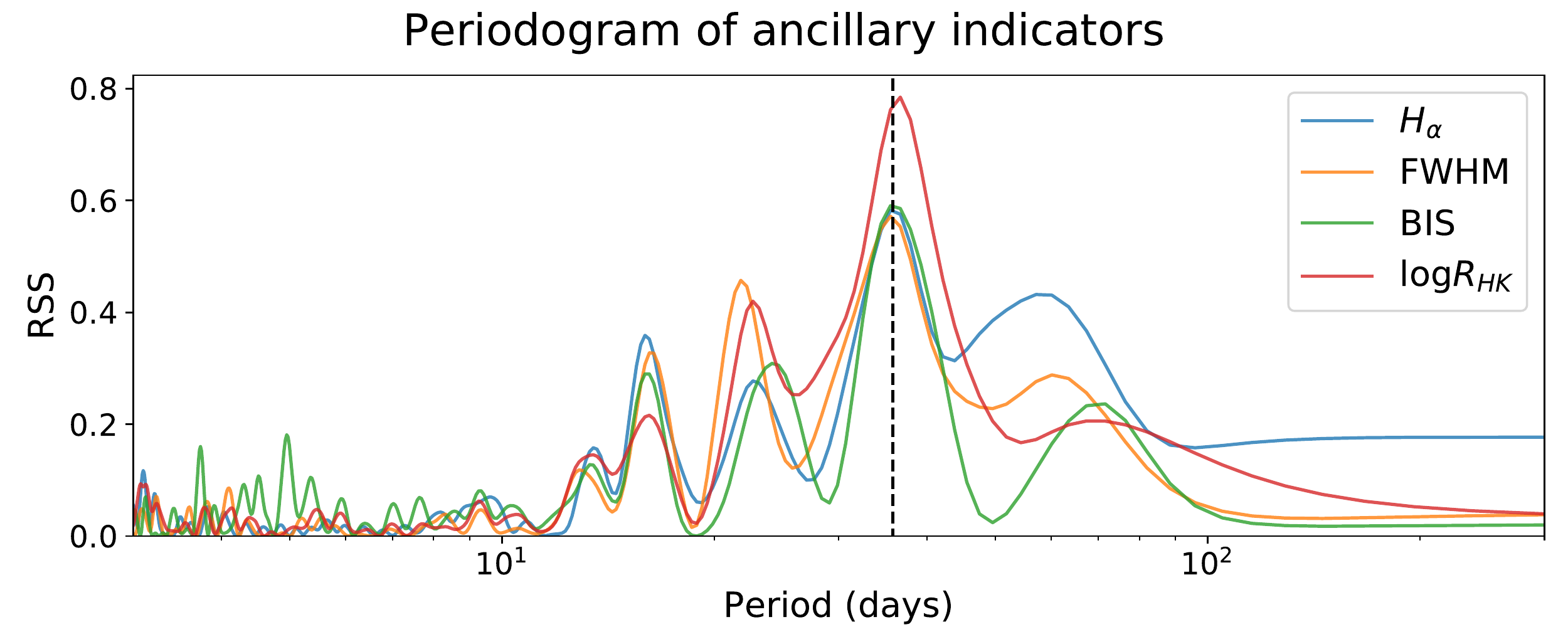}
        \includegraphics[width=1\linewidth]{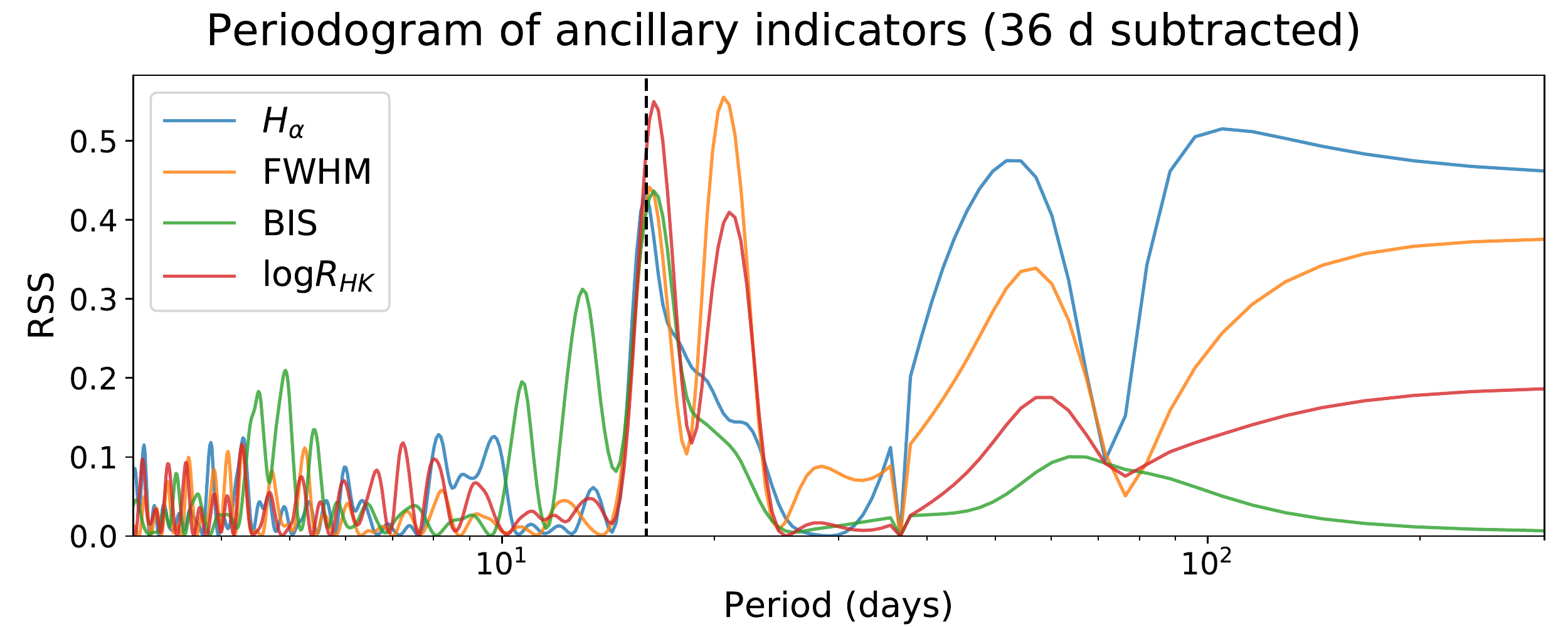}
        \caption{{Periodograms of the ancillary indicator time series ($H \alpha$, FWHM, bisector span, and $\log R'_{HK}$ shown in blue, orange, green, and red, respectively)}. Top: Periodograms of the raw ancillary indicator time series. Bottom: Periodograms of the same time series when the signal corresponding to the maximum peak is injected in the base model. }
        \label{fig:ancper}
\end{figure}

To define the alternative  RV models we, as a preliminary step, analysed the $H \alpha$, FWHM, bisector span, and $\log R'_{HK}$ time series as provided by the ESPRESSO pipeline. We computed the residual periodograms as described in~\citep{Baluev2008}. These periodograms allowed us to take into account general linear base models that are fitted along candidate frequencies. We computed the periodograms and iteratively added a sinusoidal function whose frequency corresponds to the maximum peak of the periodogram.  
Iterations 1 and 2 are shown in the top and bottom panels of Fig.~\ref{fig:ancper}, respectively. The 36-day and 16-day positions  (top and bottom panels, respectively) are marked with dotted black lines.  
Pursuing the iterations, we find signal detections with FAP <$10^{-3}$ of periods at 36, 115, and 15.9 days for $H \alpha$; 35.5, 20.8, and 145 days for the FWHM; 36 and 16 days for the bisector span; and 36.7 and 16.5  days for the $\log R'_{HK}$. The 36-day periodicities are always detected with FAP < $10^{-6}$. These results indicate that activity effects in the RV at $ \approx$ 36 and $ \approx$ 16 days are to be expected, along with low frequency effects. These signals likely stem from the rotation period of  the star, which creates signals at the fundamental frequency and the first harmonic.

We now turn to the RV and define the alternative noise models we explored. These are Gaussian, with a white component, an exponential decay, and a quasi-periodic term, as given by the formula
\small
\begin{align}
\begin{split}
V_{kl} & =  \delta_{k,l} (\sigma_{k}^2 + \sigma_{W}^2) +   \sigma_{R}^2 \e^{-\frac{(t_k-t_l)^2}{2\tau_R^2}} +
\sigma_{QP}^2\e^{-\frac{(t_k-t_l)^2}{\tau_{QP}^2}   \sin^2\left( \frac{t_k-t_l}{P_\mathrm{act}}\right) } 
\end{split},
\label{eq:kernel}
\end{align}
\normalsize
where $V_{kl}$ is the element of the covariance matrix at row $k$ and column $l$; $ \delta_{k,l}$ is the Kronecker symbol; $\sigma_{k}$ is the nominal uncertainty on the measurement $k$; and $\sigma_{W}, \sigma_{R}, \tau_R,\sigma_{QP}$, and $P_\mathrm{act}$ are the parameters of our noise model. 
A preliminary analysis on the ancillary indicators (FWHM, S-index, and $H \alpha$) showed that they all exhibit statistically significant variations at $\approx 40$ (36 days), as well as a periodogram peak at 16 d. The 36 and 16 d signals exhibit phases compatible with each other at 1 sigma, except for the 36 d signal in the FWHM which is 3$\sigma$ away from the phase fitted on the S-index and $H \alpha$. This points to a stellar rotation period of 36 d. 
We considered all the possible combinations of values for $\sigma_R$ and $\sigma_W$ in 0.0,0.5 1.0,1.25, 1.5,1.75,2 m/s, 
$\tau$ = 0, 2, 4, 6 d, $P_\mathrm{act} $ = 36.5 d, $\sigma_{QP}$ = 0,1,2,3,4 m/s, and $\tau_{QP}$ = 18, 36, or 72 d.  

The computation of the $\ell_1$-periodogram was made assuming a certain base model. By this, we mean a linear model that is assumed to be in the data by default and will thus automatically be fitted. This base model could represent, for instance, offsets, trends, or certain periodic signals. We tried the following base models: one offset, one offset and  smoothed $H \alpha$, one offset and  smoothed FWHM, and one offset and the  smoothed $H \alpha$ and FWHM time series. The smoothing of a given indicator is done via a GP. This process has a Gaussian kernel, whose parameters (timescale and amplitude) have been optimised to maximise the likelihood of the data.


\subsection{Results}
\label{sec:rvresults}
\begin{figure}
        \includegraphics[width=1\linewidth]{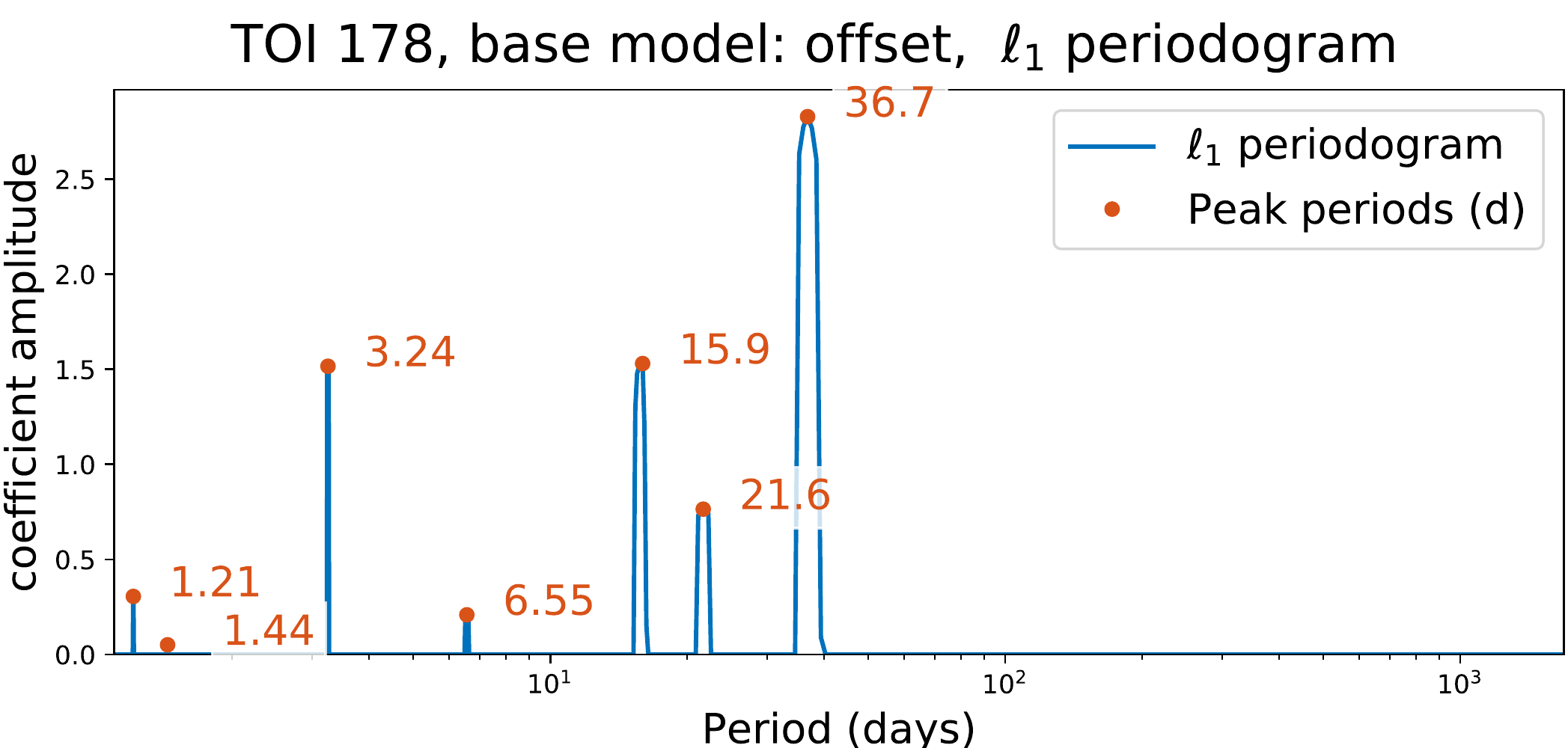}
        \includegraphics[width=1\linewidth]{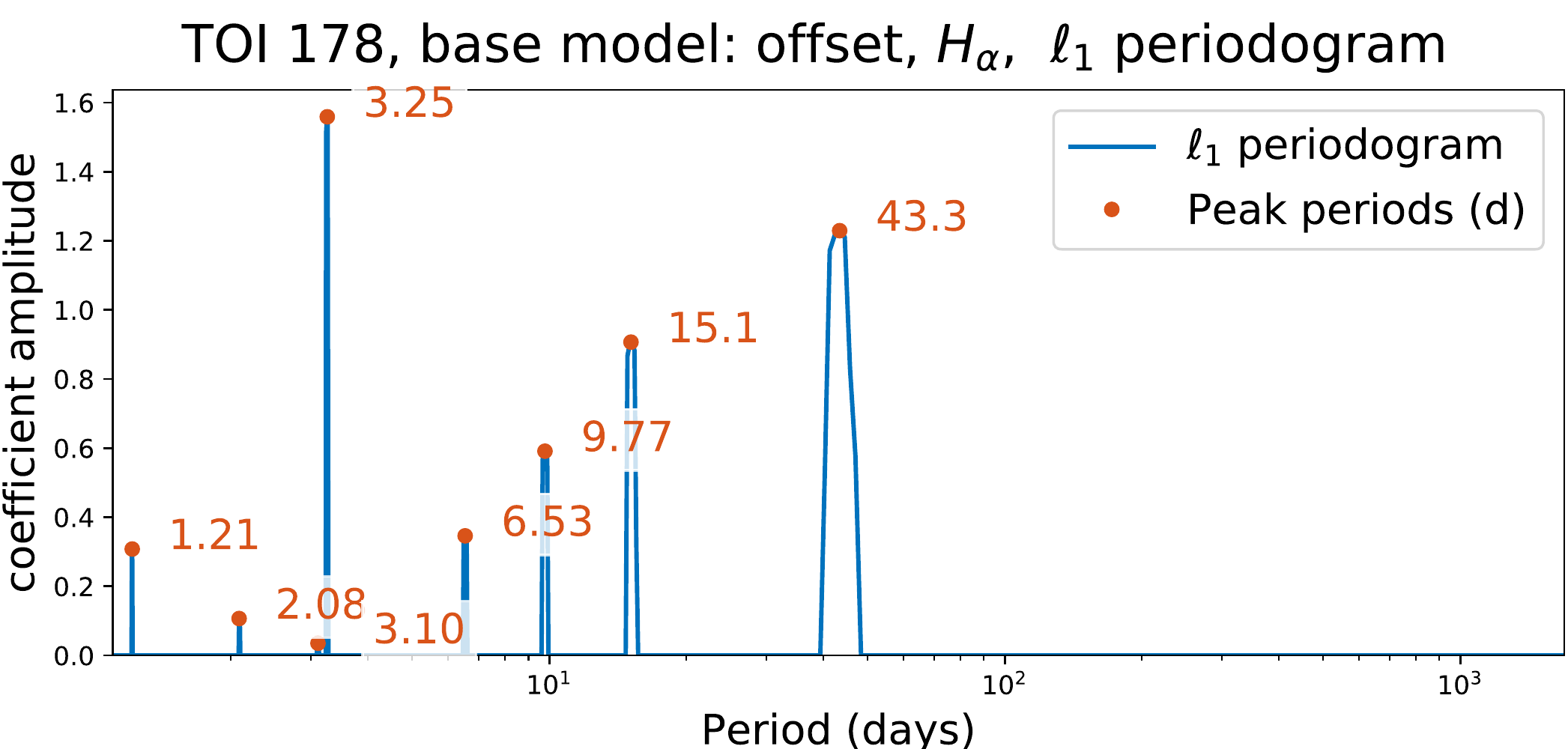}
        \includegraphics[width=1\linewidth]{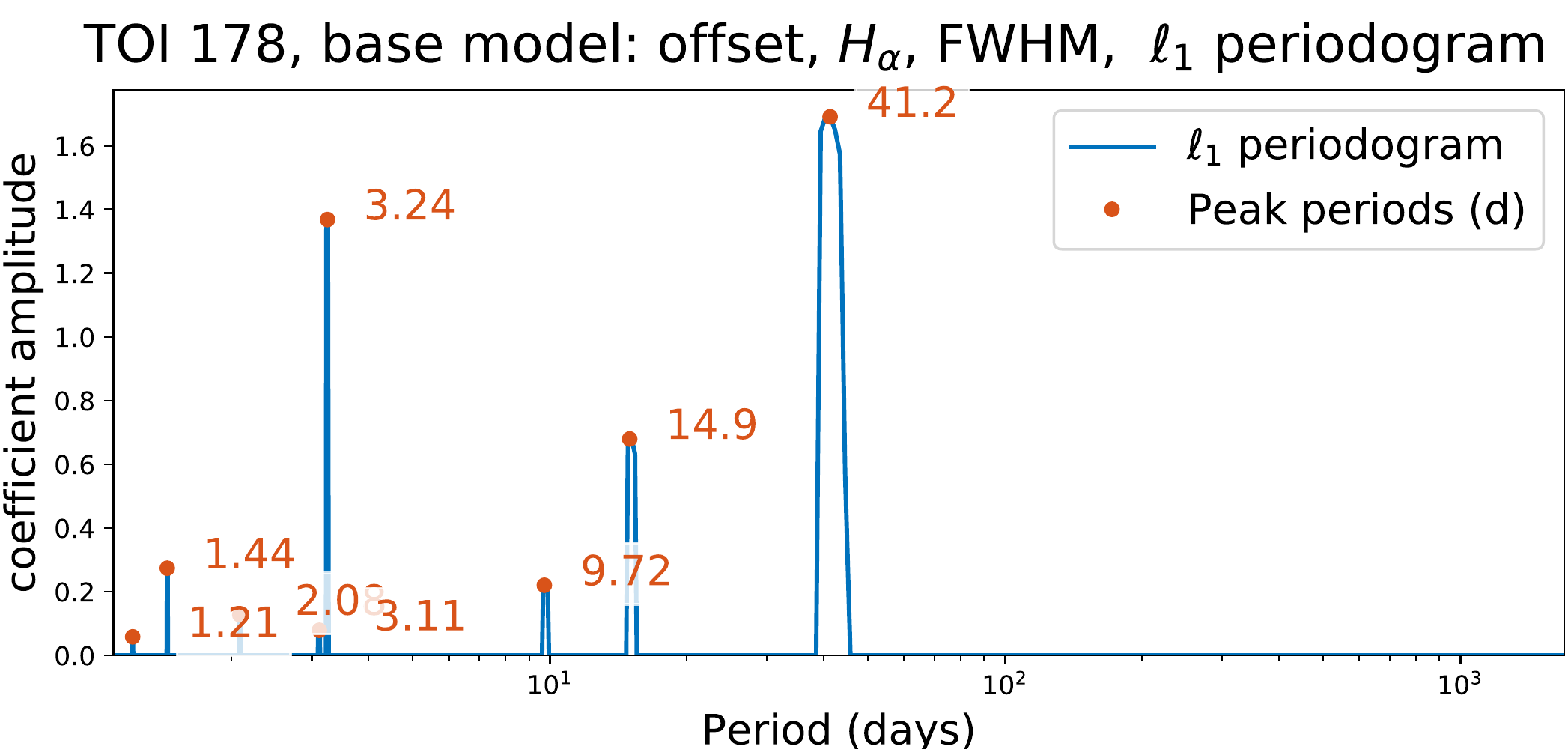}
        \caption{$\ell_1$-periodogram of the ESPRESSO RV corresponding to the best CV score with different linear base models for the data: Top: Only one offset. Middle: Offset, smoothed $H \alpha$. Bottom: Smoothed $H \alpha$ and smoothed FWHM. }
        \label{fig:l1_1}
\end{figure}

In Fig.~\ref{fig:l1_1}, we represent the $\ell_1$-periodogram obtained with the noise model with the maximum CV score with three different base models (from top to bottom: without any indicator, with smoothed $H \alpha$, and with both smoothed $H \alpha$ and smoothed FWHM). In all cases, we find signals at 35-40 d, 15 d, and 3.24 d, as well as a peak at 6.55 d. We can also find peaks at 21 d, 9.84 d, 2.08 d, 1.92 1.44, or 1.21 days, depending on the models used. We note that  2.08 and 1.92 d are aliases of each other, so that the presence of one or the other might originate from the same signal. The model with the overall highest CV score includes only the smoothed $H \alpha$ time series in the base model and with the notations of Eq.~\eqref{eq:kernel}, $\sigma_W$ =  1.75 m/s, $\sigma_R$ = 1.5 m/s, $\tau$ = 2 days, and $\sigma_{QP}$ = 0, and it corresponds to the middle panel in Fig.~\ref{fig:l1_1}.

We computed the $\ell_1$-periodogram on a grid from 0 to 0.95 cycles per day, {hence ignoring periods below one day}, with all combinations of the base and noise models. As in~\cite{Hara2020},  the models are ranked by CV. 
We then considered the 20\% highest ranked models (all noise and base models considered), which we denote with $\mathrm{CV}_{20}$, and computed the number of times a signal is included in the model. {A signal is considered to be included if a peak has a frequency} within $1/T_{obs}$ of a reference frequency $1/P_0$, where $T_{obs}$ is the observation time span and has an FAP below 0.5. We report these values in Table~\ref{tab:cv20} for the reference periods $P_0$ that correspond to signals appearing at least once in  Fig.~\ref{fig:l1_1}. We note that, due to the short time span, the frequency resolution is not good enough to distinguish between 36 and 45 days.

We find that signals at 36, 16, and 3.2 days are consistently included in the model. The 3.2-day signal presents a median FAP of 0.002 and can therefore be confidently detected.  When the base model consists of only an offset, 36 and 16 days are systematically significant with FAP <5\%, but their significance decreases when activity indicators are included in the model. Since these periodicities also appear in ancillary indicators, we conclude that they are due to activity. Signals at 6.5 and 9.9 days appear in $\approx$ 15\% of the models, but with low significance. Although detections cannot be claimed from the RV data only, we note that there are signals at 2.08 d (alias of 1.91 d) and 9.9 d. We find that they are in phase with the photometric signals within 1.5 $\sigma$, which further strengthens the detection of transiting planets at these periods.\ We finally note a small hint at 1.2 days (which can also appear at its alias, 5.6 days, in certain models).

\begin{figure}
        \includegraphics[width=9cm]{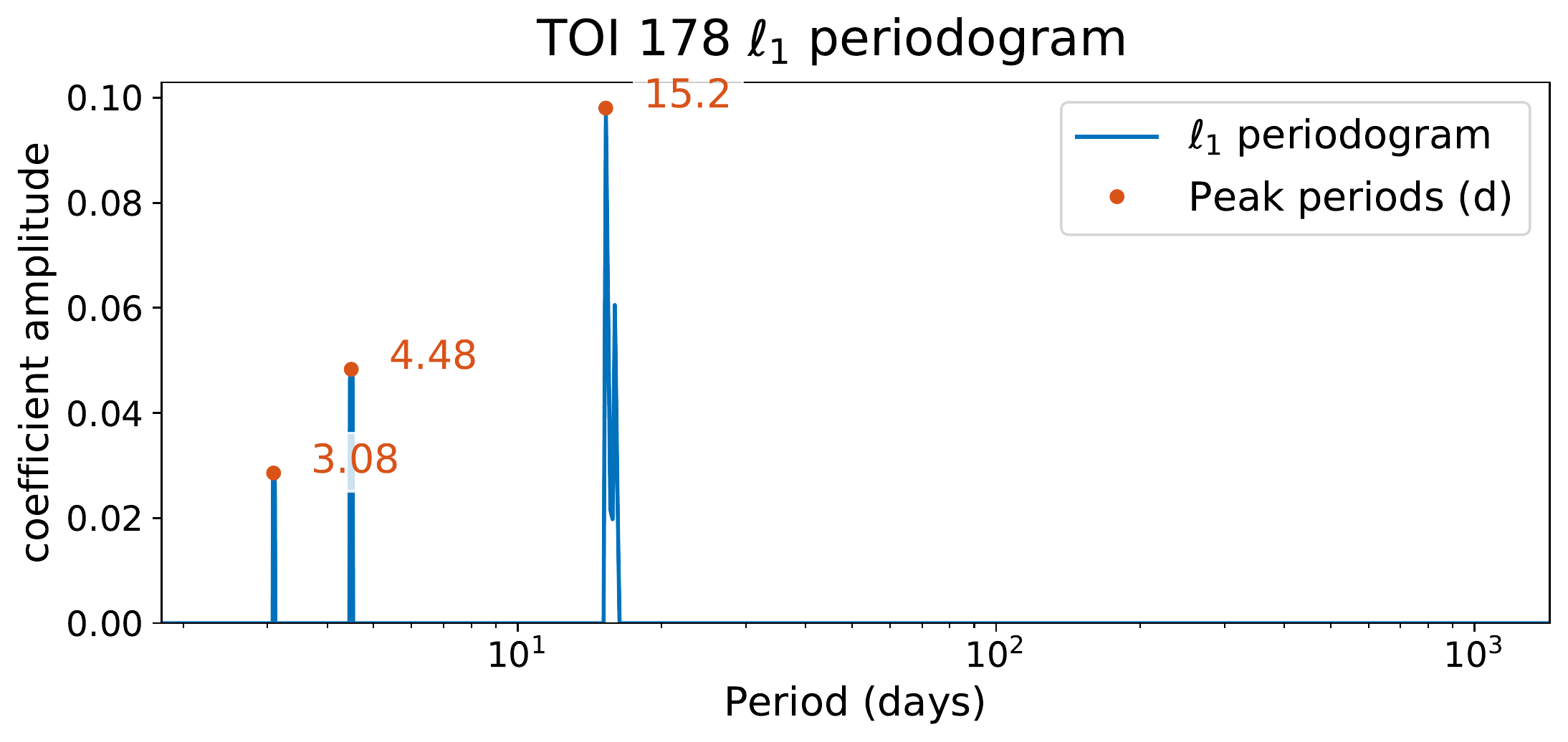}
        \caption{$\ell_1$-periodogram when all transiting planets are added to the base models, as well as a smoothed $H \alpha$ indicator and sinusoids at 15.7 and 36 days. }
        \label{fig:l1_3}
\end{figure}

The period of planet $f$ is 15.2 days, and an activity signal in that region seems to be present ($\approx 16$\,d, depending on ancillary indicators). The periods are such that $1/(1/15.23-1/16) = 316$ days, which is greater than the observation time span. To check whether our activity model allows us to detect planet $f$ in the RV (and yield a meaningful mass estimate), we added to the base model all the known transiting planets, except the 15.2-day one, as well as the smoothed $H \alpha$ indicators and sine functions at 15.7 (closest to 15.2 days) and 36 days. From that we obtain Fig.~\ref{fig:l1_3}, where a signal at 15.2 days appears with an FAP of $\approx$ 20\%.

The 20.7-day transiting planet could correspond to the peak at 21.6 d in the $\ell_1$-periodogram (Fig.~\ref{fig:l1_1}, top). When restricting the frequency grid to 0 to 0.55 cycles per day, the best CV model yields Fig.~\ref{fig:l1perio_best}, where a signal appears at 20.6 days. 
However, signals close to 20.7 days seem to disappear when the stellar activity model is changed. The planetary signature might be hidden in the RV due to stellar effects. Further observations would allow us to better disentangle stellar and planetary signals.

\subsection{Detections: Conclusion}
\label{ap:rvdetection}

In conclusion, we can claim an independent detection of the 3.2 d planet with RV. We find significant signatures at 36 and 15-16 days, which we attribute to activity. We find signals at 1.91 d (or its alias 2.08 d), 6.5 d, and 9.9 d in phase with the detected transits. We see a signal at 21 d for some models, which could correspond to the 20.7 d planet, but it seems that there are not enough points {to disentangle a planetary signal from activity at this period}. We note that modelling an activity signal at 15.7 days leaves a signal at 15.2 days, which likely stems from the presence of a planet  at this period.

Finally, we checked the consistency in the phase of the signals fitted onto the photometric  and RV data. We performed an MCMC computation of the orbital elements on the RVs with exactly the same priors as model 2, except that we set a flat prior on the phases. We find that the uncertainties on the phase corresponding to planets $b$, $c$, $d$, $e$, $f$, and $g$ correspond respectively to 7, 2, 3.3, 4, and 100\% of the period. The phases from transits are included, respectively, in the 1.5, 1, 2, 1, and 1 $\sigma$ intervals derived from the RVs, such that we deem the phases derived from RVs consistent with the transits.

\begin{table} 
        \caption{Inclusion in the top 20\% best models of different periodicities. We report the FAP associated with the best model, the frequency of inclusion in the model, and the median FAP in the top 20\% best models.}
        \label{tab:cv20}
        \centering\begin{tabular}{p{1.2cm}|p{1.5cm}|p{1.5cm}|p{1.5cm}} Period (d) & FAP (best fit) & Inclusion in the model & $\mathrm{CV}_{20}$ median FAP \\ \hline \hline 1.21& $5.33\cdot10^{-1}$ & 1.730\% &  - \\ 1.44& $1.00$ & 0.0\% &  - \\ 1.914& $1.00$ & 0.0\% &  - \\ 2.08& $2.93\cdot10^{-1}$ & 0.384\% &  - \\ 3.24& $1.33\cdot10^{-2}$ & 100.0\% & $2.20\cdot10^{-3}$\\ 6.5& $1.31\cdot10^{-1}$ & 15.57\% &  - \\ 9.9& $3.41\cdot10^{-1}$ & 16.63\% &  - \\ 15.2& $8.38\cdot10^{-2}$ & 89.32\% & $6.83\cdot10^{-2}$\\ 20.7 & $1.00$ & 0.0\% &  - \\ 36 (45)& $5.24\cdot10^{-1}$ & 95.48\% & $2.57\cdot10^{-1}$ \end{tabular}\end{table}

\subsection{Mass and density estimates}
\label{ap:mass1}

The RV measurements allowed us to obtain mass estimates of the planets. These estimates can depend on the model of stellar activity used. To account for this, we estimated masses with two different stellar activity models: (1) Activity is modelled as two sinusoidal signals plus a correlated noise, and (2) activity is modelled as a correlated Gaussian noise with a semi-periodic kernel. In both cases, we added the $H \alpha$ time series smoothed with a GP, as described  in Sect.~\ref{sec:rvresults}. In model 1, the two sinusoids have periods of $\approx 40$ days and $\approx 16$ days. This is motivated by the fact that these periodicities appeared systematically in ancillary indicators, though with different phases. We therefore allowed the phase to vary freely in the RVs. The priors on the stellar activity periods are taken as Gaussians with mean 16 days, $\sigma = 1$ day and mean 36.8 days, $\sigma = 8$ days, according to the variability of the position of the peaks appearing in the spectroscopic ancillary indicators. We further added free noise components: a white component and a correlated component with an exponential kernel. The prior on the variance is a truncated Gaussian  with $\sigma = 100$ $m^2/s^2$, and the prior on the noise timescale is $\log$-uniform between 1h and 30 days. The second model is a GP (or here, correlated Gaussian noise) with a quasi-periodic kernel $k$ of the form 
\begin{align}
    k(t;\sigma_W,\sigma_R, \lambda, \nu  ) = \sigma_W^2 + \sigma_R^2 \e^{-\frac{t}{\tau} } \cos(\nu t).
    \label{eq:kernelmass}
\end{align}
We imposed a Gaussian truncated prior on $\sigma_W^2,\sigma_R^2$ with $\sigma = 100$ $m^2/s^2$. We imposed a flat prior on $\nu$ between $2\pi/50$ and $2\pi/30$ rad/day and a $\log$-uniform prior on $\tau$ on 1h to 1000 days. 
For planets $b$, $c$, $d$, $g$, $e$, and $f,$ the priors on the periods and the times of conjunction are set as Gaussian with means and variances set according to the constraints obtained from the joint fit of the TESS and CHEOPS data. The densities are computed by combining the posterior samples of mass and the posterior samples of radii, assuming the mass and radius estimates are independent.

We ran an adaptive MCMC algorithm as described in~\cite{Delisle2018}, implementing spleaf~\citep{Delisle2020}\footnote{https://gitlab.unige.ch/Jean-Baptiste.Delisle/spleaf}, which offers optimised routines to compute the inverse of covariance matrices. We checked the convergence of the algorithm by ascertaining that 600 effective samples had been obtained for each variable~\citep{Delisle2018}. The posterior medians as well as the 1 $\sigma$ credible intervals are reported in Table~\ref{tab:rvsimp} for both activity models. The kernel~\eqref{eq:kernelmass} is close to the stochastic harmonic oscillator (SHO), as defined in~\cite{celerite}. We also tried the SHO kernel and simply imposed that the quality factor $Q$ be greater than one half, the other parameters having flat priors. We found very similar results, and, as such, they are not reported in Table~\ref{tab:rvsimp}.

 

\begin{figure}
    \centering
    \includegraphics[width=\linewidth]{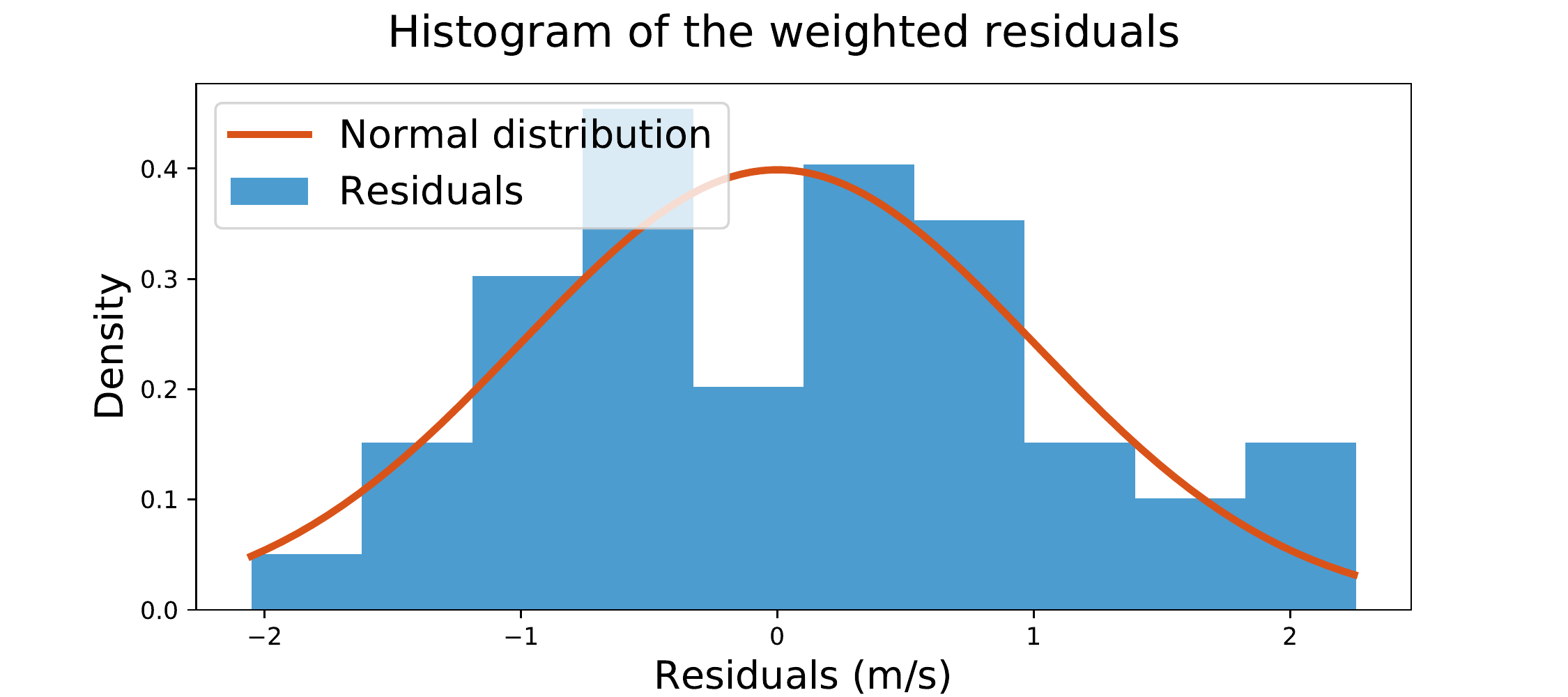}

    \includegraphics[width=\linewidth]{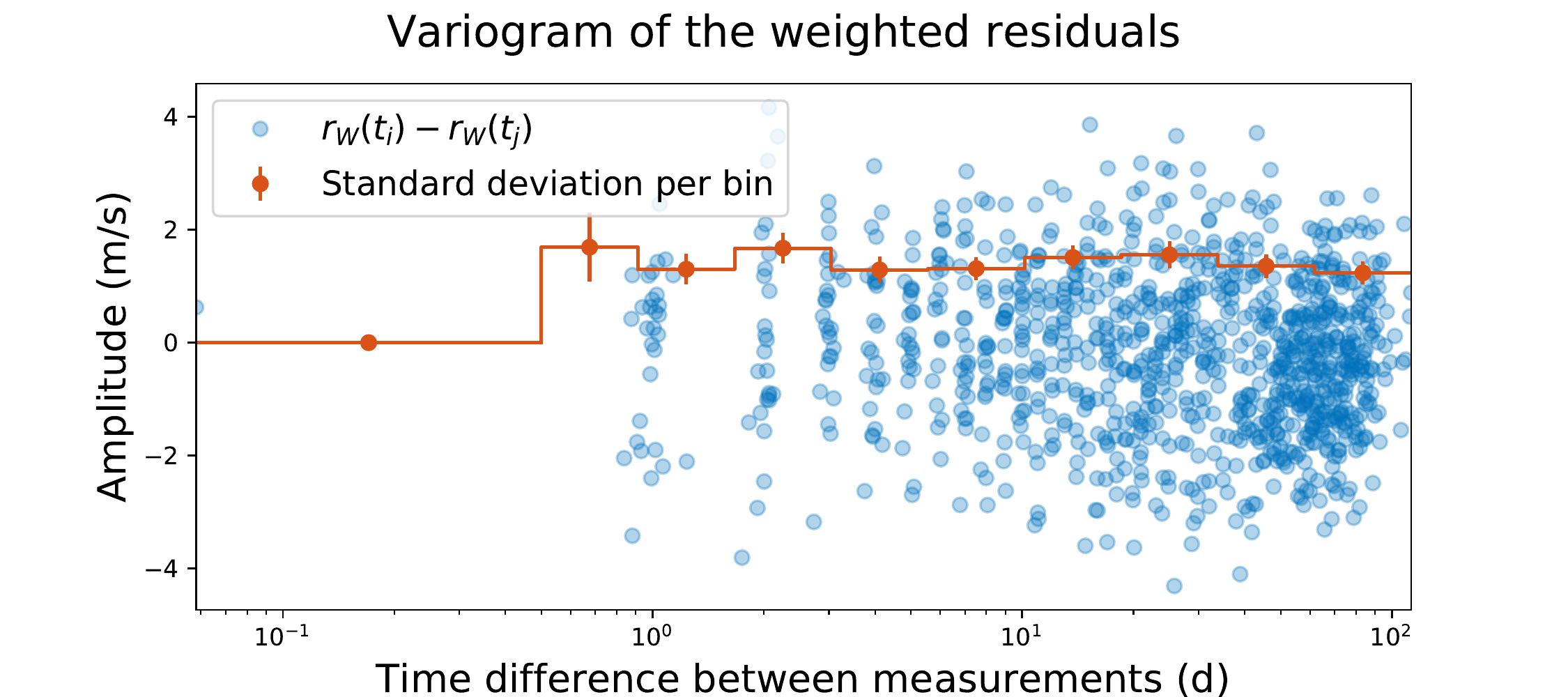}
    \caption{Histograms of the weighted residuals of the maximum likelihood model (top) and their variogram (bottom) for activity modelled as two sinusoidal functions (model 1). }
    \label{fig:weighted_rvres}
\end{figure}

Modelling errors might leave a trace in the residuals. In~\cite{Hara2019}, it is shown that if the model used for the analysis is correct, the residuals of the maximum likelihood model, appropriately weighted, should follow a normal distribution and not exhibit correlations. In Fig.~\ref{fig:weighted_rvres}, we show the histogram of the residuals (in blue) and the probability distribution function of a normal variable. The two appear to be in agreement. We further computed the Shapiro-Wilk normality test~\citep{Shapirowilk1965} and found a $p$-value of 0.78, which is compatible with normality. The variogram does not exhibit signs of correlations. The same analysis was performed with model 2, for which the residuals also exhibit neither non-normality nor correlation ($p$-value of 0.99 on the residuals). We conclude that both models are compatible with the data.  

\begin{table}

\caption{Mass estimation with activity model as two sinusoids, $H \alpha$ model, and correlated noise.}
\label{table:massRV}
\centering
\begin{tabular}{ccc}
Parameter & Estimates (1) & Estimates (2)  \\ \hline \hline
\multicolumn{3}{c}{Planets}\  \\ \hline \hline
\multicolumn{3}{c}{TOI 178b (1.9 days) }\\\hline
$K$ [m/s] &  $1.05^{+0.11}_{-0.11}$ &  $1.04^{+0.27}_{-0.28}$ \\ 
$m$ [$M_\oplus$] & $1.52^{+0.18}_{-0.22}$ & $1.49^{+0.40}_{-0.43}$\\ 
$\rho$ [$\rho_\oplus$]  & $0.99^{+0.26}_{-0.20}$ & $0.97^{+0.36}_{-0.30}$\\ \hline
\multicolumn{3}{c}{TOI 178c (3.2 days)}  \\\hline      
$K$ [m/s] &  $2.83^{+0.15}_{-0.13}$  & $2.72^{+0.27}_{-0.28}$ \\ 
$m [M_\oplus]$ &  $4.88^{+0.43}_{-0.47}$ & $4.67^{+0.62}_{-0.58}$\\
$\rho$ [$\rho_\oplus$] & $1.04^{+0.25}_{-0.21}$ & $0.99^{+0.26}_{-0.20}$ \\ \hline
\multicolumn{3}{c}{TOI 178d (6.5 days)}\\ \hline   
$K$ [m/s] &  $1.53^{+0.18}_{-0.20}$  & $1.24^{+0.33}_{-0.29}$ \\ 
$m [M_\oplus]$ &  $3.33^{+0.45}_{-0.53}$ & $2.70^{+0.71}_{-0.71}$\\   
$\rho$ [$\rho_\oplus$] &  $0.197^{+0.04}_{-0.03}$ & $0.158^{+0.047}_{-0.042}$ \\ \hline
\multicolumn{3}{c}{TOI 178e  (9.9 days)}\\ \hline      
$K$ [m/s] &  $1.38^{+0.19}_{-0.15}$  &  $1.72^{+0.31}_{-0.35}$\\ 
$m [M_\oplus]$ &  $3.44^{+0.53}_{-0.49}$ & $4.28^{+0.82}_{-0.96}$ \\
$\rho$ [$\rho_\oplus$] &  $0.32^{+0.07}_{-0.06}$ &  $ 0.4^{+0.103}_{-0.090}$ \\ \hline
\multicolumn{3}{c}{TOI 178f (15.2 days)}\\ \hline   
$K$ [m/s] &  $2.49^{+0.25}_{-0.24}$ &  $2.88^{+0.34}_{-0.37}$ \\ 
$m [M_\oplus]$ &  $7.18^{+0.89}_{-0.93}$ & $8.28^{+1.11}_{-1.34}$ \\  
$\rho$ [$\rho_\oplus$] &  $0.60^{+0.13}_{-0.10}$& $0.69^{+0.16}_{-0.13}$\\ \hline
\multicolumn{3}{c}{TOI 178g (20.7 days)}\\ \hline   
$K$ [m/s] &  $1.29^{+0.20}_{-0.21}$  & $1.19^{+0.42}_{-0.47}$ \\ 
$m$ [$M_\oplus$] &  $4.10^{+0.73}_{-0.73}$  & $3.78^{+1.47}_{-1.45}$\\ 
$\rho$ [$\rho_\oplus$] &  $0.173^{+0.044}_{-0.037}$& $0.159^{+0.073}_{-0.061}$ \\ \hline \hline
\multicolumn{3}{c}{Other signals)}\\ \hline \hline
\multicolumn{3}{c}{Signal at $\approx$ 40 days (probably $P_{\mathrm{rot}}$)} \\ \hline 
$P$ [days] &  $39.4^{+1.09}_{-3}$& -  \\ 
$K$ [m/s] &  $3.01^{+0.35}_{-0.36}$ & - \\   
\hline
\multicolumn{3}{c}{Activity signal at 16 days}\\ 
\hline
$P$ [days] &  $16.2^{+0.28}_{-0}$& - \\ 
$K$ [m/s] &  $1.07^{+0.44}_{-0.40}$ & - \\ 
\hline
\multicolumn{3}{c}{Noise parameters}\\ 
\hline
$\sigma_W$ [m/s] &  $0.51^{+0.13}_{-0.11}$ & $0.90^{+0.26}_{-0.25}$   \\ 
$\sigma_R$ [m/s]    &  $0.75^{+0.22}_{-0.11}$  & - \\ 
$\tau_{R}$   [days]  & $5.74^{+4.03}_{-5.12}$ & - \\   
$\sigma_{QP}$ [m/s]  & - & $2.90^{+0.99}_{-1.47}$ \\   
$\tau_{QP}$  [days]  & - & $350^{+192}_{-337}$\\  
$P$ [days]   & - & $42.7^{+3.27}_{-3.72}$\\  
\end{tabular}
\label{tab:rvsimp}
\end{table}

\section{Continuation of a Laplace resonant chain}
\label{ap:Px}
\begin{figure}
  \includegraphics[width = 9cm]{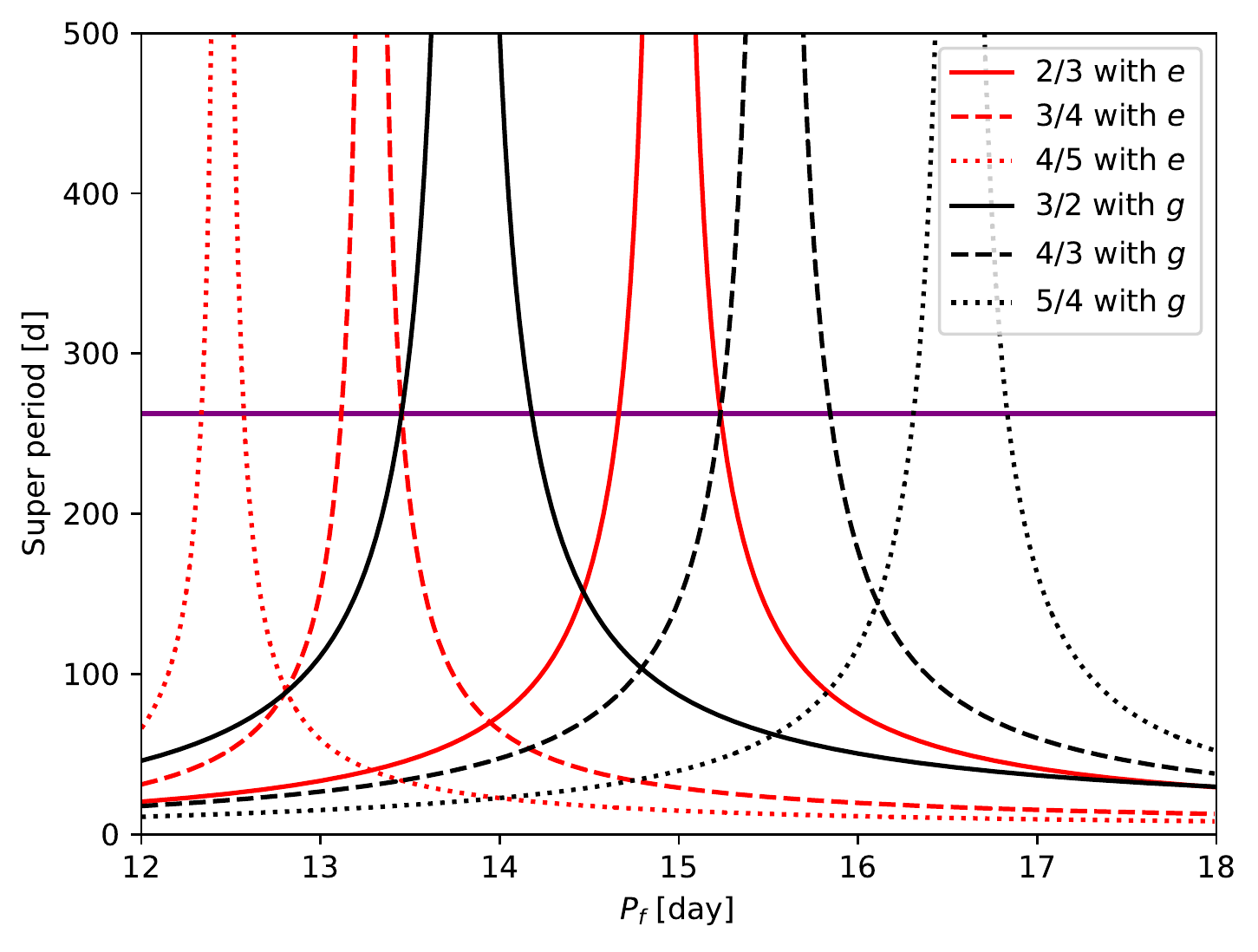}
   \caption{Super-period of planet $f$ with respect to $e$ (in red) and $g$ (in black) for a set of first-order MMRs. The purple line indicates {the value of the super-period ($260$ day) between the other pairs of the chain}. Only two possible periods allowed planet $f$ to be part of the resonant chain, which correspond to the near-intersection of the right-hand side of a red curve, the left-hand side of a black curve, and the horizontal purple line.   }
  \label{fig:Pgineq}
\end{figure}

TOI-178 is in a configuration where successive pairs of planets are at the same distance to the exact neighbouring first-order MMR. Generalising the configuration a bit, we define fictive planets $1$, $2,$ and $3$ such that planets $1$ and $2$ are close to the resonance $(k_1+q_1) : k_1$ and planets $2$ and $3$ are close to the resonance $(k_2+q_2) : k_2$, where $k_i$ and $q_i$ are integers. We hence write the near-resonant angles:
\begin{equation}
\begin{aligned}
\varphi_1  &=  k_1 \lambda_1 - (k_1+q_1)  \lambda_2\, ,\\
\varphi_2  &=   k_2 \lambda_2 - (k_2+q_2)  \lambda_3\, ,
\end{aligned}
\label{eq:fictive_neares}
\end{equation}
where $\lambda_i$ is the mean longitude of planet $i$. The associated distances to the resonances read:
\begin{equation}
\begin{aligned}
\Delta_1  &=   k_1 n_1 - (k_1+q_1)  n_2 \, ,\\
\Delta_2  &=   k_2 n_2 - (k_2+q_2)  n_3 \, ,
\end{aligned}
\label{eq:fictive_neares}
\end{equation}
where $n_i$ is the mean motion of planet $i$. A Laplace relation exists between these three planets if:
\begin{equation}
\begin{aligned}
j_1 \Delta_1 - j_2 \Delta_2 \approx 0\, ,
\end{aligned}
\label{eq:LapRel}
\end{equation}
where $j_1$ and $j_2$ are integers. In addition, the invariance by rotation of the Laplace angle (D'Alembert relation) gives:
\begin{equation}
\begin{aligned}
j_1 q_1 = j_2 q_2 \, .
\end{aligned}
\label{eq:DAlembert}
\end{equation}
As a result, the Laplace relation requires:
\begin{equation}
\begin{aligned}
n_3 \approx \frac{k_2 n_2- \frac{q_2}{q_1}\Delta_1}{k_2+q_2}\, ,
\end{aligned}
\label{eq:fictive_neares}
\end{equation}
which translates, for the period of the third planet, as:
\begin{equation}
\begin{aligned}
P_3 \approx \frac{k_2+q_2}{\frac{k_2}{P_2}-\frac{q_2}{q_1P_{1,2}}}\, ,
\end{aligned}
\label{eq:Periodx_theo}
\end{equation}
where $P_{1,2}$ is the super-period associated with $\Delta_1$. 
From this we can compute the periods of potential additional planets that would continue the Laplace resonant chain of TOI-178. Taking planets $f$ and $g$ as planets $1$ and $2$, the formula to compute the possible period of a planet $x$ that could continue the resonant chain becomes:
\begin{equation}
P_x=\frac{k+q}{\frac{k}{P_f}-\frac{q}{P_{f,g}}}\, ,
\label{eq:Periodx}
\end{equation}
where $P_{f,g}$ is the super-period between the near first-order resonances of the known chain defined by Eq. (\ref{eq:superperiod}), here $\sim 260\,$ days, and $k$ and $q$ are integers such that planet $x$ and $f$ are near a $(k+q)/k$ MMR. Some of the relevant periods are displayed in Table \ref{table:Px}. A similar computation allowed us to determine the possible period of planet $f$ prior to its confirmation by CHEOPS (see Fig. \ref{fig:Pgineq}).

%
%
%

\section{Internal structure}
\label{ap:internal_structure}

\begin{figure}
  \includegraphics[width = \linewidth]{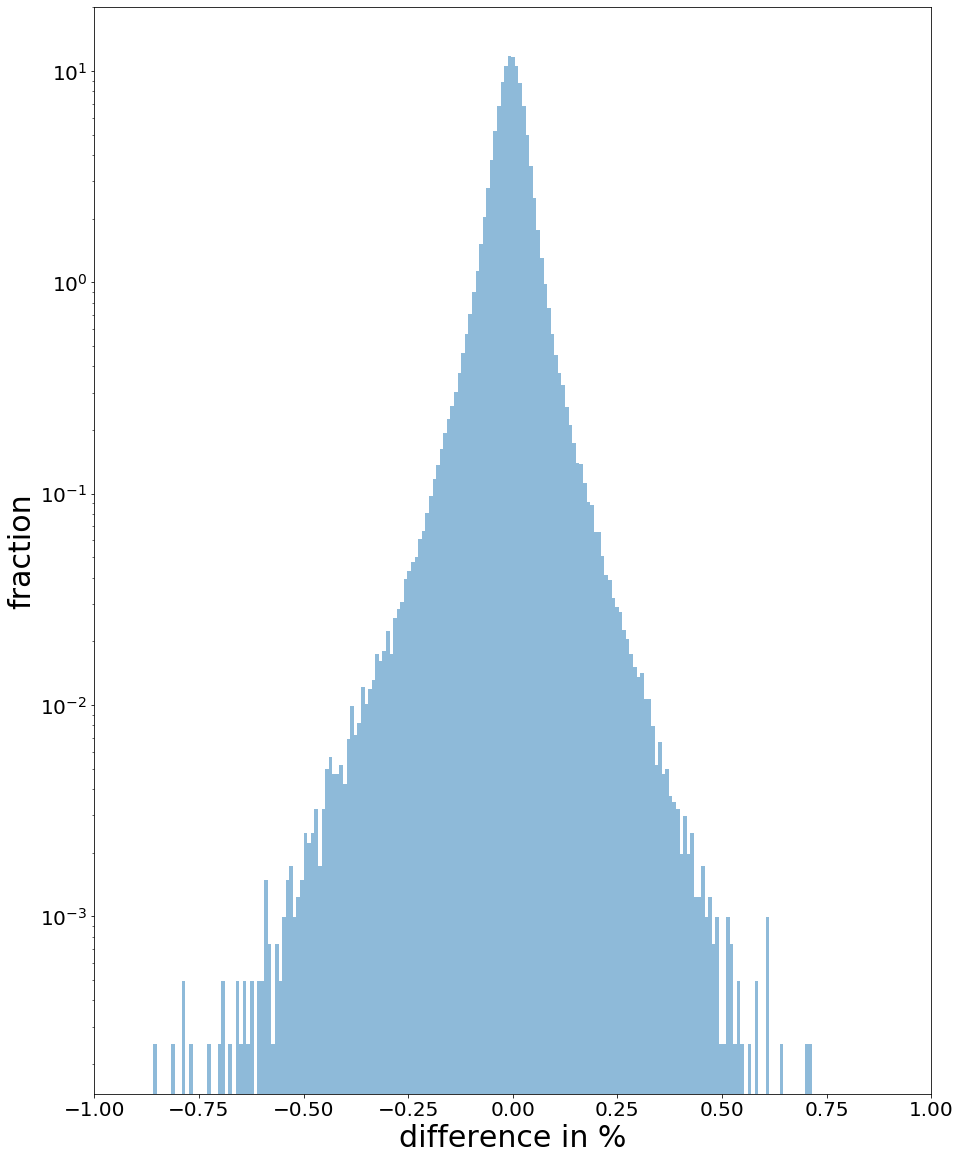}
   \caption{Histogram of the prediction error from our DNN on the test set. The y-axis is in log scale, and the x-axis covers an error from -1$\%$ to 1$\%$.}
  \label{fig:performancesDNN}
\end{figure}

We provide here the posterior distributions of the interior planetary models, as well as some more details on their calculations.
As mentioned in the main text, deriving the posterior distribution of the internal structure parameters requires computing the radius of planets millions of times  for different sets of parameters. In order to speed up this calculation, we first computed a large (5 million-point) database of internal structure models, varying the different parameters. This database was split randomly into three sets: one training set (80$\%$ of the models), one validation set, and one test set (the last two each containing 10$\%$ of the whole database). We then, in a second time, fitted a deep neural network (DNN) in order to be able to compute the radius of a planet very rapidly with a given set of internal structure parameters. The architecture of the DNN we used is made of six layers of 2048 nodes each, and we used the classical rectified linear unit (ReLU) as an activation function \citep{AlibertVenturini19}. The DNN is trained for a few hundred epochs, using a learning rate that is progressively reduced from $1.e-2$ down to $1.e-4$. Our DNN allowed us to compute these radii with an average error below 0.25$\%$  and with an increased rapidity of many orders of magnitude (a few thousand models computed per second). Figure \ref{fig:performancesDNN} shows the prediction error we reach on the test set (which was not used for training). The error on the predicted radius is lower than 0.4$\%$ -- an error much lower than the uncertainty on the radii we obtained for the TOI-178 planets -- in 99.9$\%$ of the cases. It is finally important to note that this model does not include the compression effect that would be generated by the gas envelope onto the solid part. Given the mass of the gas envelope in all planets, this approximation is justified.

The following plots show the posterior distribution of the interior structure parameters. The parameters are: the inner core, mantle, and water mass fraction relative to the mass of the solid planet; the Fe, Si, and Mg molar fraction in the mantle; the Fe molar fraction in the core; and the mass of gas (log scale). It is important to remember that since the core, mantle, and water mass fractions add up to one, they are not independent. This is also the case for the Si, Mg, and Fe molar fraction in the mantle. 

\begin{figure}
  \includegraphics[width = \linewidth]{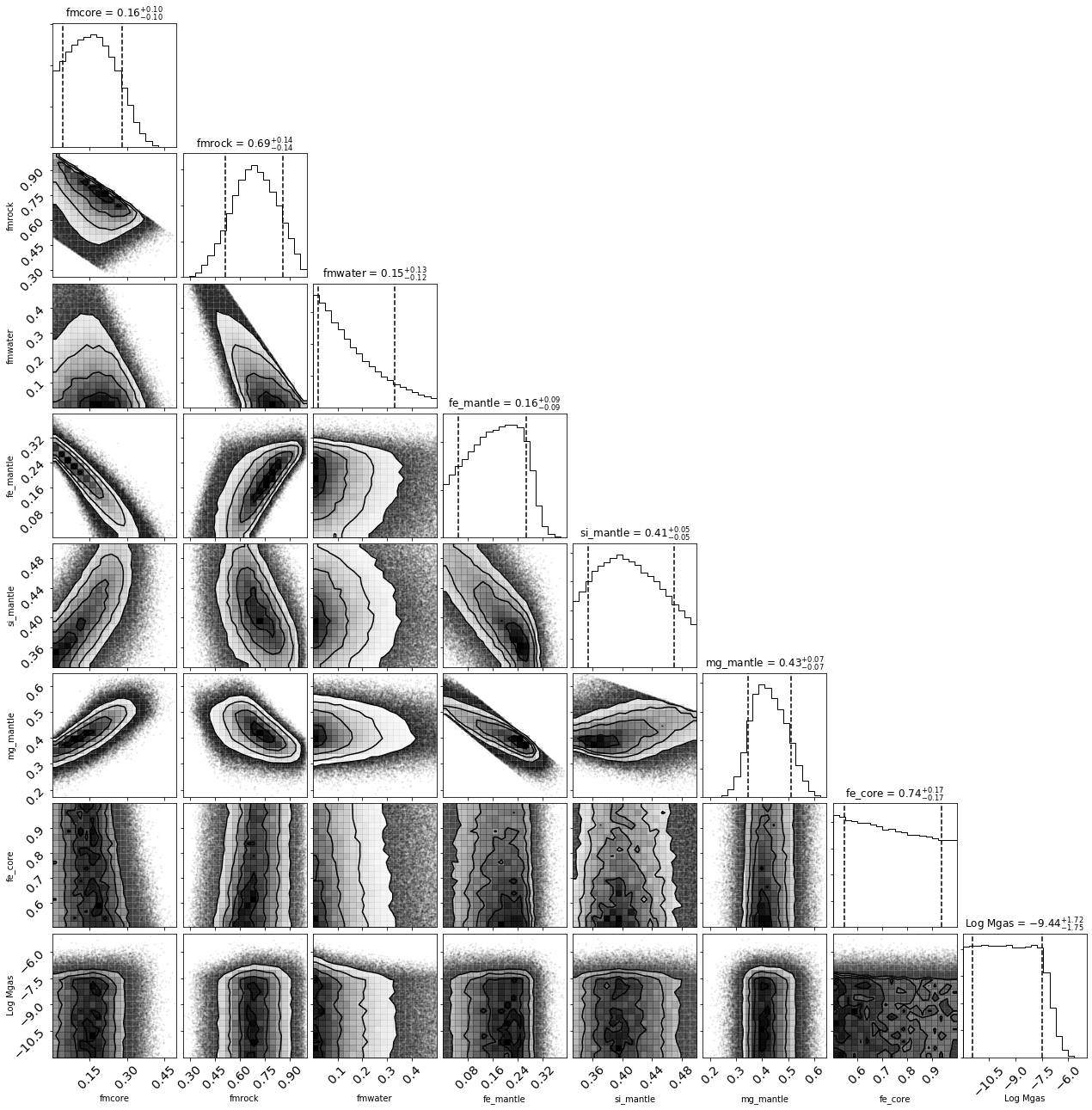}
   \caption{Corner plot showing the main parameters of the internal structure of planet $b$. The parameters are: the core mass fraction; the mantle mass fraction; the water mass fraction (all relative to the solid planet); the molar fractions of Fe, Si, and Mg in the mantle; the molar fraction of Fe in the core; and the mass of gas (log scale). The dashed lines give the positions of the $16\%$ and $84 \%$ quantiles, and the number at the top of each column gives the median and the 16\% and 84 \% quantiles. }
  \label{fig:cornerb}
\end{figure}

\begin{figure}
  \includegraphics[width = \linewidth]{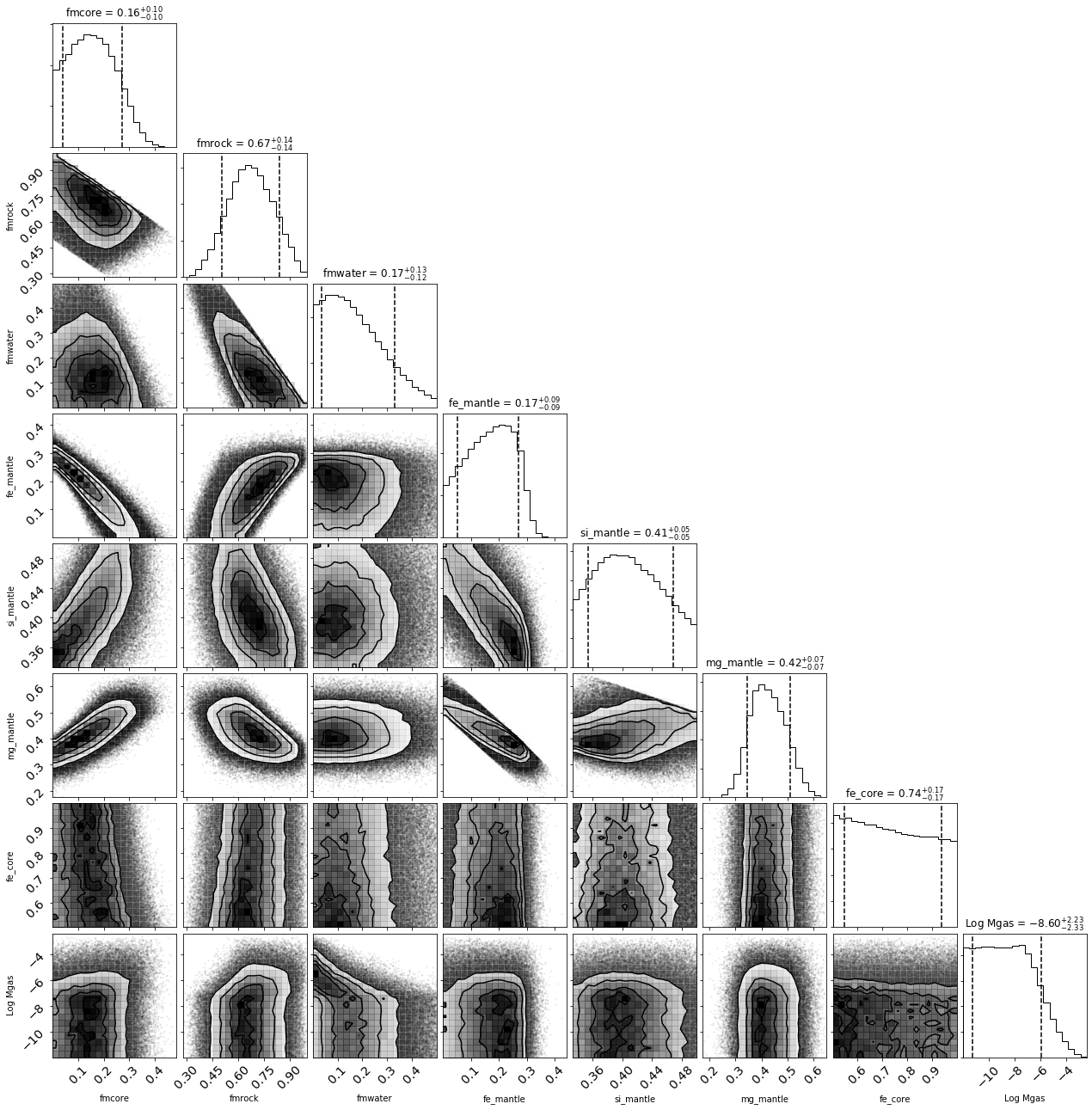}
   \caption{Same as Fig. \ref{fig:cornerb} but for planet $c$.}
  \label{fig:cornerc}
\end{figure}

\begin{figure}
  \includegraphics[width = \linewidth]{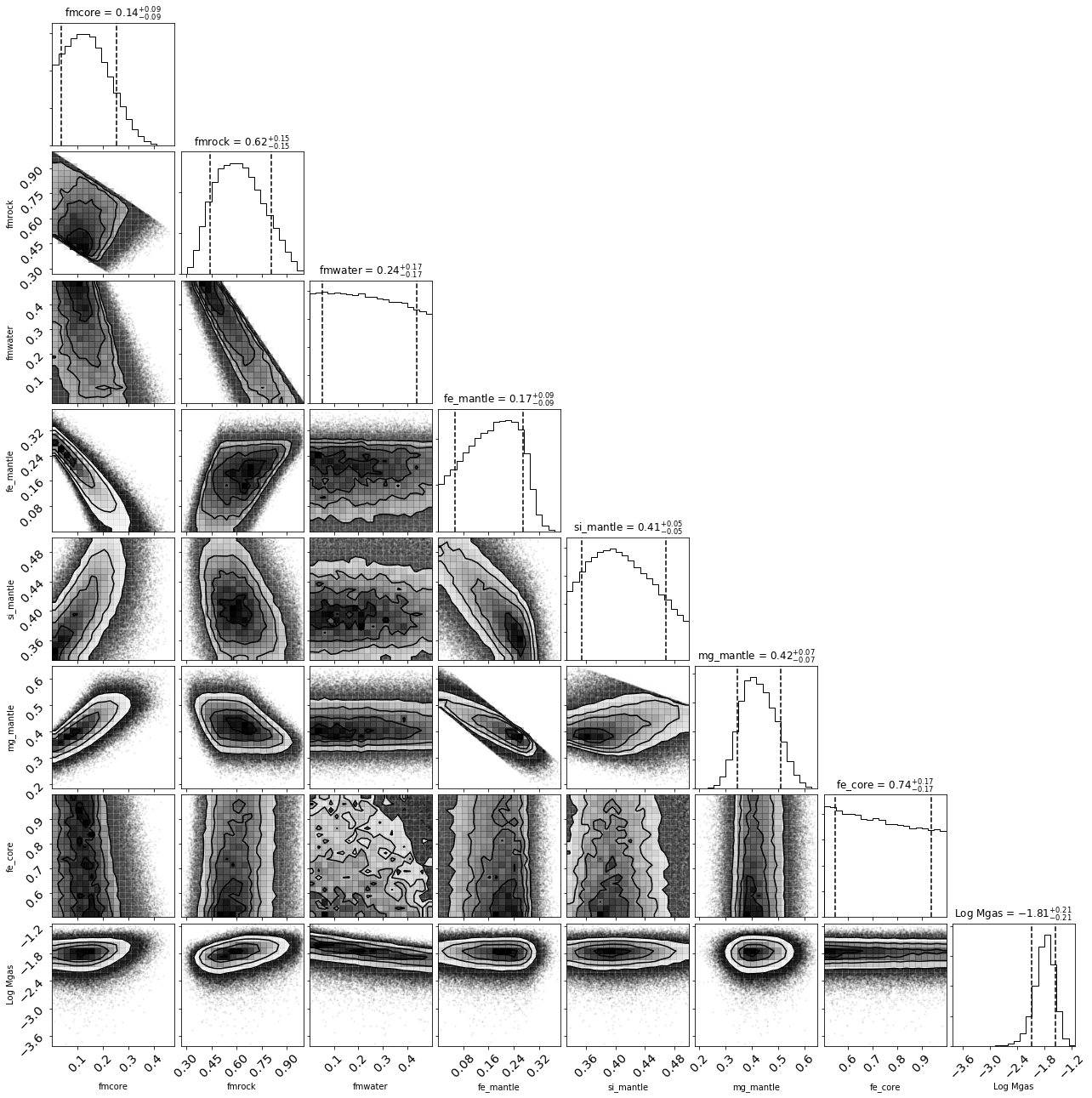}
   \caption{Same as Fig. \ref{fig:cornerb} but for planet $d$.}
  \label{fig:cornerd}
\end{figure}

\begin{figure}
  \includegraphics[width = \linewidth]{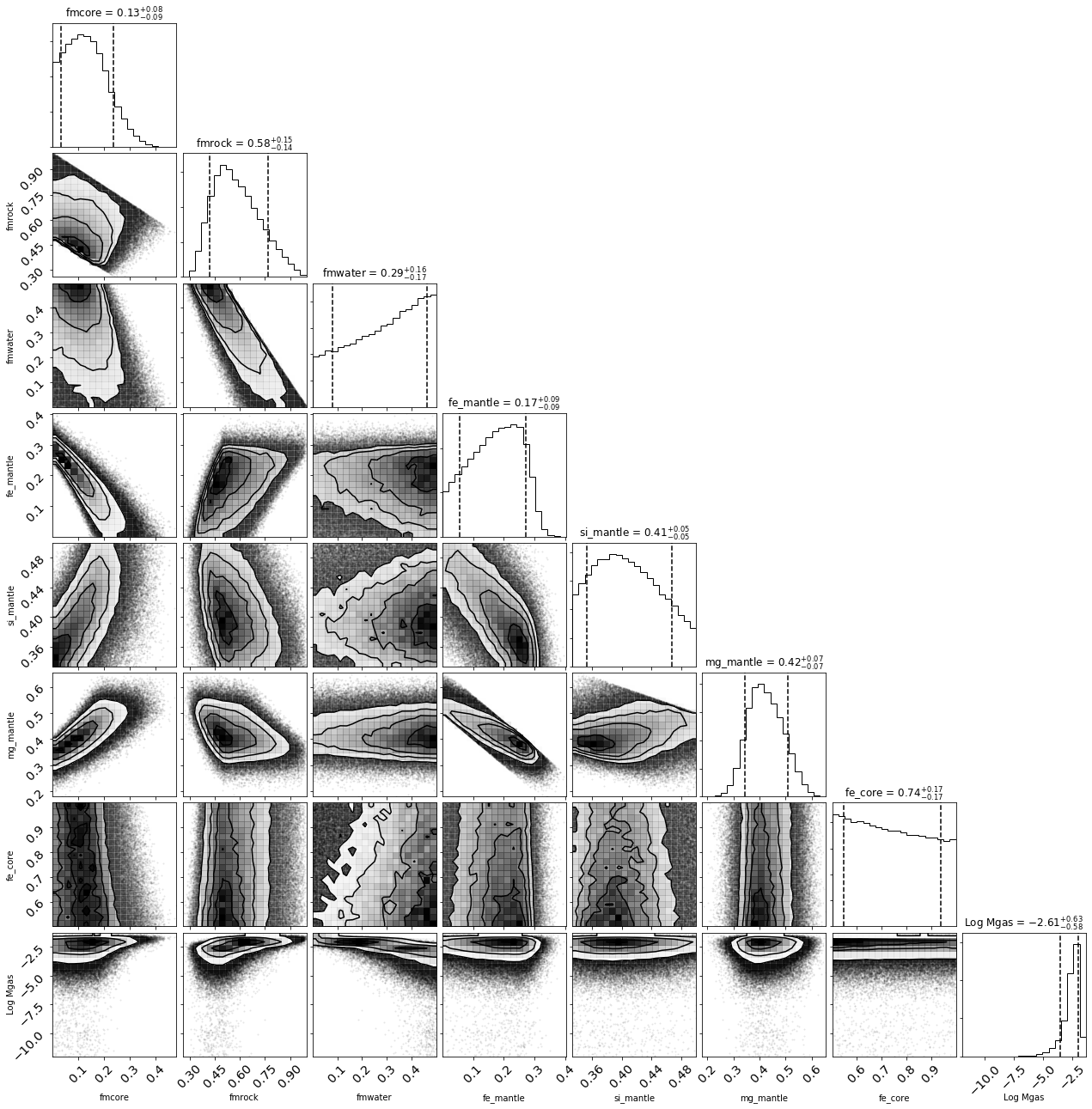}
   \caption{Same as Fig. \ref{fig:cornerb} but for planet $e$.}
  \label{fig:cornere}
\end{figure}

\begin{figure}
  \includegraphics[width = \linewidth]{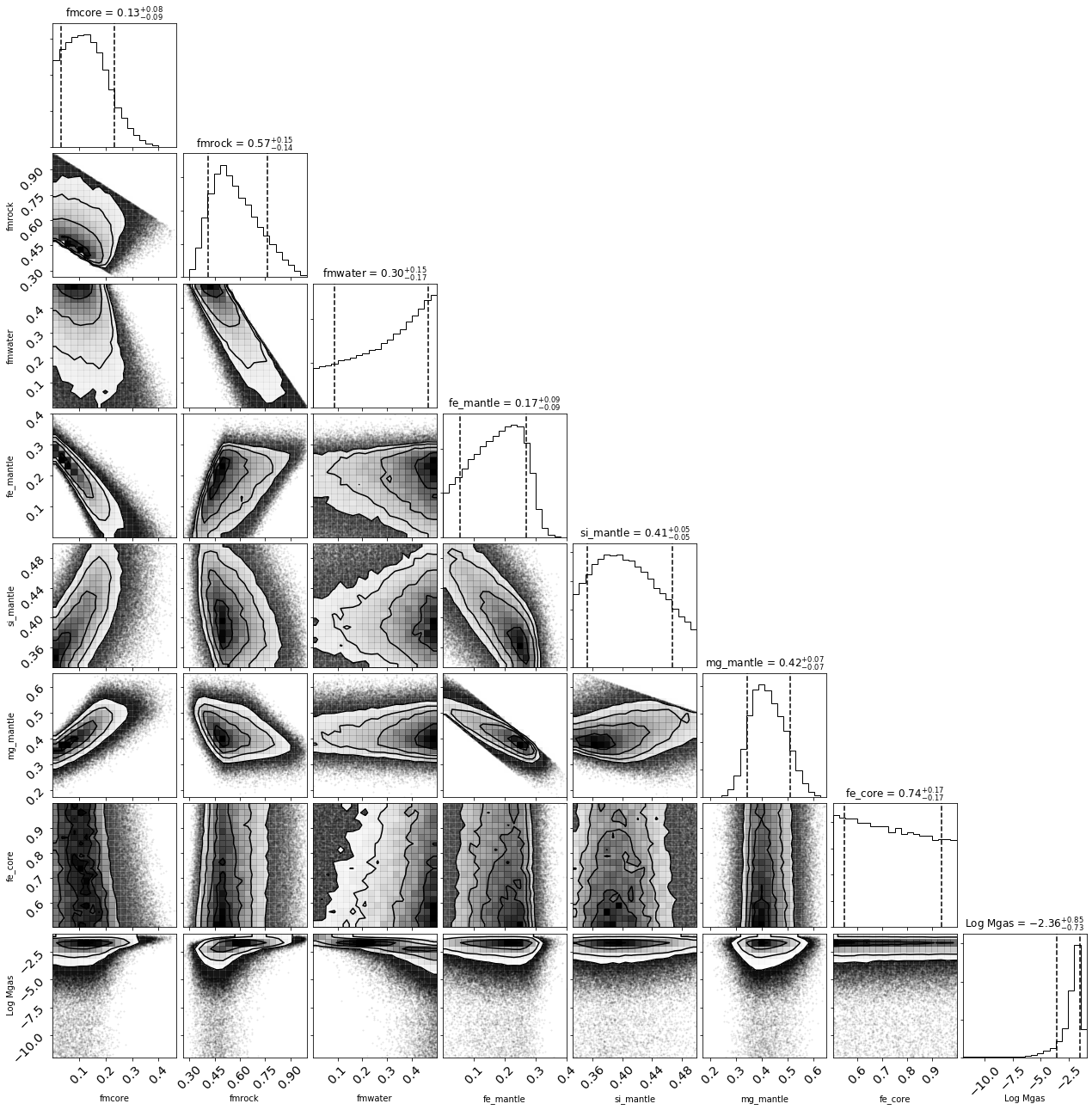}
   \caption{Same as Fig. \ref{fig:cornerb} but for planet $f$.}
  \label{fig:cornerf}
\end{figure}

\begin{figure}
  \includegraphics[width = \linewidth]{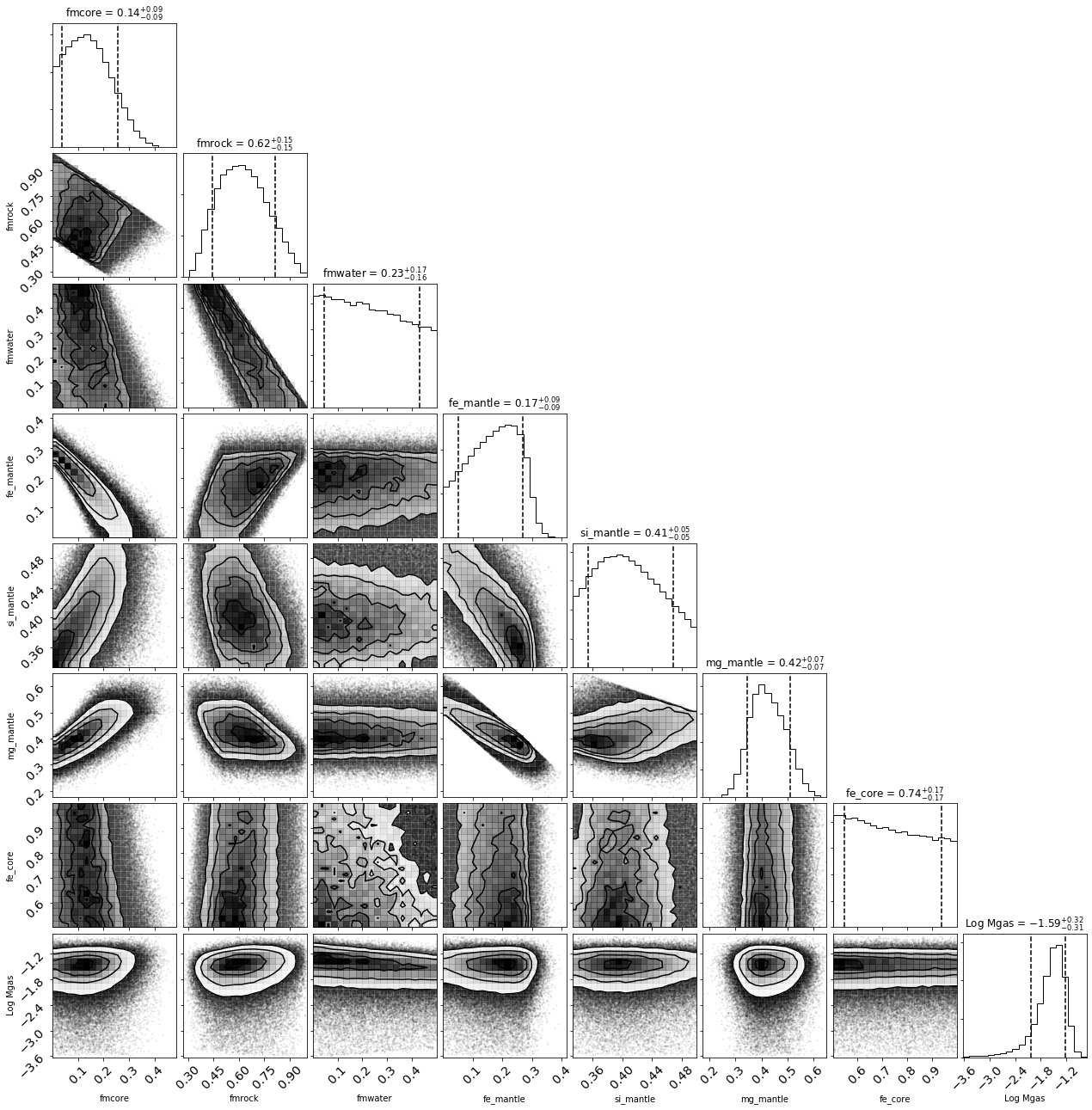}
   \caption{Same as Fig. \ref{fig:cornerb} but for planet $g$.}
  \label{fig:cornerg}
\end{figure}

\end{document}